\documentclass[aps,prd,reprint,superscriptaddress,nofootinbib,longbibliography,floatfix]{revtex4-2}

\usepackage{amsmath,amssymb,bm}
\usepackage{graphicx}
\usepackage{booktabs}
\usepackage{siunitx}
\usepackage{xcolor}
\usepackage{hyperref}
\usepackage{microtype}
\usepackage{multirow}
\usepackage{enumitem}
\hypersetup{colorlinks=true,linkcolor=blue,citecolor=blue,urlcolor=blue}
\newcommand{\GeV}{\mathrm{GeV}}
\newcommand{\TeV}{\mathrm{TeV}}
\newcommand{\keV}{\mathrm{keV}}
\newcommand{\kms}{\mathrm{km\,s^{-1}}}
\newcommand{\cmsq}{\mathrm{cm}^{2}}
\newcommand{\cmcubeds}{\mathrm{cm}^{3}\,\mathrm{s}^{-1}}

\begin{document}

\title{Testing Higgs-Coupled Minimal Dark Matter with Solar Neutrinos after the LZ High-Recoil Event}

\author{Mattia Di Mauro}
\email{dimauro.mattia@gmail.com}
\affiliation{Istituto Nazionale di Fisica Nucleare, Sezione di Torino, Via P. Giuria 1, 10125 Torino, Italy}

\date{\today}

\begin{abstract}
The LUX-ZEPLIN (LZ) Collaboration has reported a nuclear-recoil candidate at $248\,\keV$, motivating interpretations in terms of endothermic dark matter (DM). We investigate this signal in Higgs-coupled minimal dark matter (HC-MDM), where the same electroweak interaction responsible for the terrestrial recoil also induces DM capture in the Sun and a potentially observable high-energy neutrino signal.
We independently determine the mass splittings required to reproduce the LZ event and perform an improved calculation of Solar capture and post-capture evolution. Our treatment includes natural isotope mixtures, exact inelastic kinematics, nuclear form factors, finite-temperature nuclear motion near kinematic closure, orbital cooling through loop-induced elastic scattering, and an explicit capture--annihilation equilibrium test. The resulting Solar predictions are numerically stable and robust against the astrophysical and nuclear variations considered.
For the two thermal benchmarks within the released IceCube DM mass grid, $3_M2_D$ and $5_M4_D$, the predicted annihilation rates exceed the mass-matched $W^+W^-$ IceCube response by factors of approximately $2.2\times10^3$ and $91$, placing both standard-halo solutions under strong tension within the $W^+W^-$ response mapping used here. For the heavier representations, whose masses lie above the published $10\,\TeV$ IceCube grid, we construct a response-based extrapolation to higher masses. The $22\,\TeV$ $7_M6_D$ benchmark remains in strong tension with IceCube even under our conservative response-loss extrapolation, while the $48\,\TeV$ $9_M8_D$ benchmark also remains above the extrapolated response, although much closer to the sensitivity boundary. The $82\,\TeV$ $11_M10_D$ benchmark is response dependent, whereas the $130\,\TeV$ $13_M12_D$ benchmark remains below the extrapolated IceCube sensitivity.
\end{abstract}

\maketitle

\section{Introduction}
\label{sec:intro}

A wide range of astrophysical and cosmological observations requires a non-luminous, non-baryonic dark matter (DM) component.  Evidence comes from galactic and cluster dynamics, gravitational lensing, the cosmic microwave background, and the growth of large-scale structure, while precision cosmology fixes the cosmological DM abundance with high accuracy \cite{BertoneHooperSilk2005,BertoneHooper2018,Planck2020,CirelliStrumiaZupan2024}.  The microscopic identity of DM nevertheless remains unknown.  No Standard Model (SM) particle has the required combination of stability, abundance, electric neutrality, and sufficiently cold cosmological evolution, making particle DM one of the principal motivations for physics beyond the SM \cite{Jungman1996,Feng2010,Arcadi2018,Arcadi2025,CirelliStrumiaZupan2024}.

Weakly interacting massive particles (WIMPs) are among the best studied possibilities.  In the standard thermal freeze-out picture, an annihilation rate of approximately weak strength can naturally reproduce the measured abundance, while related interactions may be probed through direct detection, indirect detection, and collider searches \cite{Jungman1996,Feng2010,Arcadi2018,Arcadi2025}.  The continuing improvement of direct-detection limits has excluded large regions of simple WIMP parameter space and has increased the importance of models in which the scattering rate is suppressed by symmetry, kinematics, resonance structure, or a separation between the interaction controlling freeze-out and the portal probed in the laboratory \cite{PospelovRitzVoloshin2008,DiMauroArina2023,DiMauroXie2025,DiMauroWang2025,ShaikhDiMauro2026}.  
Endothermic inelastic scattering realizes a particularly sharp version of this idea: the incoming DM state must provide the mass splitting required to excite a heavier state, pushing the signal toward large recoil energies and the high-velocity tail of the halo, where the available DM population is strongly reduced and the expected direct-detection signal is therefore suppressed \cite{TuckerSmith2001,TuckerSmith2005,Bramante2016,DallaValleGarcia2025,Foguel2025}.

The recent LUX-ZEPLIN (LZ) analysis extending the nuclear-recoil energy window from approximately $5.4\,\keV$ to $270\,\keV$ makes this possibility especially timely \cite{LZ2026}. 
In an analysis using an exposure of $2.84$ tonne-years and reaching these high recoil energies, LZ reports one event reconstructed at
\begin{equation}
E_R=248\pm23\,(\mathrm{stat})\pm23\,(\mathrm{sys})\,\keV,
\label{eq:lzevent}
\end{equation}
with a maximum local significance of $3.4\sigma$ among the tested interactions and a global significance of $2.6\sigma$ after the look-elsewhere effect \cite{LZ2026}.  A single event is not evidence for a DM discovery.  Its unusually high recoil energy is nevertheless informative because ordinary elastic spin-independent (SI) scattering is strongly suppressed in this regime by the nuclear form factor, whereas endothermic scattering naturally generates a spectrum concentrated at finite recoil energy.

The LZ result has already generated a broad theoretical literature. Higgsino and electroweak interpretations have been studied in several supersymmetric and non-supersymmetric settings \cite{DiMauro2026,FanReece2026,FreeseTheodosopoulos2026,WuZhangZhu2026,DuWang2026,SmirnovGriffithBeacom2026}, while more general inelastic constructions include dark-photon and mixing-suppressed scenarios \cite{SuYangYang2026,Yamashita2026,LeeYoun2026}. The proximity of the preferred endothermic solutions to the terrestrial kinematic boundary also implies potentially dramatic annual modulation \cite{DiMauro2026,McCabe2026}. In addition, the continuation of the inelastic spectrum above the nominal signal region provides a high-energy sideband test \cite{RoddSafdiSlatyerXu2026}.

A particularly important complementary test is supplied by the Sun. DM particles falling into the Solar gravitational potential can reach speeds approaching $1400\,\kms$ in the core \cite{Gould1987a,NussinovWangYavin2009}.
Endothermic transitions that are strongly suppressed on Earth can therefore become efficient on heavy Solar nuclei.  Once captured, DM can lose additional orbital energy, concentrate near the Solar center, and annihilate into SM states whose decays generate high-energy neutrinos.  Searches for neutrinos from the direction of the Sun then constrain this process \cite{Gould1987a,Jungman1996,NussinovWangYavin2009,MenonMorrisPierceWeiner2010,BlennowClementzHerreroGarcia2016,Catena2018,IceCubeSolar2025}.  
Pospelov and Ramani recently emphasized the power of this argument for an inelastic thermal Higgsino and found that IceCube bounds \cite{IceCubeSolar2025} exclude splittings below approximately $566\,\keV$ in their efficiently thermalized setup \cite{PospelovRamani2026}.  A subsequent analysis showed explicitly that this conclusion is model dependent: a thermal Higgsino is excluded, whereas two-state annihilation kinetics or neutron-philic spin-dependent (SD) interactions can strongly suppress the Solar-neutrino signal, allowing the corresponding scenarios to remain compatible with the IceCube bounds \cite{DiMauroShaikh2026Solar}.  Related neutrino constraints on the Higgsino interpretation have also been obtained using Super-Kamiokande and IceCube data \cite{BoseEtAl2026}. Independently, Ref.~\cite{LeeYoun2026} emphasized a closely related kinematic point for fixed-coupling inelastic models: moving the terrestrial signal into an extreme halo tail does not automatically suppress Solar capture because gravitational acceleration in the Sun restores access to the endothermic transition.

The model studied in this work is Higgs-coupled minimal dark matter (HC-MDM), introduced as a generalized minimal electroweak-DM framework and recently applied to the LZ high-recoil event \cite{GriffithSmirnov2026,SmirnovGriffithBeacom2026}.  It contains a Majorana electroweak multiplet and an adjacent Dirac multiplet coupled through the Higgs field.  At a custodial point, the lightest neutral state has no tree-level diagonal Higgs coupling, while an off-diagonal $Z$ current generates an endothermic transition to two degenerate excited neutral states.  The inelastic interaction is representation independent, but the thermal relic mass and the loop-induced elastic cross section vary strongly with the electroweak representation.  
This produces six proposed LZ benchmarks, from a $3_M2_D$ model near $1.4\,\TeV$ to a $13_M12_D$ model near $130\,\TeV$ \cite{SmirnovGriffithBeacom2026}.  The same paper argues that Solar capture strongly constrains the two lightest Standard Halo Model (SHM) solutions but that faster halo components can rescue them, that $7_M6_D$ is borderline, and that the three largest representations survive even for an SHM velocity distribution.

It is also useful to contrast this scenario with standard Minimal Dark Matter (MDM) \cite{Cirelli:2005uq,Cirelli:2007xd}. In the usual real fermionic MDM candidates with $Y=0$, the neutral Majorana component has no tree-level coupling to the $Z$ boson, and SI scattering arises only through electroweak loops \cite{Hisano:2011cs,Bottaro:2021snn}. The resulting elastic signal is therefore strongly suppressed and does not naturally account for the high-energy LZ event considered here. Conversely, electroweak multiplets with nonzero hypercharge possess an unsuppressed tree-level coupling to the $Z$ boson and, in their minimal form, are excluded by direct-detection constraints \cite{Cirelli:2005uq}. Standard MDM is moreover increasingly constrained by complementary searches. Future large-exposure xenon experiments are expected to probe the loop-induced scattering rates of the real $Y=0$ electroweak multiplets, with exposures of order $50~{\rm tonne,yr}$ sufficient to test their predicted SI rates \cite{Bloch:2024wimps}. Even more strongly, a recent analysis of 14 years of Fermi-LAT observations of the inner Galaxy finds that thermal fermionic real MDM multiplets with $n<9$, including the wino, quintuplet, and septuplet, are strongly disfavored under the assumption that they constitute all of the DM and follow a standard cosmological history \cite{Safdi:2025mdm}. Explaining the LZ event therefore requires an additional ingredient beyond standard MDM. In HC-MDM, Higgs-induced mixing splits the neutral Dirac state into two Majorana states and converts the otherwise problematic neutral-current interaction into the endothermic transition $\chi_1 N\to\chi_2 N$, providing the kinematic structure required for the LZ event.

Here we independently reassess those Solar conclusions within the same HC-MDM framework considered in the original analysis, but using a more complete treatment of Solar capture and of the subsequent evolution of the captured DM population, following the methodology developed in Ref.~\cite{DiMauroShaikh2026Solar}: capture, post-capture evolution, and annihilation are treated as distinct physical steps and are only reduced to the usual equilibrium formula after the required assumptions have been verified.  Our calculation uses a sampled BS05(AGS,OP) Solar structure, a self-consistent escape-speed profile, natural isotope mixtures, finite-momentum nuclear form factors, exact endothermic kinematics, and finite-temperature nuclear velocities close to kinematic closure.  We explicitly test the initial-transit optical depth, radiative and collisional de-excitation, loop-induced elastic cooling, and capture-annihilation equilibrium.  For the comparison with IceCube bounds \cite{IceCubeSolar2025} we distinguish the two HC-MDM candidates that lie within the published $20\,\GeV$--$10\,\TeV$ signal-template grid from the four heavier candidates, for which only a high-mass diagnostic can presently be constructed from the public analysis \cite{IceCubeSolar2025}.

Our main results are as follows. The six LZ benchmark masses lie on the HC-MDM full-density thermal-relic trajectories by construction. The $3_M2_D$ and $5_M4_D$ SHM benchmarks have very large Solar capture rates and predict Solar annihilation rates far above the IceCube Solar-DM limits for a $W^+W^-$-like neutrino spectrum \cite{IceCubeSolar2025}. The four heavier thermal candidates, $7_M6_D$ through $13_M12_D$, have masses above the released $10\,\TeV$ IceCube DM-template grid, so no official confidence-level limit can be assigned directly. We nevertheless construct a response-matched high-mass continuation, anchored to the inferred $W^+W^-$ annihilation-rate sensitivity at the $10\,\TeV$ endpoint and using the weak logarithmic response slope of the published high-mass curve. This exploratory continuation leaves $7_M6_D$ strongly above the IceCube-equivalent sensitivity, places $9_M8_D$ above it by a factor of a few, makes $11_M10_D$ borderline, and leaves $13_M12_D$ below it. A deliberately conservative response-loss stress test does not change the robust conclusion for $7_M6_D$ and leaves $9_M8_D$ close to the boundary. For the proposed $\delta=570\,\keV$ high-velocity solutions, we find that finite-temperature effects in the Solar core substantially enhance capture. As a result, the lightest high-splitting trajectory remains significantly constrained by IceCube, whereas the heavier trajectory can remain compatible with the Solar-neutrino bounds.

The paper is organized as follows. Sec.~\ref{sec:model} reviews the HC-MDM field content, neutral-state mixing, direct-detection interaction, loop-induced elastic scattering, and annihilation channels. Sec.~\ref{sec:capture} develops the Solar-capture calculation. Sec.~\ref{sec:postcapture} studies de-excitation, orbital cooling, and annihilation equilibrium. Sec.~\ref{sec:icecube} defines our IceCube comparison and develops the response-matched high-mass continuation. Sec.~\ref{sec:results} gives the representation-by-representation results, including the high-velocity solutions, and we conclude in Sec.~\ref{sec:conclusions}. A detailed derivation of the HC-MDM model is given in Appendix~\ref{app:theory}; the finite-temperature capture kernel and post-capture orbital evolution are developed in Appendices~\ref{app:finiteT} and~\ref{app:orbitcooling}; and the numerical validation, isotope treatment, systematic variations, and optical-depth checks are collected in Appendix~\ref{app:validation}.

\section{Higgs-coupled minimal dark matter}
\label{sec:model}

\subsection{Particle content and custodial limit}

Minimal DM extends the SM by electroweak multiplets whose interactions are largely fixed by gauge invariance \cite{CirelliFornengoStrumia2006,CirelliStrumiaTamburini2007,CirelliGouttenoirePetrakiSala2019,Bottaro2022}. 
HC-MDM extends the electroweak dark sector by introducing a Majorana multiplet $M$ and a vectorlike Dirac multiplet $D$ in adjacent $SU(2)_L$ representations, allowing a renormalizable interaction with the Higgs field \cite{TaitYu2016,LopezHonorez2018,GriffithSmirnov2026,SmirnovGriffithBeacom2026}. A complete two-component formulation, including the gauge representations, Yukawa contractions, general neutral mass matrix, custodial diagonalization, and broken-phase currents used below, is given in Appendix~\ref{app:theory}. The two multiplets transform under $SU(2)_L\times U(1)_Y$ as
\begin{equation}
M\sim(2n+1,0),\qquad D\sim(2n,1/2),
\end{equation}
where $n$ is a positive integer. The first entry denotes the dimension of the $SU(2)_L$ representation and the second the weak hypercharge $Y$, with electric charge defined by $Q=T_3+Y$. Equivalently, the two multiplets carry weak isospins $j_M=n$ and $j_D=n-1/2$. The HC-MDM sequence considered in this work is therefore
\begin{equation}
3_M2_D,\ 5_M4_D,\ 7_M6_D,\ 9_M8_D,\ 11_M10_D,\ 13_M12_D.
\end{equation}

Both multiplets contain an electrically neutral component. The Majorana multiplet contains a neutral state with $T_3=0$ together with charged components satisfying $Q=T_3$, while the Dirac multiplet contains neutral Weyl components with $T_3=\mp1/2$ and a corresponding tower of charged partners. These charged states play an important role in coannihilation, Sommerfeld enhancement, bound-state formation, electroweak annihilation, and loop-induced processes. Their gauge interactions and their role in the full HC-MDM spectrum are summarized in Appendix~\ref{app:theory}.

The fermionic Lagrangian relevant for the neutral sector is \cite{SmirnovGriffithBeacom2026}
\begin{align}
\mathcal L\supset{}&
\overline D\left(i\gamma^\mu\mathcal D_\mu-m_D\right)D
+\frac{1}{2}\overline M\left(i\gamma^\mu\mathcal D_\mu-m_M\right)M
\\
&-y_1DMH^{\ast}-y_2\overline D M H+\mathrm{h.c.}.
\label{eq:hcmdm_lagrangian}
\end{align}
Here $\mathcal D_\mu$ denotes the SM electroweak covariant derivative, $m_D$ and $m_M$ are the respective gauge-invariant masses, $H$ is the SM Higgs doublet, and $y_1$ and $y_2$ are the two Yukawa couplings allowed by the adjacent representations. We work at the custodial point adopted in the LZ interpretation,
\begin{equation}
m_M=m_D\equiv m_\chi,\qquad y_1=-y_2\equiv y.
\label{eq:custodial}
\end{equation}
The condition $y_1=-y_2$ selects the custodial direction relevant for the convention used here, while $m_M=m_D$ further simplifies the neutral-state spectrum. At this point the lightest neutral state is an exact unmixed Dirac-sector combination whose tree-level diagonal Higgs coupling vanishes. Higgs-mediated elastic scattering is therefore absent at tree level; the leading elastic SI signal is generated by electroweak loops, whereas electroweak symmetry breaking generates the neutral-state splitting and the off-diagonal interaction relevant for endothermic scattering \cite{TaitYu2016,LopezHonorez2018,Hisano2011,ChenHill2020,GriffithSmirnov2026}. The symmetry structure and the relation between this statement and the general pre-custodial mass matrix are derived explicitly in Appendix~\ref{app:theory}.

\subsection{Neutral dark-matter spectrum and the endothermic transition}

Each electroweak multiplet contains several components with different electric charges. Since present-day Galactic DM must be electrically neutral, the relevant states for direct detection are the neutral components of the Majorana and Dirac multiplets. The charged partners play an important role in annihilation and radiative processes. Along the benchmark trajectories considered here the cosmologically stable state is the neutral eigenstate $\chi_1$; the charged components belong to the coannihilating electroweak spectrum and subsequently decay into the neutral sector. The charged-current structure and the stability assumption are discussed in Appendix~\ref{app:theory}.

The Majorana multiplet $M$ contains one neutral Weyl component, which we denote simply by $M$. The Dirac multiplet can be written in terms of two Weyl multiplets with opposite hypercharge, and correspondingly contains two neutral Weyl fields, denoted by $\psi^0$ and $\widetilde\psi^0$. Before electroweak symmetry breaking these two fields combine into the neutral component of the Dirac fermion. It is convenient to introduce the combinations
\begin{equation}
D_{\pm}=
\frac{\psi^0\pm\widetilde\psi^0}{\sqrt{2}}.
\end{equation}

After electroweak symmetry breaking, the Higgs interaction mixes the neutral component of the Majorana multiplet with one linear combination of the neutral Dirac fields. At the custodial point defined in Eq.~\eqref{eq:custodial}, the neutral mass matrix takes a particularly simple form. The general mass matrix before imposing the custodial relations, including the representation-independent neutral Clebsch--Gordan coefficient, is derived in Appendix~\ref{app:theory}. In the basis $(M,D_-,D_+)$ it is
\begin{equation}
\mathcal M_0=
\begin{pmatrix}
m_\chi & yv/\sqrt{2} & 0\\
yv/\sqrt{2} & -m_\chi & 0\\
0 & 0 & m_\chi
\end{pmatrix},
\label{eq:massmatrix}
\end{equation}
where $v=246\,\GeV$ is the Higgs vacuum expectation value. The combination $D_+$ does not mix with $M$ and therefore remains an eigenstate with mass $m_\chi$. This state is the lightest neutral particle and constitutes the DM candidate; we denote it by
\begin{equation}
\chi_1\equiv D_+.
\end{equation}

The remaining neutral states arise from the mixing of $M$ and $D_-$. The corresponding two eigenvalues have equal magnitude,
\begin{equation}
M_{\ast}=
\sqrt{m_\chi^2+\frac{1}{2}y^2v^2},
\end{equation}
with opposite signs at the level of the Majorana mass matrix. After the standard field redefinition that renders both physical masses positive, these states form two degenerate excited Majorana fermions, which we denote by $\chi_2$ and $\chi_3$, with
\begin{equation}
m_{\chi_2}=m_{\chi_3}=M_{\ast}.
\end{equation}

The physical DM spectrum therefore consists of a light neutral state $\chi_1$ and two heavier neutral states $\chi_2$ and $\chi_3$. Their mass splitting is
\begin{equation}
\delta
\equiv
m_{\chi_{2,3}}-m_{\chi_1}
=
\sqrt{m_\chi^2+\frac{1}{2}y^2v^2}-m_\chi
\simeq
\frac{y^2v^2}{4m_\chi},
\label{eq:splitting}
\end{equation}
where the last expression holds for $yv\ll m_\chi$.

The interaction responsible for direct detection follows from the neutral current of the Dirac multiplet. In the $(D_+,D_-)$ basis, the coupling to the $Z$ boson is purely off diagonal,
\begin{equation}
\mathcal L_Z=
-\frac{g}{2c_W}Z_\mu
\left(
D_+^\dagger\bar\sigma^\mu D_-
+
D_-^\dagger\bar\sigma^\mu D_+
\right),
\label{eq:zcurrent}
\end{equation}
where $g$ is the $SU(2)_L$ gauge coupling, $c_W\equiv\cos\theta_W$, and $\bar\sigma^\mu=(1,-\boldsymbol\sigma)$ in two-component notation. Since the DM state is $\chi_1=D_+$, whereas $D_-$ is distributed among the two heavier neutral mass eigenstates $\chi_2$ and $\chi_3$, the $Z$ interaction does not mediate elastic scattering of $\chi_1$. Instead, it induces the inelastic transitions
\begin{equation}
\chi_1+A\rightarrow\chi_{2,3}+A,
\label{eq:inel_process}
\end{equation}
where $A$ denotes a target nucleus. The incoming DM particle must therefore supply the mass splitting $\delta$ required to produce one of the heavier neutral states, giving rise to the endothermic scattering relevant for the LZ signal.

Unitarity of the rotation from $D_-$ to the mass eigenstates $\chi_2$ and $\chi_3$ implies
\begin{equation}
\sum_{a=2,3}
\left|g_{\chi_1\chi_a Z}\right|^2
=
\left(\frac{g}{2c_W}\right)^2,
\end{equation}
where $g_{\chi_1\chi_a Z}$ is the off-diagonal $Z$ coupling between the DM ground state $\chi_1$ and the excited neutral eigenstate $\chi_a$, with $a=2,3$.
Consequently, after summing over the two excited final states, the inclusive neutral-current interaction retains the standard pseudo-Dirac normalization. The corresponding zero-momentum DM--neutron cross section is
\begin{equation}
\sigma_n^{\rm inel}
=
\frac{G_F^2\mu_{\chi n}^2}{2\pi}
\simeq
7\times10^{-39}\,\cmsq,
\label{eq:sigmaninel}
\end{equation}
where $G_F$ is the Fermi constant and
\begin{equation}
\mu_{\chi n}
=
\frac{m_\chi m_n}{m_\chi+m_n}
\end{equation}
is the DM--neutron reduced mass. The numerical value in Eq.~\eqref{eq:sigmaninel} corresponds to the heavy-DM limit $m_\chi\gg m_n$.

For a nucleus with mass number $A$, proton number $Z$, and mass $m_A$, the coherent vector weak charge is
\begin{equation}
Q_V
=
(A-Z)
-
\left(1-4\sin^2\theta_W\right)Z,
\label{eq:weakcharge}
\end{equation}
so that the zero-momentum nuclear cross section is
\begin{equation}
\sigma_A^0
=
\sigma_n^{\rm inel}
\frac{\mu_{\chi A}^2}{\mu_{\chi n}^2}
Q_V^2,
\label{eq:sigmaA0}
\end{equation}
with
\begin{equation}
\mu_{\chi A}
=
\frac{m_\chi m_A}{m_\chi+m_A}.
\end{equation}
The dominant coherent contribution therefore arises from the neutron content of the nucleus, since the proton vector coupling is suppressed by the factor $1-4\sin^2\theta_W$.

The normalization in Eq.~\eqref{eq:sigmaninel} can be seen directly by integrating out the $Z$ boson. A more complete derivation, including the inclusive sum over the two degenerate heavy neutral states and the elastic-limit check of the nuclear recoil cross section, is given in Appendix~\ref{app:theory}. For scattering on a neutron, the resulting effective interaction is
\begin{equation}
\mathcal L_{\rm eff}^{n}
=
C_n J_{\chi}^{\mu}\,\bar n\gamma_{\mu}n,
\qquad
C_n
=
\frac{g^2}{8c_W^2m_Z^2}
=
\frac{G_F}{\sqrt{2}},
\label{eq:Cn}
\end{equation}
where $J_{\chi}^{\mu}$ denotes the off-diagonal DM transition current. In the nonrelativistic limit this gives
\begin{equation}
\sigma_n
=
\frac{\mu_{\chi n}^2 C_n^2}{\pi}
=
\frac{G_F^2\mu_{\chi n}^2}{2\pi},
\label{eq:sigman_derivation}
\end{equation}
in agreement with Eq.~\eqref{eq:sigmaninel} and with the zero-momentum limit obtained by integrating Eq.~\eqref{eq:dsigma}.\footnote{An expression proportional to $G_F^2\mu_n^2/(8\pi)$ has appeared in several recoil-level treatments of split electroweak fermions, including the first public version and the current public arXiv text of Ref.~\cite{SmirnovGriffithBeacom2026}, as well as an earlier version of our LZ analysis. With the vector transition current and weak-charge convention adopted here, direct integration instead gives Eq.~\eqref{eq:sigman_derivation}. The factor-of-four difference is therefore a normalization issue rather than a difference in the underlying particle-physics interaction. We use the $1/(2\pi)$ normalization consistently throughout this work, as in Refs.~\cite{PospelovRamani2026,DiMauroShaikh2026Solar}.}

The endothermic nature of the transition modifies the recoil kinematics. For a nuclear recoil energy $E_R$, the minimum incident DM speed in the detector frame is
\begin{equation}
v_{\min}(E_R)
=
\frac{1}{\sqrt{2m_AE_R}}
\left(
\frac{m_AE_R}{\mu_{\chi A}}
+
\delta
\right),
\label{eq:vmin}
\end{equation}
where $\delta>0$ is the mass splitting between the incoming state $\chi_1$ and the excited states $\chi_{2,3}$. This function reaches its minimum at
\begin{equation}
E_R^\star
=
\frac{\mu_{\chi A}}{m_A}\delta.
\end{equation}
A splitting of a few hundred keV therefore shifts the preferred xenon recoil energies well above the conventional low-energy direct-detection region and can naturally place the signal close to the high-energy LZ event \cite{TuckerSmith2001,Bramante2016,SmirnovGriffithBeacom2026}.

\subsection{Thermal benchmarks and interactions relevant for Solar evolution}

The purpose of this subsection is to identify the ingredients of HC-MDM that are needed once the endothermic capture process has been specified. Three ingredients are particularly important. First, the thermal relic-density condition fixes a relation between the Higgs coupling and the DM mass for each electroweak representation, thereby selecting the benchmark points relevant for the LZ interpretation. Second, loop-induced elastic scattering controls the subsequent cooling and thermalization of DM particles captured by the Sun after inelastic scattering becomes inefficient. Third, annihilation of the captured ground-state population into electroweak final states determines the high-energy neutrino signal that can be tested with IceCube. Although these processes arise from the same electroweak structure, they probe different stages of the DM evolution and should therefore be distinguished clearly.

The late-time neutral-current interaction discussed above is essentially representation independent, but the thermal relic calculation is not. Before electroweak symmetry breaking, the Higgs interaction affects annihilation, coannihilation, Sommerfeld enhancement, and bound-state formation. As a consequence, each pair of Majorana and Dirac representations defines a different thermal relic trajectory,
\begin{equation}
m_\chi=m_{\rm th}^{(R_M,R_D)}(y),
\label{eq:thermaltrajectory}
\end{equation}
where $R_M$ and $R_D$ denote the dimensions of the Majorana and Dirac $SU(2)_L$ representations, respectively, and $m_{\rm th}^{(R_M,R_D)}(y)$ is the DM mass that reproduces the observed relic abundance for a given Higgs coupling $y$. These trajectories were computed in Ref.~\cite{GriffithSmirnov2026}. Their intersections with the parameter region selected by the LZ recoil rate define the six thermal benchmark points listed in Tab.~\ref{tab:benchmarks}. Appendix~\ref{app:theory} summarizes how the Yukawa interaction enters the coannihilation, Sommerfeld, and bound-state problem and clarifies which parts of the relic calculation are imported from Ref.~\cite{GriffithSmirnov2026} rather than recomputed here.

The same benchmark models also exhibit loop-induced elastic scattering, which plays an important role in the evolution of DM after Solar capture. At the custodial point, the diagonal tree-level Higgs coupling vanishes, so elastic spin-independent scattering arises at loop level. The central per-nucleon cross sections adopted in Ref.~\cite{SmirnovGriffithBeacom2026} are
\begin{align}
\sigma_{\rm SI}^{\rm loop}={}&
(1.3\times10^{-47},\ 1.1\times10^{-46},\ 4.5\times10^{-46},\nonumber\\
&1.2\times10^{-45},\ 2.6\times10^{-45},\ 5.0\times10^{-45})\,\cmsq,
\label{eq:loopsix}
\end{align}
ordered from $3_M2_D$ to $13_M12_D$. Here $\sigma_{\rm SI}^{\rm loop}$ denotes the zero-momentum elastic SI cross section on a single nucleon generated by electroweak loops after the custodial cancellation of the tree-level Higgs amplitude. The increase with the representation dimension reflects the larger electroweak group factors entering the loop amplitudes. These cross sections are predictions of the benchmark model rather than parameters fitted to the LZ event.

Their precise theoretical uncertainty requires some care. Ref.~\cite{GriffithSmirnov2026} uses the gauge-loop SI cross sections of the corresponding pure minimal-DM multiplets and includes the lattice-QCD uncertainty inherited from those calculations, but Higgs-induced loop corrections specific to the mixed HC-MDM states have not yet been evaluated. The values in Eq.~\eqref{eq:loopsix} should therefore be regarded as the best available benchmark estimates rather than as exact predictions with a universal fractional error.

This uncertainty is particularly relevant because heavy-WIMP electroweak amplitudes can exhibit substantial cancellations. Electroweak matching generates scalar-quark, scalar-gluon, and spin-two operators whose contributions to the nucleon amplitude can partially cancel \cite{Hisano2011,Hisano2015,ChenHill2020,ChenDingHill2023}. In pure electroweak multiplets this can suppress the final SI cross section by several orders of magnitude relative to the individual loop contributions. For example, the NLO QCD calculation for a heavy wino gives $\sigma_{\rm SI}^p=2.3^{+0.2}_{-0.3}{}^{+0.5}_{-0.4}\times10^{-47}\,\cmsq$ \cite{Hisano2015}, while more general heavy-WIMP calculations find that the residual SI rate can vary strongly with the electroweak quantum numbers because of amplitude cancellations \cite{ChenHill2020,ChenDingHill2023}. Close to such a cancellation, formally subleading corrections can become numerically important.

For HC-MDM, there is consequently no well-motivated single fractional uncertainty that can be assigned to all six values in Eq.~\eqref{eq:loopsix}. Away from a cancellation, uncertainties from perturbative matching and nucleon matrix elements are naturally expected at the tens-of-percent level, whereas the relative uncertainty can become much larger if the leading amplitude is accidentally suppressed. We therefore do not treat the quoted loop cross sections as Gaussian theory measurements.

For the Solar analysis, however, the relevant question is not whether the loop-induced elastic cross section is known at the percent level, but whether it is large enough to cool captured DM particles within the age of the Sun. A captured particle can cross the Solar interior many times, so efficient cooling can occur even if the elastic cross section is substantially smaller than its nominal value. In Sec.~\ref{sec:postcapture} we test this explicitly by repeating the capture-generated orbital evolution after suppressing each cross section in Eq.~\eqref{eq:loopsix} by factors of $10$, $100$, and down to the $\mathcal{O}(10^{-3})$ level relative to its nominal value. Even at $1\%$ of the quoted cross section, $94.1$--$99.7\%$ of the capture-weighted population reaches the thermal region within the Solar age. The median capture-weighted cooling time becomes comparable to $t_\odot$ only when the elastic rate is reduced to approximately $(0.21$--$0.32)\%$ of the nominal value, while thermalizing the $99$th-percentile orbit requires approximately $(0.7$--$2.0)\%$. These thresholds are not intended as precision limits on $\sigma_{\rm SI}^{\rm loop}$; instead, they quantify how strongly the elastic interaction can be reduced before post-capture cooling becomes inefficient.

In the cooling calculation we include only the loop-induced SI interaction. Loop-induced SD elastic scattering could provide an additional cooling channel, in particular through scattering on Solar hydrogen, but a dedicated representation-by-representation HC-MDM calculation is not presently available. Studies of Higgsino- and wino-like neutralinos indicate that radiative corrections to SD scattering can be significant in mixed systems \cite{Bisal2024SD}, but those results cannot be transferred quantitatively to HC-MDM. Neglecting this additional channel is conservative for the thermalization problem, since any nonzero SD contribution would only shorten the cooling time.

The final ingredient is the annihilation of the captured DM population. The ground state $\chi_1$ remains part of a nontrivial electroweak multiplet and retains gauge interactions, so present-day annihilation does not require a thermally populated excited state. Gauge interactions therefore generate electroweak final states such as $W^+W^-$ and $ZZ$, together with additional representation-dependent channels involving charged partners and electroweak radiation. The representation dependence of the charged-current couplings and charged spectrum underlying these channels is reviewed in Appendix~\ref{app:theory}. The complete broken-phase annihilation spectrum has not been provided in an IceCube-ready form for all of the large HC-MDM representations considered here. We therefore do not assign an artificially precise branching fraction into $W^+W^-$. Instead, in Sec.~\ref{sec:icecube} we compare the predicted equilibrium Solar annihilation rate with the IceCube $W^+W^-$ limit where the published templates are directly applicable, and use a separately identified response-based diagnostic for the heavier benchmarks that lie beyond the released IceCube mass grid.

\begin{figure}[t]
\centering
\includegraphics[width=\columnwidth]{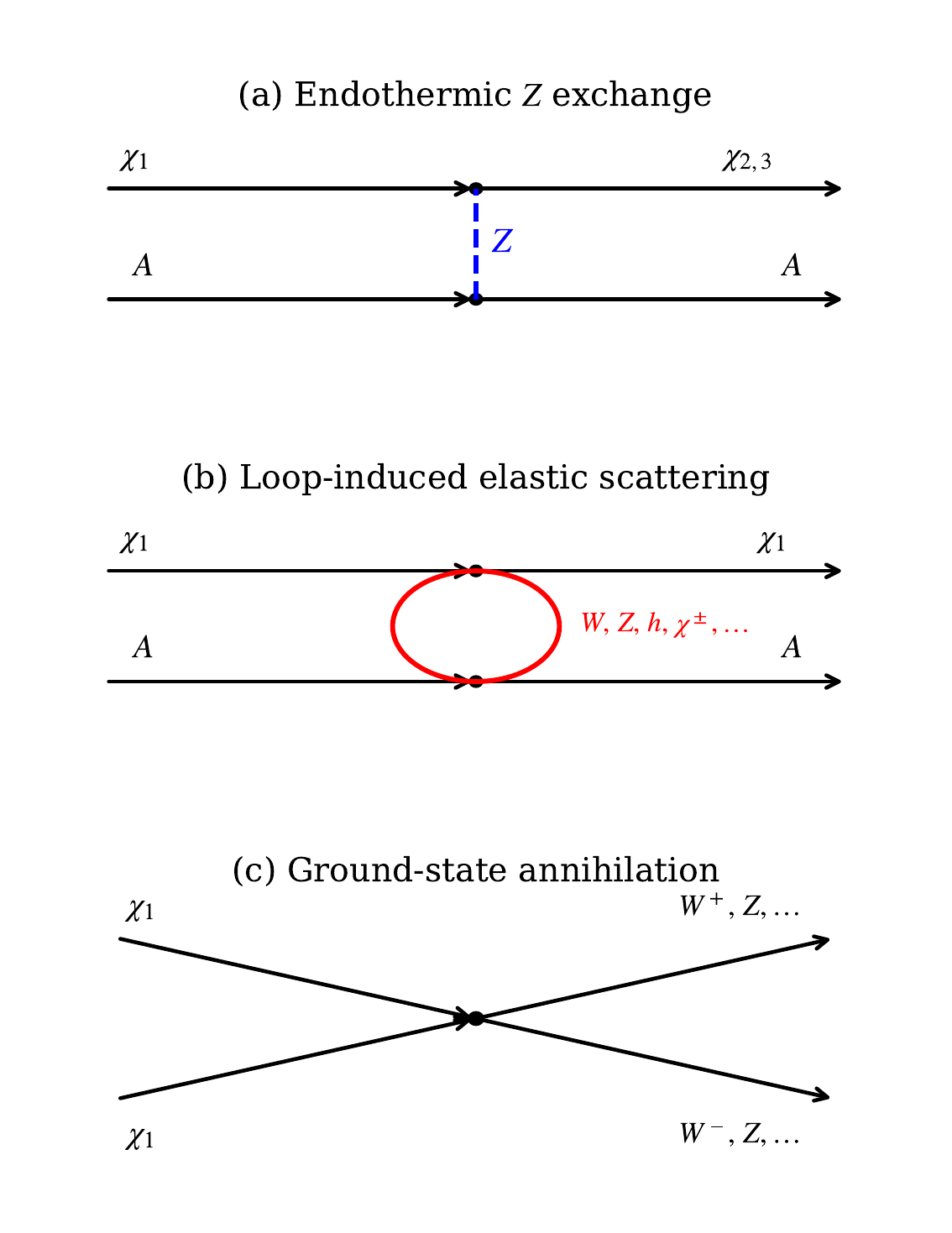}
\caption{Schematic interactions entering the LZ--Solar analysis. Panel~(a) shows the tree-level endothermic transition $\chi_1 A\to\chi_{2,3}A$ mediated by $Z$ exchange, which produces the LZ recoil and initiates Solar capture. Panel~(b) represents loop-induced elastic $\chi_1A\to\chi_1A$ scattering, which governs the subsequent orbital cooling of the captured DM population. Panel~(c) shows annihilation of the ground-state population into electroweak final states, which generate the high-energy neutrino signal tested by IceCube.}
\label{fig:interactions}
\end{figure}

The three processes relevant for connecting the LZ interpretation to the Solar-neutrino test are summarized in Fig.~\ref{fig:interactions}. Panel~(a) shows the tree-level endothermic $Z$-mediated transition responsible both for the terrestrial LZ recoil and for the initial Solar capture. Panel~(b) represents the loop-induced elastic scattering that cools the captured DM population after inelastic transitions become inefficient. Panel~(c) illustrates annihilation of the resulting ground-state population into electroweak final states, which ultimately produces the high-energy neutrinos searched for by IceCube. The important point is that capture, cooling, and annihilation are not introduced as independent effective interactions: they all follow from the same HC-MDM electroweak structure.

\begin{table*}[t]
\caption{HC-MDM benchmark points used in this work. The thermal DM masses $m_{\chi,\star}$ are those obtained from the relic-density trajectories of Ref.~\cite{SmirnovGriffithBeacom2026}. At each fixed thermal mass, we re-evaluate the endothermic splitting $\delta_\star$ required by the LZ event rate using the corrected neutral-current normalization $\sigma_n^{\rm inel}=G_F^2\mu_{\chi n}^2/(2\pi)$. The quoted uncertainties on $\delta_\star$ are $68.27\%$ central, count-only Poisson intervals obtained by mapping the exact one-event Poisson interval through the predicted LZ rate $N_{\rm sig}(\delta)$. They do not include backgrounds, recoil-energy information, detector systematics, or astrophysical uncertainties. The Higgs coupling $y_\star$ is obtained from the exact neutral-state mass-splitting relation at the quoted mass. The loop-induced SI cross sections and elastic LZ yields are those adopted in Ref.~\cite{SmirnovGriffithBeacom2026}. All six thermal masses reproduce $\Omega_\chi h^2\simeq0.12$ by construction of the relic-density trajectories.}
\label{tab:benchmarks}
\begin{ruledtabular}
\begin{tabular}{lcccccc}
Model &
$y_\star$ &
$m_{\chi,\star}$ [TeV] &
$\delta_\star$ [keV] &
$\sigma_{\rm SI}^{\rm loop}$ [$\cmsq$] &
$N_{\rm el}$ &
$\Omega_\chi h^2$
\\
\hline
$3_M2_D$ &
$5.92\times10^{-3}$ &
$1.4$ &
$379.3^{+1.6}_{-1.1}$ &
$1.3\times10^{-47}$ &
$0.5$ &
$0.12$
\\
$5_M4_D$ &
$1.407\times10^{-2}$ &
$7.7$ &
$389.1^{+3.3}_{-2.7}$ &
$1.1\times10^{-46}$ &
$0.8$ &
$0.12$
\\
$7_M6_D$ &
$2.370\times10^{-2}$ &
$22$ &
$386.2^{+4.4}_{-3.6}$ &
$4.5\times10^{-46}$ &
$1.2$ &
$0.12$
\\
$9_M8_D$ &
$3.483\times10^{-2}$ &
$48$ &
$382.3^{+5.3}_{-4.5}$ &
$1.2\times10^{-45}$ &
$1.5$ &
$0.12$
\\
$11_M10_D$ &
$4.529\times10^{-2}$ &
$82$ &
$378.5^{+6.3}_{-5.5}$ &
$2.6\times10^{-45}$ &
$1.8$ &
$0.12$
\\
$13_M12_D$ &
$5.680\times10^{-2}$ &
$130$ &
$375.4^{+7.1}_{-6.3}$ &
$5.0\times10^{-45}$ &
$2.4$ &
$0.12$
\\
\end{tabular}
\end{ruledtabular}
\end{table*}

\section{Solar capture calculation}
\label{sec:capture}

\subsection{Solar and halo inputs}

We use a publicly available sampled BS05(AGS,OP) Standard Solar Model (SSM) \cite{Bahcall2005}, as in the Solar-capture analyses of Refs.~\cite{PospelovRamani2026,DiMauroShaikh2026Solar}.  The calculation interpolates the radial density, temperature, enclosed mass, and the directly tabulated H, He, C, N, and O mass fractions.  Heavier elements are assigned AGS-like normalizations and follow the mild common radial diffusion trend of oxygen.  We include Ne, Mg, Si, S, Ca, Ti, Cr, Fe, Ni, Zn, Ge, Pb, and U.  Natural isotope abundances are used for the coherent weak response.  This is relevant near extreme kinematic closure, where small isotope-mass differences can change which nuclei remain accessible.

The escape-speed profile is derived from the same enclosed-mass table,
\begin{equation}
v_{\rm esc}^2(r)=2G\left[\frac{M_\odot}{R_\odot}
+\int_r^{R_\odot}dr'\frac{M(r')}{r'^2}\right],
\label{eq:vesc}
\end{equation}
where $G$ is Newton's constant, $M(r)$ is the Solar mass enclosed within radius $r$, and $M_\odot$ and $R_\odot$ are the Solar mass and radius. The resulting central escape speed is approximately $1385\,\kms$ and the surface value is $617.7\,\kms$.  For the Galactic boundary condition we take the SHM reference values
\begin{align}
\rho_\chi&=0.30\,\GeV\,\mathrm{cm}^{-3},\qquad
v_0=220\,\kms,\\
v_\odot&=232\,\kms,\qquad
v_{\rm esc}^{\rm Gal}=544\,\kms.
\label{eq:halo}
\end{align}
Here $\rho_\chi$ is the local DM mass density, $v_0$ is the SHM velocity-dispersion parameter, $v_\odot$ is the Solar speed through the halo, and $v_{\rm esc}^{\rm Gal}$ is the Galactic escape speed. The Galactic Maxwellian is truncated before boosting to the Solar frame.  We later vary these inputs to assess their impact.

\subsection{Endothermic single-scatter capture}
\label{subsec:singlecapture}

We first determine the kinematic conditions under which an incoming halo DM particle can undergo an endothermic collision with a Solar nucleus and lose enough energy in a single scatter to become gravitationally bound to the Sun. A DM particle with asymptotic Solar-frame speed $u$ reaches radius $r$ with speed
\begin{equation}
w^2=u^2+v_{\rm esc}^2(r).
\label{eq:w}
\end{equation}
Here $u$ is the DM speed at infinity in the Solar rest frame and $w$ is its local speed after gravitational acceleration. For a target nucleus of mass $m_A$, the endothermic up-scatter is kinematically possible when
\begin{equation}
w^2>\frac{2\delta}{\mu_{\chi A}}.
\end{equation}
For a stationary nucleus, the speed associated with the outgoing relative motion is
\begin{equation}
w'=\sqrt{w^2-\frac{2\delta}{\mu_{\chi A}}}.
\label{eq:wprime}
\end{equation}
The allowed recoil interval is
\begin{equation}
E_R^{\pm}=\frac{\mu_{\chi A}^2}{2m_A}(w\pm w')^2.
\label{eq:erpm}
\end{equation}
Capture additionally requires the outgoing DM particle to be gravitationally bound.  Energy conservation gives
\begin{equation}
E_R+\delta>\frac{1}{2}m_\chi u^2,
\end{equation}
so the lower integration boundary is
\begin{equation}
E_R^{\rm low}=\max\left[E_R^-,\frac{1}{2}m_\chi u^2-\delta\right].
\label{eq:elow}
\end{equation}

For coherent neutral-current scattering we write
\begin{equation}
\frac{d\sigma_A}{dE_R}
=\frac{G_F^2m_AQ_V^2}{4\pi w^2}F_A^2(q),
\qquad q=\sqrt{2m_AE_R},
\label{eq:dsigma}
\end{equation}
within the kinematically allowed interval.  The nuclear form factor is the Helm response,
\begin{equation}
F_A(q)=3\frac{j_1(qR_1)}{qR_1}\exp\left[-\frac{1}{2}(qs)^2\right],
\label{eq:helm}
\end{equation}
where $j_1$ is the spherical Bessel function of order one, $q$ is the momentum transfer defined in Eq.~\eqref{eq:dsigma}, $s=0.9\,\mathrm{fm}$ is the Gaussian surface-thickness parameter, and $R_1$ is the effective Helm diffraction radius. In the numerical implementation we use $R_1=[c^2+7\pi^2a^2/3-5s^2]^{1/2}$ with $c=1.23A^{1/3}-0.60\,\mathrm{fm}$ and $a=0.52\,\mathrm{fm}$.  Finite-$q$ suppression is numerically important because capture near a several-hundred-keV endothermic threshold preferentially involves large energy transfer to Fe and Ni.

The local capture kernel is
\begin{equation}
\Omega_A^-(w,r)=n_A(r)w\int_{E_R^{\rm low}}^{E_R^+}dE_R\frac{d\sigma_A}{dE_R},
\label{eq:omega}
\end{equation}
where $n_A(r)$ is the local number density of isotope $A$ and $\Omega_A^-$ is the rate per incident DM particle for scatters on that isotope that leave the outgoing state bound. The capture rate is
\begin{equation}
\frac{dC_A}{dr}=4\pi r^2\frac{\rho_\chi}{{m_\chi}}
\int du\frac{f_\odot(u)}{u}w\Omega_A^-(w,r),
\label{eq:capture}
\end{equation}
where $f_\odot(u)$ is the normalized DM speed distribution in the Solar frame. The factor $\rho_\chi/m_\chi$ converts the local DM mass density into number density. The total rate $C_\odot$ is obtained by integrating over radius and summing over all isotopes and elements.  Eqs.~\eqref{eq:w}--\eqref{eq:capture} are evaluated numerically with nested Gauss--Legendre quadratures.  The single-scatter treatment is verified independently in Appendix~\ref{app:validation}, where the optical depth is shown to remain well below unity.

\subsection{Finite-temperature nuclear motion}
\label{subsec:finiteT}

The capture calculation of Sec.~\ref{subsec:singlecapture} assumes that Solar nuclei are at rest in the Solar frame, so that the relative speed entering the inelastic-scattering kinematics is simply the DM speed $w$. This stationary-target approximation is extremely accurate when the endothermic transition is well above threshold, because the thermal velocities of Solar nuclei are then small compared with the relative velocity required to excite the heavier DM state. Close to kinematic closure, however, this approximation can underestimate the capture rate. In that regime the DM speed $w$ is slightly below the value required for scattering on a stationary nucleus, while the thermal motion of the target can either increase or decrease the DM--nucleus relative speed. Nuclei moving toward the incoming DM particle populate the high-relative-velocity tail and can therefore reopen an otherwise forbidden endothermic transition.

This effect is especially important for heavy Solar elements such as Fe and Ni, which dominate the coherent capture rate at large mass splitting. Although their thermal velocities are much smaller than the DM speed itself, they can be comparable to the small velocity deficit that separates an allowed from a forbidden inelastic collision. The proposed $\delta=570\,\keV$ high-velocity benchmarks lie precisely in this near-threshold regime. We therefore relax the stationary-target approximation and explicitly average the capture probability over the thermal velocity distribution of the Solar nuclei.

To account for this effect, we generalize the stationary-target calculation by treating each Solar nuclear species as a thermal population with a local Maxwell--Boltzmann velocity distribution,
\begin{equation}
f_A(\bm v_A;r)=
\left(\frac{m_A}{2\pi T(r)}\right)^{3/2}
\exp\left[-\frac{m_Av_A^2}{2T(r)}\right],
\label{eq:targetMB}
\end{equation}
where $\bm v_A$ is the velocity of a target nucleus of species $A$ and $T(r)$ is the local Solar temperature. We use natural units with $k_B=1$. Thus, instead of assuming $\bm v_A=0$, the relative velocity entering the inelastic collision is determined event by event from the sampled DM and nuclear velocities.

For each nuclear species and Solar radius, we sample $\bm v_A$ from Eq.~\eqref{eq:targetMB} and combine it with the incident DM velocity $\bm w$. The scattering is then evaluated using the full leading-nonrelativistic two-body kinematics in the center-of-mass frame, retaining the endothermic energy cost $\delta$ exactly. We first require that the available center-of-mass kinetic energy is sufficient to overcome the endothermic splitting $\delta$. For kinematically allowed configurations, we sample the scattering angle, determine the outgoing momentum of the excited DM state, and transform it back to the Solar frame. A scattering event contributes to capture only if the final DM speed satisfies
\begin{equation}
v_\chi'(r)<v_{\rm esc}(r),
\label{eq:finiteTcapture}
\end{equation}
so that the outgoing particle is gravitationally bound to the Sun. The same coherent weak charge and Helm nuclear form factor used in the stationary-target calculation are applied to each scattering configuration.

The thermal average involves both the three-dimensional nuclear velocity distribution and the scattering angle. We evaluate these integrals numerically using a Sobol quasi-Monte-Carlo sequence, which provides an efficient sampling of the multidimensional phase space, including the small region in which thermal nuclear motion reopens an otherwise forbidden endothermic transition. We verify numerical convergence by repeating the calculation with independently scrambled Sobol sequences. For the $\delta=570\,\keV$ benchmarks, the resulting capture rates have relative dispersions of only $0.47\%$ for $m_\chi=1.47\,\TeV$ and $0.34\%$ for $m_\chi=8.5\,\TeV$. These dispersions are much smaller than the finite-temperature enhancement itself and demonstrate that the thermally reopened capture region is numerically well resolved. The explicit finite-temperature kernel and its stationary-target limit are derived in Appendix~\ref{app:finiteT}.

\section{Post-capture evolution}
\label{sec:postcapture}

\subsection{Radiative and collisional de-excitation}

The endothermic capture process does not initially leave the DM particle in the ground state. Instead, the first inelastic collision produces one of the excited neutral states, $\chi_2$ or $\chi_3$. The subsequent fate of this excited state is important for the post-capture evolution: if it de-excites rapidly to $\chi_1$, the captured population quickly returns to the ground state and its later orbital cooling is governed by the elastic interactions discussed below. If, instead, the excited state were long lived, repeated inelastic transitions could modify both the energy loss per orbit and the annihilation dynamics.

In HC-MDM, the excited neutral states can de-excite radiatively through
\begin{equation}
\chi_{2,3}\rightarrow\chi_1+\gamma,
\end{equation}
and can also undergo collisional de-excitation in the Solar medium. We therefore estimate both processes and compare their characteristic timescales with the orbital timescale of the newly captured DM population. Following the one-loop result quoted in Ref.~\cite{SmirnovGriffithBeacom2026}, the radiative lifetime is approximately
\begin{equation}
\tau_\gamma\simeq0.90\,\mathrm{s}
\left(\frac{m_\chi}{7.63\,\TeV}\right)^2
\left(\frac{386\,\keV}{\delta}\right)^3
\left(\frac{2}{C_{\rm HC}}\right)^2,
\label{eq:taugamma}
\end{equation}
where $\tau_\gamma$ is the radiative lifetime and $C_{\rm HC}$ is the dimensionless electroweak loop coefficient controlling the transition dipole amplitude. At leading order,
\begin{equation}
C_{\rm HC}=2-\frac{5\pi}{4}\frac{m_W}{m_\chi}+\mathcal O\left(\frac{m_W^2}{m_\chi^2}\right),
\end{equation}
where $m_W$ is the $W$-boson mass. At the Tab.~\ref{tab:benchmarks} benchmarks this gives lifetimes of order $10^{-2}\,\mathrm{s}$ for $3_M2_D$ and up to a few hundred seconds for $13_M12_D$.  The light representations therefore radiatively return to the ground state well before a Solar orbit is completed.

For the larger electroweak representations, the radiative lifetime of the excited states becomes sufficiently long that $\chi_{2,3}$ may undergo an exothermic collision with Solar nuclei before decaying through photon emission. Such a collision converts the excited state back into $\chi_1$ while releasing the mass splitting $\delta$ as kinetic energy. We therefore estimate whether collisional de-excitation can occur during the lifetime of the excited state.

To obtain a conservative upper bound, we make the extreme assumption that the newly produced excited particle remains at the Solar-core density during its entire radiative lifetime. We then integrate the corresponding exothermic weak-scattering rate over that time interval. This deliberately maximizes the amount of Solar material encountered by the excited state: in reality, a newly captured particle generally follows an extended orbit and spends only a small fraction of each orbital period in the dense central region. The resulting estimate should therefore be regarded as an upper bound on the probability of collisional de-excitation.

With this prescription, the collisional probability is negligible for the two benchmarks lying within the published IceCube mass range and remains below a few percent through the $7_M6_D$ model. For the largest representations, the longer excited-state lifetime allows the conservative core-fixed probability to become appreciable. In this regime, however, the approximation becomes increasingly pessimistic because it replaces the true orbit-averaged Solar density by the much larger core density throughout the entire lifetime.

Collisional de-excitation also differs qualitatively from radiative decay because the transition
\begin{equation}
\chi_{2,3}+A\rightarrow\chi_1+A
\end{equation}
is exothermic. The released energy $\delta$ is shared between the outgoing DM particle and the recoiling nucleus and can therefore increase the DM kinetic energy. If the original capture orbit is only weakly bound, this additional kinetic energy can in principle make the outgoing $\chi_1$ unbound from the Sun. We consequently do not interpret collisional de-excitation as an additional mechanism that increases the retained Solar DM population. Instead, for the high-mass benchmarks where its probability may be non-negligible, we treat it conservatively as a one-sided uncertainty on the fraction of initially captured particles that remain gravitationally bound.

\subsection{Elastic cooling and thermal radius}

After radiative de-excitation the ground state remains gravitationally bound.  Repeated inelastic up-scattering can remove additional orbital energy until the transition becomes kinematically blocked, while the loop-induced elastic interaction remains available at all lower velocities.  We therefore treat elastic cooling explicitly rather than assuming instantaneous thermalization.  The calculation uses the loop-induced SI cross sections quoted in Ref.~\cite{SmirnovGriffithBeacom2026}, a coherent SI nuclear response with the same Helm form factor as in the capture calculation, and orbit averaging through the sampled Solar model.

A useful subtlety is that the first capture scatter does not in general place the particle on an orbit contained inside the Sun. The positive binding energy immediately after capture is
\begin{equation}
E_B=E_R+\delta-\frac{1}{2}m_\chi u^2,
\label{eq:bindingaftercapture}
\end{equation}
where $E_B>0$, $E_R$ is the nuclear recoil energy in the capture scatter, and $u$ is the incident DM speed at infinity. For an approximately radial orbit whose apocenter lies outside the Sun, the exterior Kepler segment has
\begin{equation}
r_{\rm max}\simeq\frac{G M_\odot m_\chi}{E_B},
\label{eq:outerapocenter}
\end{equation}
where $r_{\rm max}$ is the first-orbit apocenter, $G$ is Newton's constant, and $M_\odot$ is the Solar mass. This Kepler approximation is used only when $r_{\rm max}>R_\odot$.  We therefore construct the capture-weighted distribution of $E_B$ using the same $(r,u,E_R)$ integration that gives $C_\odot$. The corresponding probability measure and the exterior-to-interior cooling evolution are given explicitly in Appendix~\ref{app:orbitcooling}.  We evolve the exterior Kepler segment down to $r_{\rm max}=R_\odot$ and then continue with the orbit-averaged Solar-interior slowing equation.  Along the exterior segment, the Solar-transit SI slowing column varies by only about $1$--$2\%$ through the $99$th percentile of the capture-weighted binding-energy distribution for the light and high-splitting benchmarks, so we evaluate that segment in the weak-binding limit.  This removes an otherwise hidden assumption about the first bound orbit while introducing only a percent-level approximation in the exterior cooling time.

Near the Solar center the density can be approximated as constant and the thermalized distribution has characteristic size
\begin{equation}
r_\chi=\left(\frac{3T_c}{2\pi G\rho_c m_\chi}\right)^{1/2},
\label{eq:rthermal}
\end{equation}
where $T_c$ and $\rho_c$ are the Solar central temperature and density. This is the Gaussian scale radius of a nonrelativistic thermalized DM population in the approximately harmonic central Solar potential. For the six thermal-relic benchmarks $r_\chi/R_\odot$ decreases from $2.90\times10^{-3}$ to $3.01\times10^{-4}$.  The first-capture orbits are substantially more extended: for example, the capture-weighted median cold-target apocenters are approximately $9.3R_\odot$ and $53R_\odot$ for $3_M2_D$ and $5_M4_D$, respectively.  Nevertheless, their orbital periods are short compared with the Solar age, and the large loop-induced elastic rates predicted by HC-MDM cool them efficiently.

With the nominal loop-induced SI cross sections, the capture-weighted median time from the first bound orbit to the thermal radius ranges from $9.6$ to $14.8$ Myr across the six thermal benchmarks.  The $99$th percentile of the rate-matched first-capture distribution cools in approximately $33$--$91$ Myr.  As a conservative stress test for the $\delta=570\,\keV$ trajectories, we use the stationary-target first-capture binding distribution before applying the same exterior-plus-interior loop cooling.  This gives median cooling times of $10.0$ and $12.6$ Myr and $99$th-percentile times of $44$ and $42$ Myr for $3_M2_D$ and $5_M4_D$, respectively.  Since the finite-temperature correction changes the capture phase space but not the existence of the elastic loop channel, these values are sufficient to show that the high-splitting solutions do not generate a long-lived nonthermal population at the nominal loop rates.

The theoretical uncertainty in $\sigma_{\rm SI}^{\rm loop}$ therefore matters only if it is large enough to compete with the roughly three-orders-of-magnitude dynamical margin between the nominal cooling time and the Solar age. A conventional tens-of-percent uncertainty from nucleon matrix elements or higher-order matching is negligible for this purpose. Even a cancellation that reduces the cross section by two orders of magnitude leaves nearly the entire capture-weighted population thermalized. Only when the cross section is driven to the $\mathcal{O}(10^{-3})$ level of the quoted central prediction does the weakly bound tail start to remain extended for a Solar age. Thus we do not claim that a mathematically vanishing loop cross section thermalizes the population; rather, we show that the equilibrium conclusion survives an extremely small but nonzero loop interaction, far below the nominal HC-MDM values.

The orbit-averaged slowing equation and the treatment of the weakly bound exterior part of the first orbit are detailed in Appendix~\ref{app:orbitcooling}. We also perform a deliberately severe loop-rate stress test.  Cooling times scale approximately as $1/\sigma_{\rm SI}^{\rm loop}$.  If every quoted SI loop cross section is reduced to $1\%$ of its nominal value, the capture-weighted median cooling time becomes approximately $0.96$--$1.48$ Gyr for the six thermal benchmarks.  Even then, approximately $94.1$--$99.7\%$ of the stationary-target capture-weighted population cools within the Solar age; for the two $570\,\keV$ trajectories the corresponding fractions are both about $99\%$.  Suppressions at the $\mathcal{O}(10^{-3})$ level begin to make the weakly bound tail important.  We use only the SI loop contribution in this test, so any additional elastic electroweak channel, including an SD contribution, can only accelerate the cooling.

\begin{table*}[t]
\caption{Post-capture cooling and equilibrium diagnostics for the six HC-MDM thermal-relic benchmarks.  The loop-induced SI nucleon cross sections are the central benchmark values quoted in Ref.~\cite{SmirnovGriffithBeacom2026}; their perturbative and hadronic uncertainty is discussed in Sec.~\ref{sec:model}, and the $1\%$ stress test spans a much larger variation than the expected perturbative or hadronic uncertainty. The cooling calculation is evaluated at the independently rate-matched splittings of Eq.~\eqref{eq:deltarematched}.  The third and fourth columns are the median and $99$th-percentile cooling times after including the capture-generated exterior orbit and the subsequent Solar-interior evolution.  The fifth column gives the capture-weighted fraction that still reaches the thermal region within the Solar age if the loop SI rate is artificially reduced to $1\%$ of its nominal value.  The last column gives the capture--annihilation equilibration time for $\langle\sigma v\rangle=2.2\times10^{-26}\,\mathrm{cm^3\,s^{-1}}$.}
\label{tab:postcapture}
\begin{ruledtabular}
\begin{tabular}{lccccc}
Model & $\sigma_{\rm SI}^{\rm loop}$ [$\cmsq$] & $t_{\rm cool}^{50}$ [Myr] & $t_{\rm cool}^{99}$ [Myr] & $f_{\rm therm}(0.01\sigma_{\rm loop})$ & $\tau_{\rm eq}$ [Myr]\\
\hline
$3_M2_D$ & $1.3\times10^{-47}$ & $9.65$ & $39.8$ & $0.997$ & $9.56$\\
$5_M4_D$ & $1.1\times10^{-46}$ & $12.1$ & $33.1$ & $0.996$ & $14.3$\\
$7_M6_D$ & $4.5\times10^{-46}$ & $12.8$ & $35.5$ & $0.994$ & $17.8$\\
$9_M8_D$ & $1.2\times10^{-45}$ & $14.1$ & $90.0$ & $0.967$ & $20.7$\\
$11_M10_D$ & $2.6\times10^{-45}$ & $13.9$ & $91.5$ & $0.963$ & $22.8$\\
$13_M12_D$ & $5.0\times10^{-45}$ & $14.7$ & $89.4$ & $0.941$ & $24.9$\\
\end{tabular}
\end{ruledtabular}
\end{table*}

\subsection{Capture--annihilation equilibrium}

Once the captured DM population has lost sufficient orbital energy through the loop-induced elastic interactions discussed above and has thermalized in the Solar interior, its subsequent evolution is controlled by the competition between continuous capture and pair annihilation. The purpose of this subsection is to determine whether these two processes have reached equilibrium over the age of the Sun. This point is essential because only in the equilibrium regime can the Solar annihilation rate be inferred directly from the capture rate.

Neglecting evaporation, which is completely negligible for TeV-scale DM, the number $N$ of captured ground-state particles evolves according to
\begin{equation}
\frac{dN}{dt}
=
C_\odot-C_A N^2,
\label{eq:number}
\end{equation}
where $C_\odot$ is the total Solar capture rate and $C_A$ is the annihilation coefficient. The annihilation term is quadratic in $N$ because two captured DM particles are required for each annihilation event.

For a thermalized population,
\begin{equation}
C_A
=
\frac{\langle\sigma v\rangle}{V_{\rm eff}},
\qquad
V_{\rm eff}
=
(2\pi)^{3/2}r_\chi^3,
\label{eq:ca}
\end{equation}
where $\langle\sigma v\rangle$ is the velocity-averaged annihilation cross section and $V_{\rm eff}$ is the effective volume occupied by the thermal DM distribution in the Solar core.

The total annihilation rate is
\begin{equation}
\Gamma_A
=
\frac{1}{2}C_A N^2.
\label{eq:gamma_definition}
\end{equation}
The factor $1/2$ appears because each annihilation event removes two DM particles. Eq.~\eqref{eq:number} can therefore equivalently be written as
\begin{equation}
\frac{dN}{dt}
=
C_\odot-2\Gamma_A.
\end{equation}

Starting from a negligible initial captured population, the solution of Eq.~\eqref{eq:number} is
\begin{align}
N(t)
&=
\sqrt{\frac{C_\odot}{C_A}}
\tanh\left(\frac{t}{\tau_{\rm eq}}\right),
\\
\tau_{\rm eq}
&=
\frac{1}{\sqrt{C_\odot C_A}},
\end{align}
where $\tau_{\rm eq}$ is the characteristic capture--annihilation equilibration time. Substituting $N(t)$ into Eq.~\eqref{eq:gamma_definition} gives
\begin{equation}
\Gamma_A(t)
=
\frac{C_\odot}{2}
\tanh^2\left(\frac{t}{\tau_{\rm eq}}\right).
\label{eq:gammaA}
\end{equation}

Eq.~\eqref{eq:gammaA} makes the physical meaning of equilibrium transparent. If the Solar age satisfies
\begin{equation}
t_\odot\gg\tau_{\rm eq},
\end{equation}
then
\begin{equation}
\tanh^2\left(\frac{t_\odot}{\tau_{\rm eq}}\right)\rightarrow1,
\end{equation}
and the annihilation rate approaches
\begin{equation}
\Gamma_A
\simeq
\frac{C_\odot}{2}.
\label{eq:equilibrium}
\end{equation}
Equivalently, in this regime $dN/dt\simeq0$, so capture and annihilation balance,
\begin{equation}
C_\odot
\simeq
C_A N^2
=
2\Gamma_A.
\end{equation}
Thus one half of the capture rate appears as annihilation events because every annihilation consumes two captured DM particles.

We verify explicitly that HC-MDM lies in this regime rather than assuming equilibrium. We invert Eq.~\eqref{eq:gammaA} and determine the annihilation cross section required for the annihilation rate to reach $99\%$ of its equilibrium value at the Solar age. Across the six benchmarks of Tab.~\ref{tab:benchmarks}, we find
\begin{equation}
\langle\sigma v\rangle_{99}
\simeq
(0.86-5.86)\times10^{-30}\,\cmcubeds.
\label{eq:sv99}
\end{equation}
These values are several orders of magnitude below the annihilation rates naturally expected for electroweak DM. For illustration, even the much smaller reference value
\begin{equation}
\langle\sigma v\rangle
=
10^{-28}\,\cmcubeds
\end{equation}
gives
\begin{equation}
\tau_{\rm eq}
\simeq
0.142-0.370\,{\rm Gyr},
\end{equation}
well below the Solar age, while
\begin{equation}
\langle\sigma v\rangle
=
2.2\times10^{-26}\,\cmcubeds
\end{equation}
gives only
\begin{equation}
\tau_{\rm eq}
\simeq
9.6-24.9\,{\rm Myr}.
\end{equation}

We can therefore safely use
\begin{equation}
\Gamma_A
\simeq
\frac{C_\odot}{2}
\end{equation}
for the HC-MDM benchmarks considered here. This relation is not an assumption but the consequence of the explicitly verified hierarchy $t_\odot\gg\tau_{\rm eq}$. It allows the Solar capture calculation to be translated directly into the annihilation rate relevant for the IceCube neutrino limits. This behavior differs qualitatively from models in which annihilation requires a depleted excited-state population, for which capture--annihilation equilibrium can be strongly delayed or absent \cite{DiMauroShaikh2026Solar}.

\begin{figure}[t]
\centering
\includegraphics[width=\columnwidth]{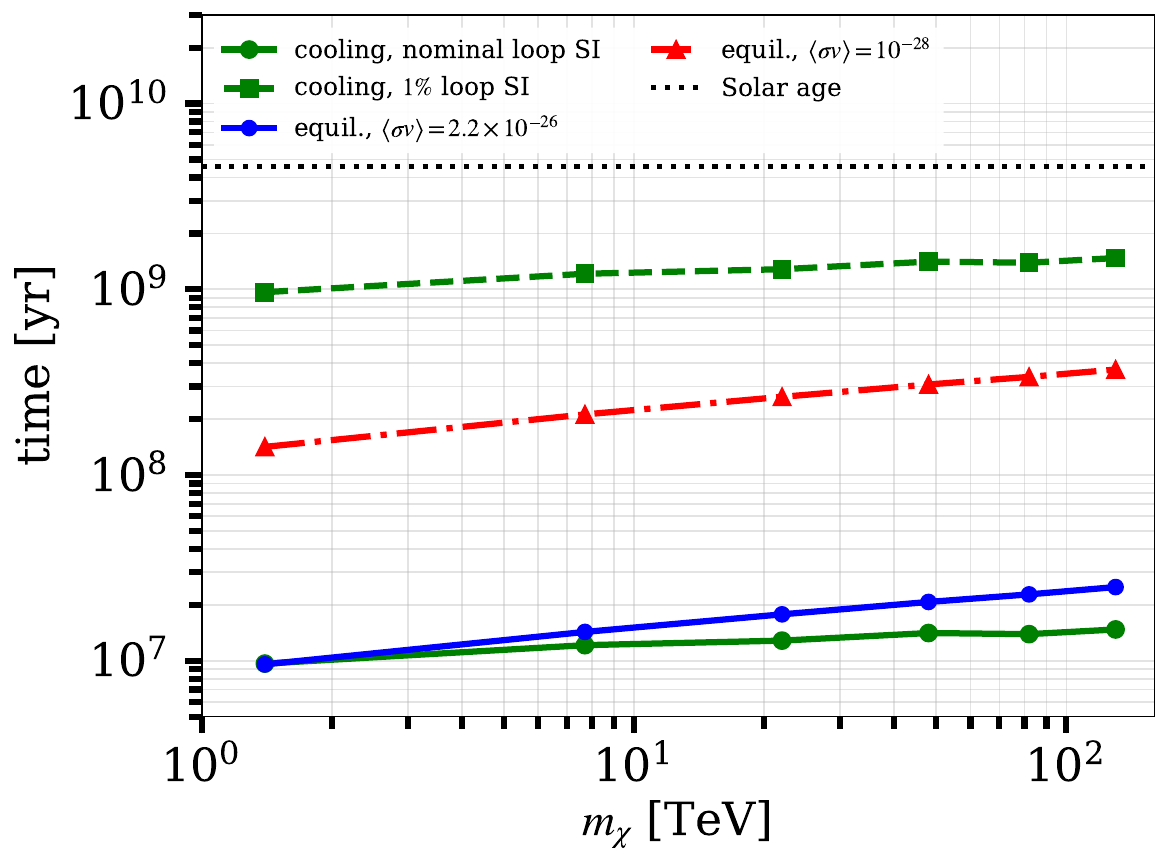}
\caption{Post-capture cooling and capture--annihilation equilibration times for the six HC-MDM thermal benchmarks. Green solid circles show the capture-weighted median cooling time for the nominal loop-induced SI cross section, while green dashed squares show the result after suppressing this cross section by a factor $100$. Blue circles and red triangles show the capture--annihilation equilibration time for $\langle\sigma v\rangle=2.2\times10^{-26}$ and $10^{-28}\,\mathrm{cm^3\,s^{-1}}$, respectively. The black dotted line denotes the Solar age, $t_\odot=4.57\,\mathrm{Gyr}$. Both cooling and equilibration remain faster than the Solar age, even under a two-orders-of-magnitude suppression of the elastic cooling interaction.}
\label{fig:postcapture}
\end{figure}

Fig.~\ref{fig:postcapture} compares the two timescales that determine whether Solar capture can be translated into an equilibrium annihilation rate. The green curves describe the cooling of the initially captured population, while the blue and red curves show the subsequent capture--annihilation equilibration time. The black horizontal line indicates the Solar age.

The solid green curve gives the median time required for a newly captured particle to lose sufficient orbital energy to reach the thermal region, using the nominal loop-induced SI cross sections of Eq.~\eqref{eq:loopsix}. These cooling times are much shorter than the Solar age for all six benchmarks. To test how strongly this conclusion depends on the theoretical prediction for the loop-induced elastic interaction, the dashed green curve repeats the calculation after reducing $\sigma_{\rm SI}^{\rm loop}$ by a factor $100$. Even under this deliberately extreme suppression, the median cooling time remains below $t_\odot$ for every representation. More detailed information about the tail of the cooling-time distribution is given in Tab.~\ref{tab:postcapture}: at $1\%$ of the nominal loop-induced SI rate, $94.1$--$99.7\%$ of the capture-weighted population still reaches the thermal region within the Solar age.

Once the population has thermalized, capture--annihilation equilibrium is reached even more rapidly. The blue and red curves show $\tau_{\rm eq}$ for two illustrative annihilation cross sections. Even for $\langle\sigma v\rangle=10^{-28}\,\mathrm{cm^3\,s^{-1}}$, well below a typical electroweak annihilation rate, the equilibration time remains far shorter than the Solar age. Equivalently, Eq.~\eqref{eq:sv99} shows that an annihilation cross section of only order $10^{-30}\,\mathrm{cm^3\,s^{-1}}$ is sufficient to reach $99\%$ of capture--annihilation equilibrium.

We have also extended the cooling stress test below the range shown in the figure. At $0.3\%$ of the nominal loop-induced SI rate, the median cooling times are approximately $3.2$ and $4.0\,\mathrm{Gyr}$ for $3_M2_D$ and $5_M4_D$, respectively, while about $88\%$ and $68\%$ of their capture-weighted populations still thermalize within the Solar age. Thus, for the benchmarks directly tested with the released IceCube templates, the equilibrium conclusion is not sensitive to the precise normalization of the loop-induced elastic cross section. Only when this interaction is suppressed to the $\mathcal{O}(10^{-3})$ level relative to its nominal prediction does incomplete cooling become important. We therefore conclude that, for the HC-MDM benchmark interactions, both post-capture cooling and subsequent capture--annihilation equilibration occur within the Solar age, justifying the use of $\Gamma_A\simeq C_\odot/2$. We do not extend this statement to the exact limit $\sigma_{\rm SI}^{\rm loop}=0$, which would require a separate treatment of a nonthermal captured population.

\subsection{Independent LZ rate matching and uncertainty on the mass splitting}
\label{subsec:lzrematching}

Before confronting the Solar predictions with IceCube, it is useful to clarify how the endothermic splittings used in our analysis are determined. The thermal masses of the six HC-MDM representations are taken from the relic-density trajectories calculated in Refs.~\cite{GriffithSmirnov2026,SmirnovGriffithBeacom2026}. We do not recompute the full representation-dependent thermal freeze-out calculation, including Sommerfeld enhancement and bound-state effects. Instead, at each of these thermal masses we independently determine the value of the endothermic splitting required to reproduce the observed LZ event rate.

For fixed $m_\chi$, we calculate the expected number of accepted inelastic LZ events as a function of $\delta$ and determine the central rate-matched splitting from
\begin{equation}
N_{\rm sig}(m_\chi,\delta_\star)=1.
\label{eq:lzratematch}
\end{equation}
The calculation uses the corrected neutral-current normalization in Eq.~\eqref{eq:sigman_derivation}
together with the xenon isotope composition, weak nuclear charges, finite-momentum nuclear form factors, the Standard Halo Model velocity distribution, and the LZ exposure and recoil acceptance. The resulting rate-matched splittings are
\begin{equation}
\delta_\star=
\left(
379.3,\,
389.1,\,
386.2,\,
382.3,\,
378.5,\,
375.4
\right)\,\keV,
\label{eq:deltarematched}
\end{equation}
for $3_M2_D$, $5_M4_D$, $7_M6_D$, $9_M8_D$, $11_M10_D$, and $13_M12_D$, respectively.

Because the LZ high-energy signal currently consists of a single candidate event, the statistical uncertainty on the rate is intrinsically non-Gaussian. To provide an estimate of the corresponding uncertainty on $\delta$, we use the exact central Poisson interval for one observed event. Neglecting backgrounds for this count-only diagnostic, the $68.27\%$ confidence interval for the expected signal mean is
\begin{equation}
0.173<\mu<3.300.
\label{eq:poissonone}
\end{equation}
Since the accepted endothermic event rate decreases monotonically with increasing $\delta$, we map the upper and lower Poisson limits into the splitting through
\begin{equation}
N_{\rm sig}(m_\chi,\delta_-)=3.300,
\qquad
N_{\rm sig}(m_\chi,\delta_+)=0.173.
\label{eq:deltapoissonmapping}
\end{equation}
This gives the values reported in Tab.~\ref{tab:benchmarks}.
The increasing uncertainty for the heavier representations reflects the progressively weaker dependence of the accepted event rate on $\delta$ along the thermal sequence.

These values can be compared with the splittings
quoted for the LZ--thermal-relic intersections in Ref.~\cite{SmirnovGriffithBeacom2026}. Our independently rate-matched values are systematically larger by
\begin{equation}
\Delta\delta
\equiv
\delta_\star-\delta_{\rm Ref.}
=
\left(
1.3,\,
3.1,\,
4.2,\,
5.3,\,
6.5,\,
7.4
\right)\,\keV.
\label{eq:deltashifts}
\end{equation}
The direction of this shift is expected. With the larger $1/(2\pi)$ neutral-current normalization adopted here, a fixed value of $\delta$ produces a larger inelastic recoil rate. Matching the rate back to one event therefore requires a slightly larger splitting, which moves the scattering farther into the suppressed high-velocity tail of the halo distribution.

The difference should nevertheless not be interpreted as a statistically significant disagreement with Ref.~\cite{SmirnovGriffithBeacom2026}. The published splittings lie just below, or very close to, the lower edge of our $68\%$ count-only intervals and are all contained within the corresponding $90\%$ Poisson intervals. Moreover, the uncertainties quoted in Tab.~\ref{tab:benchmarks}, obtained through Eqs.~\eqref{eq:poissonone} and \eqref{eq:deltapoissonmapping}, include only the Poisson fluctuation associated with observing one event. They do not include the recoil-energy information of the $248\,\keV$ candidate, the expected LZ background, detector-response uncertainties, or uncertainties in the Galactic velocity distribution. A full experimental determination of $\delta$ would require the multidimensional LZ likelihood including these ingredients. We therefore refer to the values in Eq.~\eqref{eq:deltarematched} as our \emph{rate-matched splittings}, rather than as formal LZ best-fit values.

Although the shifts relative to Ref.~\cite{SmirnovGriffithBeacom2026} are only of order a few keV, they are potentially relevant for the Solar calculation. Endothermic capture becomes increasingly sensitive to $\delta$ close to kinematic closure, particularly for the heavier HC-MDM representations. We therefore use the independently rate-matched values of Eq.~\eqref{eq:deltarematched} as our primary benchmark splittings in the Solar analysis below, while retaining the published values as a useful comparison.

\begin{figure}[t]
\centering
\includegraphics[width=\columnwidth]{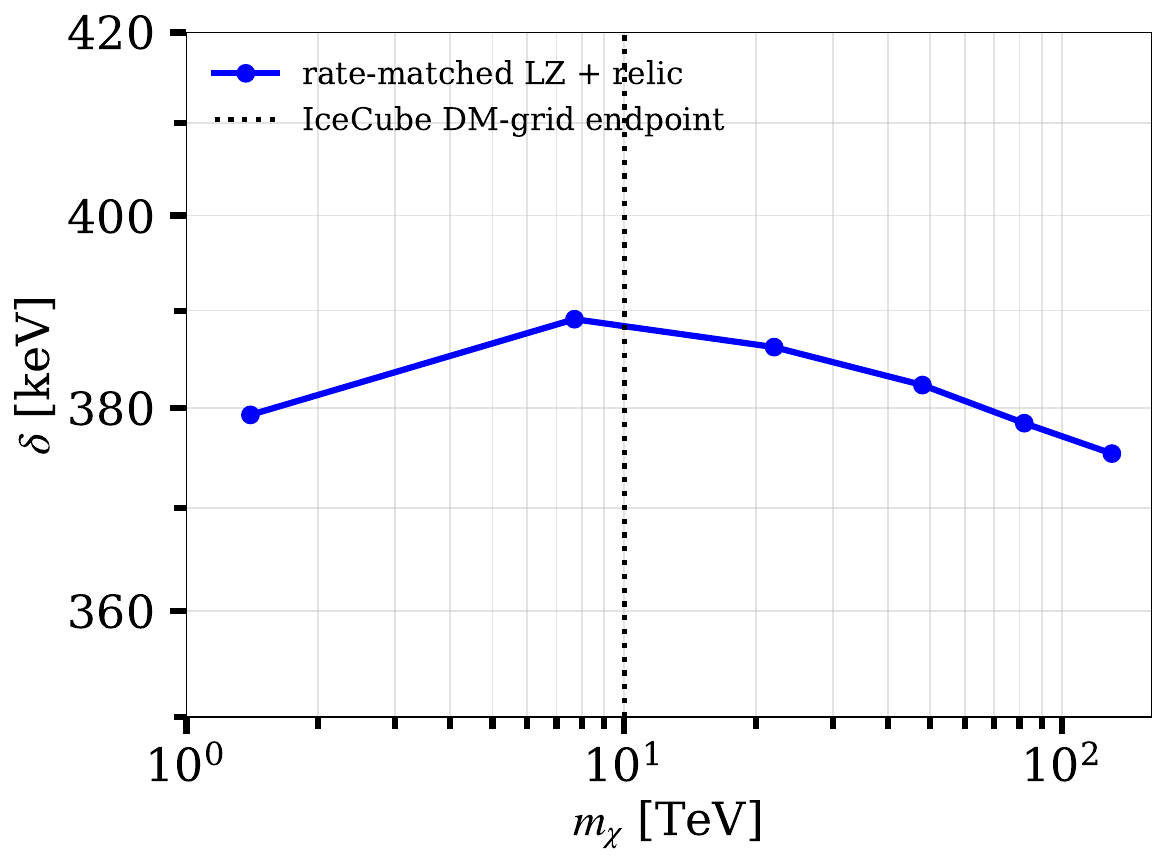}
\caption{Independently rate-matched endothermic splitting $\delta_\star$ as a function of DM mass for the six HC-MDM thermal representations. Blue circles connected by a solid line show the values obtained by requiring one expected accepted LZ inelastic event at each thermal-relic mass, using the corrected neutral-current normalization adopted in this work. The vertical black dotted line marks $m_\chi=10\,\TeV$, the upper endpoint of the DM signal-hypothesis grid released with the 2025 IceCube Solar analysis.}
\label{fig:bookkeeping}
\end{figure}

Fig.~\ref{fig:bookkeeping} summarizes the independently determined rate-matched splittings along the HC-MDM thermal-relic sequence. The required splitting is not constant with the DM mass: it first increases from approximately $379\,\keV$ for $3_M2_D$ to about $389\,\keV$ for $5_M4_D$ and then gradually decreases toward approximately $375\,\keV$ for the largest representation. This behavior reflects the interplay between the DM mass, the endothermic kinematics, the high-velocity tail of the halo distribution, and the xenon recoil acceptance entering the LZ event rate. The vertical line at $10\,\TeV$ also makes clear that only the first two thermal benchmarks lie inside the mass range covered by the released IceCube DM templates, while the four heavier representations require the high-mass response treatment developed below.

\section{Mapping the Solar signal to IceCube}
\label{sec:icecube}

The IceCube Collaboration recently searched for high-energy neutrinos from the Sun using ten years of data and found no significant evidence for a DM signal \cite{IceCubeSolar2025}. The released analysis provides DM signal hypotheses for masses between $20\,\GeV$ and $10\,\TeV$ and for representative annihilation channels including $b\bar b$, $W^+W^-$, $\tau^+\tau^-$, and direct neutrinos. The calculation includes electroweak corrections to the annihilation spectra and neutrino propagation through the Sun, including charged- and neutral-current interactions, regeneration, and flavor evolution \cite{Charon2020,Blennow2008,Bauer2021,IceCubeSolar2025}. Among the six HC-MDM thermal benchmarks considered here, only $3_M2_D$ and $5_M4_D$ lie inside the released IceCube mass grid; the remaining four candidates have masses between $22$ and $130\,\TeV$.

For a Solar DM annihilation rate $\Gamma_A$ and branching fractions $B_f$ into final states $f$, the expected number of detected neutrino events can be written schematically as
\begin{equation}
N_\nu=
\frac{\Gamma_A T_{\rm live}}{4\pi D_\odot^2}
\sum_f B_f
\sum_\alpha
\int dE_\nu\,
\frac{dN_{\nu_\alpha}^{f}}{dE_\nu}
\mathcal A_{\rm eff}^{\alpha}(E_\nu),
\label{eq:icecubeevents}
\end{equation}
where $T_{\rm live}$ is the detector livetime, $D_\odot$ is the Sun--Earth distance, $\alpha$ labels neutrino flavor, $dN_{\nu_\alpha}^{f}/dE_\nu$ is the propagated neutrino yield per annihilation, and $\mathcal A_{\rm eff}^{\alpha}$ is the IceCube effective area for the relevant event selection. Eq.~\eqref{eq:icecubeevents} makes clear that an IceCube limit on the Solar annihilation rate is not a universal number: it depends on both the DM mass and the annihilation spectrum through the energy-dependent neutrino yield and detector response.

For the HC-MDM benchmarks lying inside the released grid, we use the $W^+W^-$ template as a reference because the ground-state DM particle has electroweak interactions and annihilates into energetic electroweak final states. Refs.~\cite{PospelovRamani2026,DiMauroShaikh2026Solar}, using the same IceCube data set for a thermal Higgsino, infer a $90\%$ C.L. upper limit on the $W^+W^-$-equivalent Solar annihilation rate of approximately
\begin{equation}
\Gamma_{\rm IC}^{90}(WW;1.08\,\TeV)
\simeq
1.5\times10^{19}\,\mathrm{s}^{-1}.
\label{eq:icwwanchor}
\end{equation}
Here $\Gamma_{\rm IC}^{90}(WW;m_\chi)$ denotes the IceCube upper limit for a pure $W^+W^-$ final state evaluated at the corresponding DM mass. In particular, Eq.~\eqref{eq:icwwanchor} should not be interpreted as a mass-independent limit.

For later convenience, we define the response-matched comparison ratio
\begin{equation}
\mathcal R_{\rm IC,resp}^{WW}(m_\chi)
\equiv
\frac{C_\odot/2}
{\Gamma_{\rm IC,resp}^{90}(WW;m_\chi)}.
\label{eq:icratio}
\end{equation}
This quantity provides a compact measure of how the predicted equilibrium Solar annihilation rate compares with the IceCube sensitivity for a $W^+W^-$-like neutrino spectrum. It can therefore be used to rank the different HC-MDM benchmarks according to their expected tension with IceCube and to estimate how strongly the effective neutrino yield would need to be suppressed for a given model to remain compatible with the data.
Since Sec.~\ref{sec:postcapture} establishes capture--annihilation equilibrium, the numerator $C_\odot/2$ is the predicted total Solar annihilation rate. A value
\begin{equation}
\mathcal R_{\rm IC,resp}^{WW}>1
\end{equation}
therefore means that the predicted annihilation normalization exceeds the IceCube response corresponding to a pure $W^+W^-$ spectrum. If only an effective fraction $f_{WW}^{\rm eff}$ of the HC-MDM annihilation spectrum produces an IceCube response comparable to the $W^+W^-$ template, the relevant comparison becomes
\begin{equation}
f_{WW}^{\rm eff}\,
\mathcal R_{\rm IC,resp}^{WW}.
\end{equation}
A complete HC-MDM likelihood would require the representation-dependent annihilation spectrum and the corresponding IceCube detector templates. We therefore use $\mathcal R_{\rm IC,resp}^{WW}$ as a transparent common diagnostic.

\subsection{Response-matched extrapolation above the released
$10\,\TeV$ grid}
\label{subsec:highmass}

Four of the six HC-MDM thermal benchmarks have masses above $10\,\TeV$, whereas the DM signal hypotheses released with the 2025 IceCube Solar analysis extend only up to $10\,\TeV$ \cite{IceCubeSolar2025}. This upper boundary is a limit of the published DM template grid rather than a detector-energy cutoff: the main IceCube array remains sensitive to neutrinos at considerably higher energies. It is therefore useful to investigate how the Solar-neutrino sensitivity may continue into the mass range relevant for the heavier HC-MDM representations. Since no IceCube likelihood or DM signal templates have been released above $10\,\TeV$, however, this continuation must be regarded as an exploratory recast rather than as an official IceCube limit.

An important point is that the quantity to be extrapolated is the IceCube response to a given Solar annihilation rate, rather than the published upper limit on a DM--proton scattering cross section. A scattering-cross-section limit contains two conceptually different ingredients: the efficiency with which DM is captured by the Sun and the efficiency with which the resulting annihilation neutrinos are detected by IceCube. In the present work the HC-MDM capture rate $C_\odot$ is calculated independently. Extrapolating the mass dependence of the published scattering bound directly would therefore reintroduce part of the Solar-capture suppression that is already included in our calculation.

To make this distinction explicit, consider a pure annihilation channel $f$. At the level of a simple event-counting description, the detector response per Solar annihilation can be written schematically as
\begin{equation}
{\cal K}_f(m_\chi)
=
\sum_\alpha
\int dE_\nu\,
{\cal A}_{\rm eff}^{\alpha}(E_\nu)
\frac{dN_{\nu_\alpha}^{f}}{dE_\nu}(m_\chi).
\label{eq:highmassK}
\end{equation}
Here $f$ denotes the DM annihilation final state, such as $W^+W^-$, and $\alpha$ labels the neutrino flavor. The function ${\cal A}_{\rm eff}^{\alpha}(E_\nu)$ denotes schematically the energy-dependent IceCube acceptance for neutrinos of flavor $\alpha$ and energy $E_\nu$. Thus ${\cal K}_f$ measures how efficiently one Solar DM annihilation into channel $f$ is converted into detectable events.
Its dependence on $m_\chi$ is determined by two competing effects. Increasing the DM mass changes the neutrino spectrum produced in the annihilation, but it also changes how that spectrum is modified during propagation through the Sun and how it overlaps with the energy-dependent detector response. In particular, at multi-TeV energies the emerging spectrum does not simply scale with the injected DM energy. Charged-current interactions absorb the highest-energy neutrinos, neutral-current interactions reduce their energy, and $\nu_\tau$ regeneration repopulates lower-energy states. As a result, part of the injected high-energy flux is redistributed toward the sub-TeV and TeV range rather than following a universal $m_\chi^{-2}$ suppression \cite{Charon2020,IceCubeSolar2025,MaityEtAl2023}.

At fixed experimental sensitivity, a simple counting-level extrapolation would therefore scale approximately as
\begin{equation}
\Gamma_{A,90}^{f}(m_\chi)
\simeq
\Gamma_{A,90}^{f}(m_0)
\frac{{\cal K}_f(m_0)}
{{\cal K}_f(m_\chi)}.
\label{eq:responsematched}
\end{equation}
The actual IceCube analysis is more sophisticated than Eq.~\eqref{eq:responsematched}, since its likelihood retains information on reconstructed energy, angular distance from the Sun, and the event selection. Eq.~\eqref{eq:highmassK} is therefore introduced only to identify the physical quantity that controls the mass dependence of the response. We do not perform an independent convolution with the published effective area in the present analysis. Instead, we infer the annihilation-rate sensitivity directly from the response already encoded in the IceCube result.

The IceCube Solar-DM analysis presents its principal constraints as upper limits on the DM--proton scattering cross section for each DM mass and annihilation channel. In particular, strong limits are reported on the spin-dependent proton cross section, $\sigma_{p,{\rm IC}}^{\rm SD,90}(m_\chi)$ \cite{IceCubeSolar2025}. The use of the SD limit here should not be confused with the interaction responsible for Solar capture in HC-MDM. In HC-MDM the initial capture is produced by the coherent inelastic $Z$-mediated interaction calculated in the previous sections. The IceCube SD cross section is used only as an intermediate quantity with which to reconstruct the underlying neutrino sensitivity of the experiment.

The reason this conversion is possible is that the IceCube interpretation assumes capture--annihilation equilibrium in the Sun. Under this assumption $\Gamma_A= C_\odot/2$
so that an upper limit on the scattering cross section can be mapped into an upper limit on the Solar annihilation rate by evaluating the corresponding capture rate at the excluded cross section. For the published SD limit,
\begin{equation}
\Gamma_{\rm IC}^{90}(WW;m_\chi)
=
\frac{1}{2}
C_\odot^{\rm SD}
\left[
m_\chi,
\sigma_{p,{\rm IC}}^{\rm SD,90}(WW;m_\chi)
\right],
\label{eq:icecube_reverse}
\end{equation}
where $C_\odot^{\rm SD}$ is the Solar capture rate associated with SD proton scattering using the capture prescription entering the IceCube interpretation. In the optically thin regime relevant here, $C_\odot^{\rm SD}$ is proportional to $\sigma_p^{\rm SD}$, so Eq.~\eqref{eq:icecube_reverse} directly converts the published scattering limit into the corresponding annihilation-rate sensitivity.

Once this conversion has been performed, the reference SD interaction no longer enters our HC-MDM comparison. The resulting quantity $\Gamma_{\rm IC}^{90}$ characterizes the Solar annihilation rate that would produce an IceCube signal at the experimental upper-limit level for a specified annihilation spectrum. It can therefore be compared directly with our independently calculated HC-MDM prediction,
\begin{equation}
\Gamma_A^{\rm HC-MDM}
\simeq
\frac{C_\odot^{\rm HC-MDM}}{2},
\end{equation}
whose validity was established in Sec.~\ref{sec:postcapture}.

Ref.~\cite{PospelovRamani2026} performs this reverse engineering using the SD limits and the Solar-capture prescription associated with the 2025 IceCube analysis. This procedure has the advantage that the neutrino production, propagation, detector acceptance, and likelihood response already contained in the IceCube result are retained without attempting to reconstruct them independently. We use the $1.08\,\TeV$ anchor in Eq.~\eqref{eq:icwwanchor}; the same inferred $W^+W^-$ response curve reaches approximately
\begin{equation}
\Gamma_{\rm IC}^{90}(WW;10\,\TeV)
\simeq
1.2\times10^{19}\,\mathrm{s}^{-1}
\label{eq:ic10anchor}
\end{equation}
at the upper endpoint of the released IceCube DM grid.

The inferred annihilation-rate sensitivity therefore changes only weakly between approximately $1$ and $10\,\TeV$. We characterize this last published mass decade by the logarithmic slope
\begin{equation}
\alpha_{\rm resp}
=
\frac{
\ln\left[
\frac{\Gamma_{\rm IC}^{90}(WW;10\,\TeV)}{
\Gamma_{\rm IC}^{90}(WW;1.08\,\TeV)}
\right]
}{
\ln(10/1.08)
}
\simeq
-0.10.
\label{eq:alpharesponse}
\end{equation}

As our central exploratory extrapolation, we continue this endpoint slope above $10\,\TeV$,
\begin{align}
\Gamma_{A,90}^{WW,\rm resp}(m_\chi)
={}&1.2\times10^{19}\,\mathrm{s}^{-1}
\left(\frac{m_\chi}{10\,\TeV}\right)^{-0.10},\nonumber\\
& m_\chi>10\,\TeV.
\label{eq:highmasslimit}
\end{align}
This prescription should be interpreted as a continuation of the weak mass dependence of the IceCube annihilation-rate response observed over the last released decade. It is not an official IceCube limit and is not assigned a formal IceCube confidence level above $10\,\TeV$.

To assess how strongly the HC-MDM conclusions depend on this extrapolation, we also introduce a deliberately conservative response-loss stress prescription,
\begin{align}
\Gamma_{A,90}^{WW,\rm stress}(m_\chi)
={}&1.2\times10^{19}\,\mathrm{s}^{-1}
\left(\frac{m_\chi}{10\,\TeV}\right)^{1/2},\nonumber\\
& m_\chi>10\,\TeV.
\label{eq:highmassstress}
\end{align}
This second scaling is not inferred from the IceCube data. Instead, it allows the annihilation-rate sensitivity to deteriorate appreciably above the published mass range and is used only to test the robustness of the high-mass conclusions. The results obtained with the central response continuation and with this conservative stress prescription are presented and compared in Sec.~\ref{sec:results}.
\section{Results}
\label{sec:results}

We now present the Solar-capture results in a fixed sequence. We first specify which HC-MDM benchmark definitions are being compared, then show the corresponding Solar annihilation rates and their elemental composition. We subsequently confront these predictions with the IceCube response defined in Sec.~\ref{sec:icecube}. Finally, we analyze separately the proposed $\delta=570\,\keV$ high-velocity trajectories and compare our treatment with the results obtained in Ref.~\cite{SmirnovGriffithBeacom2026}.

\subsection{Solar capture at the thermal-relic benchmarks}

We first evaluate the capture rates at the primary benchmark points using the independently rate-matched splittings of Eq.~\eqref{eq:deltarematched}. After the post-capture cooling and capture--annihilation equilibrium checks of Sec.~\ref{sec:postcapture}, the corresponding Solar annihilation rate is $\Gamma_A=C_\odot/2$. We obtain
\begin{align}
\frac{C_\odot}{2}
={}&
3.232\times10^{22},\,
1.121\times10^{21},\,
1.500\times10^{20},
\nonumber\\
&
3.424\times10^{19},\,
1.271\times10^{19},\,
5.310\times10^{18}
\ \mathrm{s}^{-1},
\label{eq:ratesprimary}
\end{align}
for $3_M2_D$, $5_M4_D$, $7_M6_D$, $9_M8_D$, $11_M10_D$, and $13_M12_D$, respectively. The rates decrease rapidly along the thermal sequence because of the decreasing incident DM number density and the progressively more restrictive kinematics for capturing very heavy DM. Relative to the isotope-resolved rates evaluated at the splittings quoted in Ref.~\cite{SmirnovGriffithBeacom2026}, the independently rate-matched splittings reduce $C_\odot/2$ by approximately $3\%$, $6\%$, $8\%$, $10\%$, $12\%$, and $13\%$, respectively. Although the change in $\delta$ is only a few keV, its impact becomes increasingly important for the heavier representations because endothermic Solar capture becomes more sensitive to the splitting near kinematic closure.

\begin{figure}[t]
\centering
\includegraphics[width=\columnwidth]{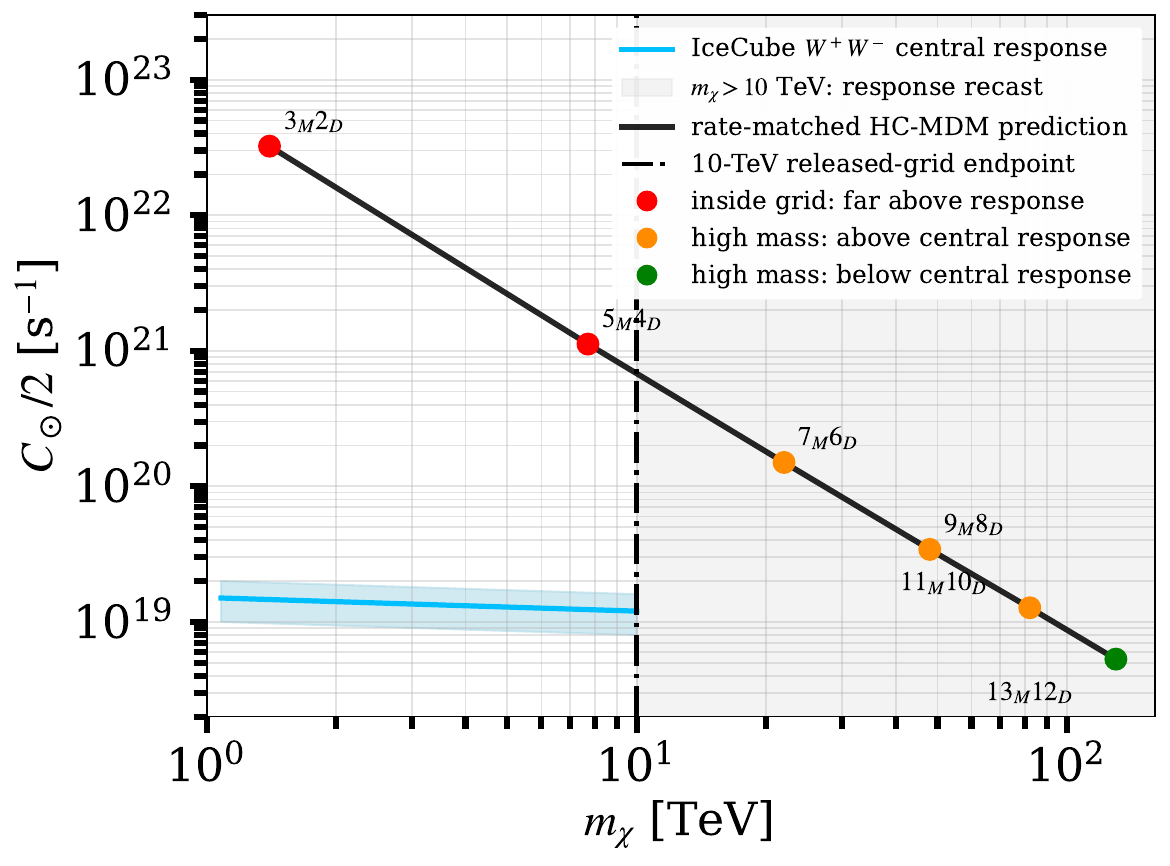}
\caption{Predicted equilibrium Solar DM annihilation rate $C_\odot/2$ for the six HC-MDM thermal benchmarks using the independently rate-matched splittings. The cyan curve and band show the mass-dependent IceCube $W^+W^-$ response over the released $1.08$--$10\,\TeV$ grid. The vertical line marks the $10\,\TeV$ endpoint; the gray region requires a response extrapolation. Red, orange, and green markers indicate benchmarks inside the released grid, above the central high-mass response, and below it, respectively.}
\label{fig:rates}
\end{figure}

Fig.~\ref{fig:rates} summarizes the predicted equilibrium Solar annihilation rates for the six HC-MDM thermal benchmarks using the independently rate-matched splittings determined in Sec.~\ref{subsec:lzrematching}. The black line connects the HC-MDM predictions, which decrease rapidly with increasing $m_\chi$ because of the lower incident DM number density and the increasingly restrictive capture kinematics for very heavy DM.

The cyan curve shows the mass-dependent IceCube $W^+W^-$ annihilation-rate response over the released $1.08$--$10\,\TeV$ DM grid, while the surrounding cyan band indicates the corresponding response reference range. The two lightest HC-MDM benchmarks lie within the released IceCube mass range and predict annihilation rates far above this response. The vertical line at $10\,\TeV$ marks the endpoint of the released DM-template grid, while the gray region identifies the four heavier benchmarks for which the IceCube response must be extrapolated. Their phenomenological interpretation is discussed below using the response-matched continuation introduced in Sec.~\ref{subsec:highmass}.

\subsection{Which Solar nuclei dominate the capture rate?}

\begin{figure}[t]
\centering
\includegraphics[width=\columnwidth]{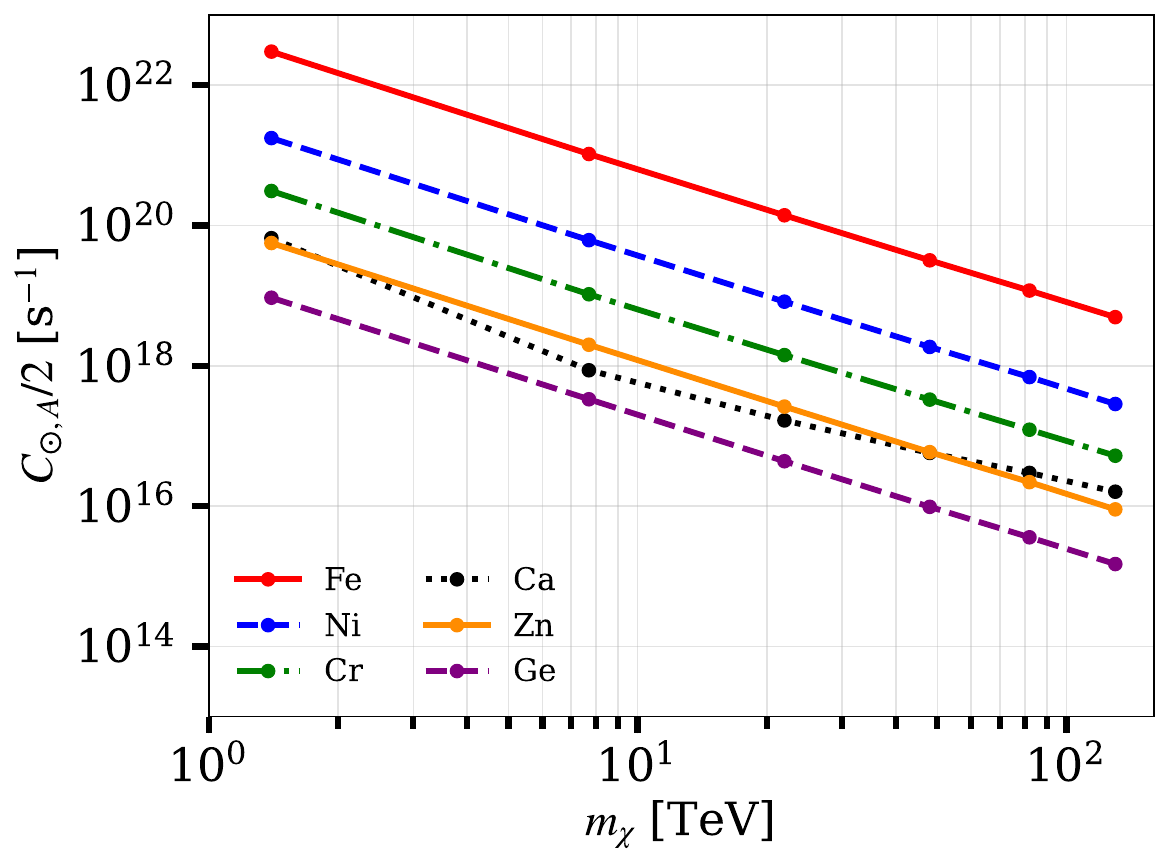}
\caption{Element-resolved contributions $C_{\odot,A}/2$ to the Solar annihilation normalization at the six primary HC-MDM benchmarks. The calculation uses the same SHM distribution, natural isotope abundances, coherent weak charge, and Helm form factor as the total capture calculation. Fe dominates the rate for all six representations, with Ni providing most of the remaining contribution.}
\label{fig:elements}
\end{figure}

The elemental composition of the capture rate is shown in Fig.~\ref{fig:elements}. The numerical convergence of the isotope-resolved capture calculation and the importance of retaining natural isotope mixtures near kinematic closure are documented in Appendix~\ref{app:validation}. Iron dominates all six primary benchmarks, contributing approximately $93\%$ of the total rate, while Ni supplies most of the remainder. Contributions from Cr, Ca, Zn, Ge, and heavier trace species are subdominant in this benchmark region.

This hierarchy is important for two reasons. First, the Solar-composition dependence is controlled primarily by the Fe and Ni abundances rather than by an incoherent sum over many nuclear species. Second, the same two elements also control the finite-temperature reopening of the endothermic channel at larger mass splitting, discussed below.

\subsection{Comparison with the IceCube response}

We now compare the Solar annihilation rates obtained with our independently rate-matched splittings with the mass-dependent IceCube $W^+W^-$ response introduced in Sec.~\ref{sec:icecube}. The first two HC-MDM benchmarks, $3_M2_D$ and $5_M4_D$, lie within the released IceCube DM mass grid. The four heavier benchmarks lie above $10\,\TeV$ and therefore require the response-matched continuation of Eq.~\eqref{eq:highmasslimit}. For these states we also use the conservative stress prescription of Eq.~\eqref{eq:highmassstress} to quantify the dependence on the high-mass extrapolation.

\begin{figure}[t]
\centering
\includegraphics[width=\columnwidth]{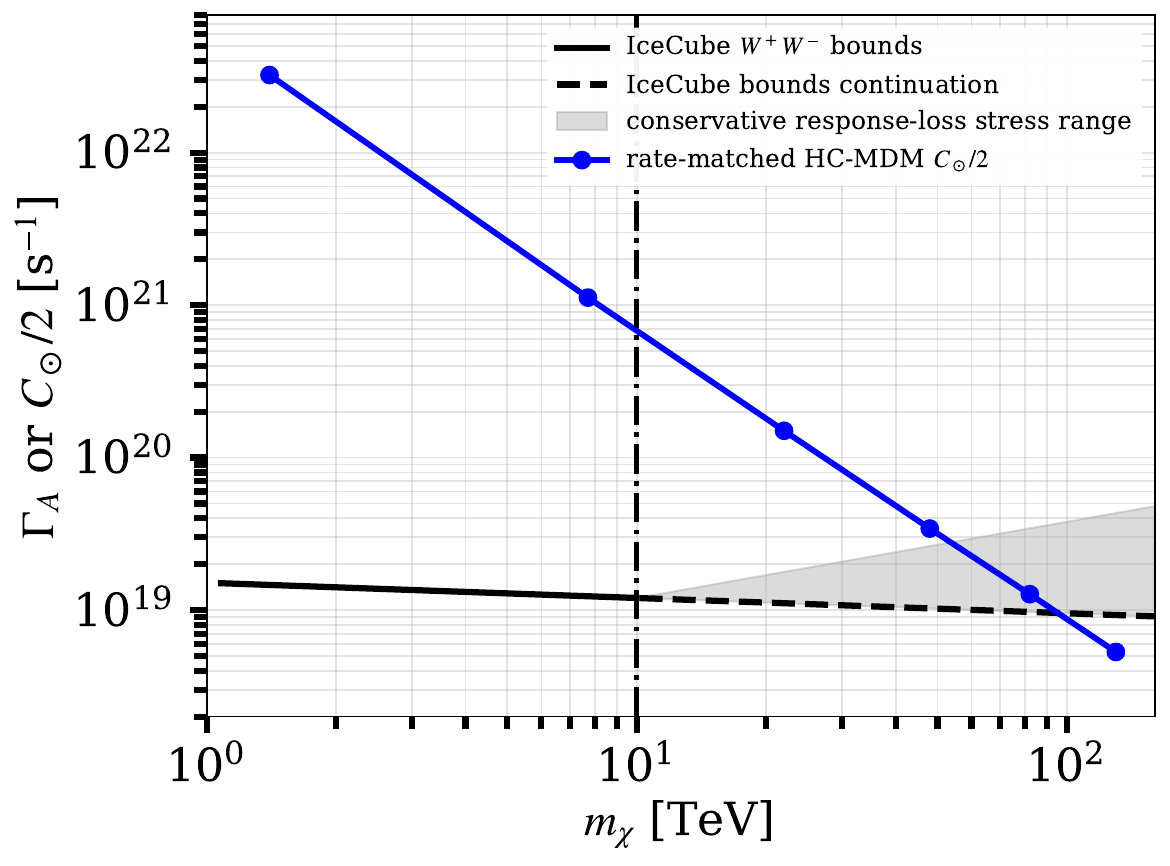}
\caption{Comparison between the predicted HC-MDM Solar annihilation rate and the IceCube $W^+W^-$ response. The blue curve shows $C_\odot/2$ for the six thermal benchmarks using the independently rate-matched splittings. The solid black curve denotes the IceCube response within the released mass grid, while the dashed curve shows the response-matched continuation above $10\,\TeV$. The gray band indicates the conservative response-loss stress range. The vertical dash-dotted line marks the $10\,\TeV$ endpoint of the released IceCube DM templates.}
\label{fig:highmassresponse}
\end{figure}

Fig.~\ref{fig:highmassresponse} shows the comparison directly in Solar-annihilation-rate space. The blue curve gives the HC-MDM prediction $C_\odot/2$ for the six thermal benchmarks. The solid black curve shows the inferred IceCube $W^+W^-$ response within the released mass range, while the dashed black curve gives its response-matched continuation above $10\,\TeV$. The gray band indicates how much the high-mass IceCube sensitivity is allowed to weaken under the conservative response-loss stress prescription. The vertical line marks the endpoint of the released IceCube DM grid.

For the two benchmarks inside the released grid, the predicted annihilation rates lie far above the corresponding $W^+W^-$ response. Using Eq.~\eqref{eq:icratio}, we find
\begin{equation}
\mathcal R_{\rm IC,resp}^{WW}
\simeq
2.21\times10^{3}
\end{equation}
for $3_M2_D$, and
\begin{equation}
\mathcal R_{\rm IC,resp}^{WW}
\simeq
90.9
\end{equation}
for $5_M4_D$. The Solar-neutrino prediction is therefore well above the IceCube $W^+W^-$ response normalization for both light thermal benchmarks.

For the four benchmarks above $10\,\TeV$, the central response continuation gives
\begin{equation}
\Gamma_{A,90}^{WW,\rm resp}
=
\left(11.1,10.3,9.72,9.28\right) \times 10^{18}\,
\mathrm{s}^{-1},
\end{equation}
at $m_\chi=22$, $48$, $82$, and $130\,\TeV$, respectively. Comparing these values with the Solar annihilation rates calculated using our independently rate-matched splittings gives
\begin{equation}
\mathcal R_{\rm IC,resp}^{WW}
=
13.53,\,
3.34,\,
1.31,\,
0.57.
\label{eq:highmassratiosI}
\end{equation}

\begin{figure}[t]
\centering
\includegraphics[width=\columnwidth]{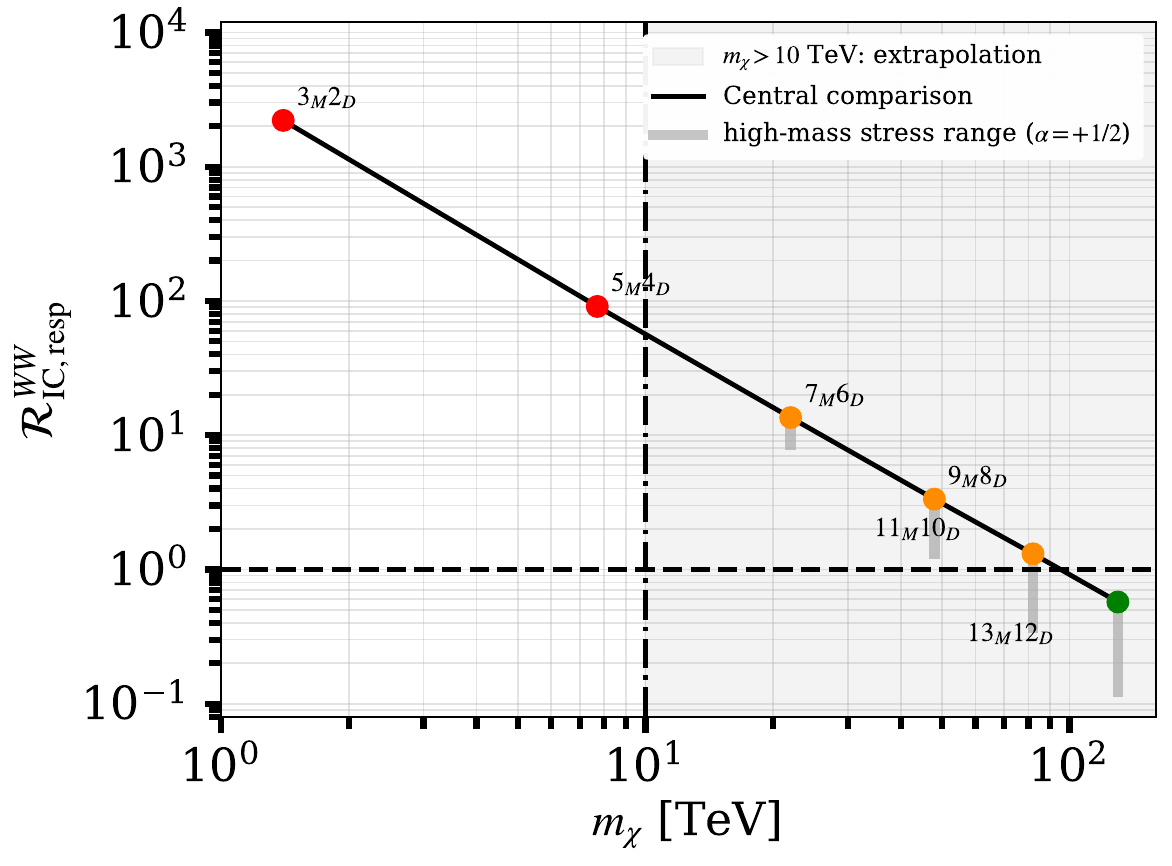}
\caption{Ratio $\mathcal R_{\rm IC,resp}^{WW}$ between the predicted HC-MDM Solar annihilation rate and the mass-dependent IceCube $W^+W^-$ response. The black curve shows the central comparison for the six thermal benchmarks using the independently rate-matched splittings. The horizontal dashed line marks $\mathcal R_{\rm IC,resp}^{WW}=1$, while the vertical dash-dotted line indicates the $10\,\TeV$ endpoint of the released IceCube DM grid. For $m_\chi>10\,\TeV$, the gray intervals show the range between the central response continuation and the conservative response-loss stress prescription.}
\label{fig:icratio}
\end{figure}

Fig.~\ref{fig:icratio} presents the same comparison in dimensionless form. The horizontal dashed line at
\begin{equation}
\mathcal R_{\rm IC,resp}^{WW}=1
\end{equation}
corresponds to equality between the predicted HC-MDM annihilation rate and the $W^+W^-$ IceCube response. The first two benchmarks are far above this line. Among the heavier states, $7_M6_D$ and $9_M8_D$ remain above the central response continuation, $11_M10_D$ lies only moderately above it, and $13_M12_D$ lies below it.

For the four benchmarks above $10\,\TeV$, the gray vertical intervals show how the ratios change when the central continuation is replaced by the conservative response-loss stress prescription. These values use the rate-matched splittings of Eq.~\eqref{eq:deltarematched}; they should not be confused with the slightly larger ratios obtained if the original splittings quoted in Ref.~\cite{SmirnovGriffithBeacom2026} are used instead. In this case we obtain
\begin{equation}
\mathcal R_{\rm IC,stress}^{WW}
=
8.43,\,
1.30,\,
0.370,\,
0.123.
\label{eq:highmassstressratios}
\end{equation}

The resulting hierarchy is clear. The $7_M6_D$ benchmark remains well above the extrapolated IceCube response even under the conservative stress prescription. The $9_M8_D$ benchmark also remains slightly above the response throughout the adopted extrapolation range, although the margin is considerably smaller. The $11_M10_D$ benchmark is genuinely sensitive to the assumed high-mass detector response: it lies modestly above the central continuation but below the conservative stress curve. Finally, $13_M12_D$ remains below the extrapolated response in both treatments.

We stress that the statements for $m_\chi>10\,\TeV$ are response-based recast diagnostics and should not be interpreted as formal IceCube confidence-level exclusions. A definitive test of these heavy HC-MDM benchmarks would require dedicated IceCube signal templates extending beyond $10\,\TeV$ together with the representation-dependent HC-MDM annihilation spectrum.

\subsection{The \texorpdfstring{$\delta=570\,\keV$}{delta = 570 keV}
high-velocity solutions}
\label{subsec:highdelta}

Ref.~\cite{SmirnovGriffithBeacom2026} considers an alternative interpretation of the LZ event in which the mass splitting is increased to $\delta=570\,\keV$
for the two lightest HC-MDM representations. This value is not a special prediction of the model. Rather, it is chosen because such a large splitting raises the minimum speed required to produce the $248\,\keV$ LZ recoil to approximately $10^3\,\kms$. The terrestrial signal must therefore originate from an additional high-velocity DM component rather than from the ordinary SHM population. Along the corresponding thermal-relic trajectories, Ref.~\cite{SmirnovGriffithBeacom2026} finds
\begin{equation}
(m_\chi,y)
=
(1.47\,\TeV,7.45\times10^{-3})
\end{equation}
for $3_M2_D$ and
\begin{equation}
(m_\chi,y)
=
(8.5\,\TeV,1.79\times10^{-2})
\end{equation}
for $5_M4_D$.

The crucial point for Solar capture is that the terrestrial and Solar velocity requirements are very different. If the fast population represents a subdominant additional component with local density fraction $f_s\ll1$, the Solar-frame distribution contains both the ordinary Galactic population and the fast component,
\begin{equation}
f_\odot(u)
=
(1-f_s)f_{\rm MW}(u)
+
f_s f_{\rm fast}(u).
\label{eq:halomixture}
\end{equation}
The ordinary halo therefore remains available for capture. Moreover, DM falling into the Solar potential is strongly accelerated, so particles that are too slow to produce the terrestrial recoil can nevertheless cross the endothermic threshold inside the Sun.

\begin{figure}[t]
\centering
\includegraphics[width=\columnwidth]{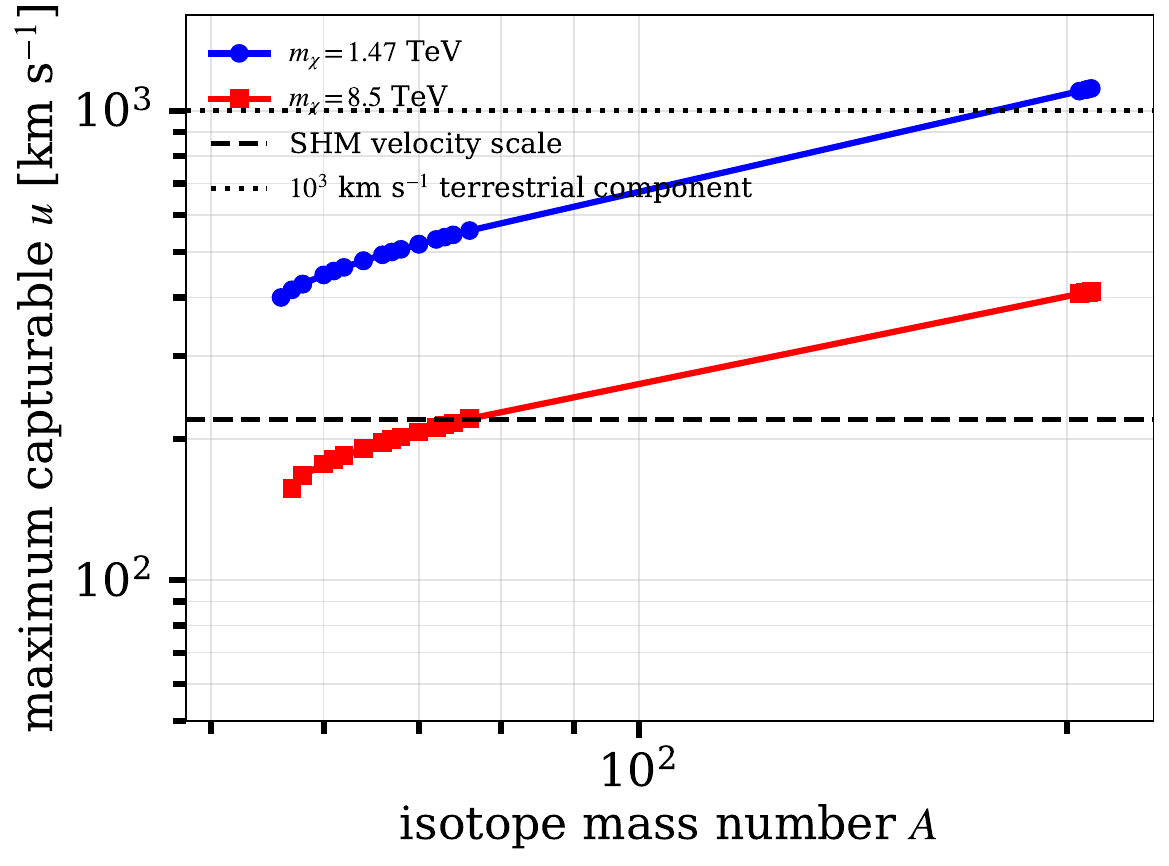}
\caption{Maximum Solar-frame asymptotic speed $u_{\rm max}$ allowing a single cold-target capture scatter at the Solar center for $\delta=570\,\keV$. Blue circles correspond to $m_\chi=1.47\,\TeV$ and red squares to $m_\chi=8.5\,\TeV$. The horizontal reference lines indicate $220\,\kms$ and $10^3\,\kms$. The dominant Fe and Ni targets preferentially capture particles from the ordinary Galactic halo rather than the extreme-velocity population relevant for the terrestrial recoil.}
\label{fig:highdeltawindows}
\end{figure}

This difference is illustrated in Fig.~\ref{fig:highdeltawindows}, which shows the maximum asymptotic Solar-frame speed that can be captured in a single collision at the Solar center. For $m_\chi=1.47\,\TeV$, the abundant Fe and Ni isotopes efficiently capture particles with asymptotic speeds of a few hundred $\kms$. For $m_\chi=8.5\,\TeV$, the capturable velocities are even smaller. The $\sim10^3\,\kms$ population invoked for the terrestrial LZ event is therefore not the population that dominates Solar capture; for the heavier trajectory it is generally too energetic to become gravitationally bound in one collision with Fe or Ni.

In the stationary-target approximation, using natural isotope mixtures, we obtain
\begin{align}
C_\odot^{T_A=0}(1.47\,\TeV)
&=
1.61\times10^{19}\,\mathrm{s}^{-1},
\\
C_\odot^{T_A=0}(8.5\,\TeV)
&=
7.12\times10^{17}\,\mathrm{s}^{-1}.
\label{eq:highdelta_stationary}
\end{align}
Close to kinematic closure, however, nuclear thermal motion becomes important. Including the local Maxwell--Boltzmann velocity distribution of each Solar isotope gives
\begin{align}
C_\odot^{T_A\neq0}(1.47\,\TeV)
&=
(1.766\pm0.008)\times10^{20}\,\mathrm{s}^{-1},
\\
C_\odot^{T_A\neq0}(8.5\,\TeV)
&=
(8.73\pm0.15)\times10^{18}\,\mathrm{s}^{-1},
\label{eq:highdelta_thermal}
\end{align}
where the quoted uncertainties quantify the numerical dispersion of the Sobol integration. Thermal nuclear motion therefore increases the capture rate by approximately a factor $11$--$12$ at $\delta=570\,\keV$. The independent $T_A\rightarrow0$ recovery test of the finite-temperature implementation is given in Appendix~\ref{app:validation}. The enhancement is dominated by Fe and Ni: for the $1.47\,\TeV$ trajectory they contribute approximately $82.5\%$ and $14.8\%$ of the finite-temperature rate, respectively.

\begin{figure}[t]
\centering
\includegraphics[width=\columnwidth]{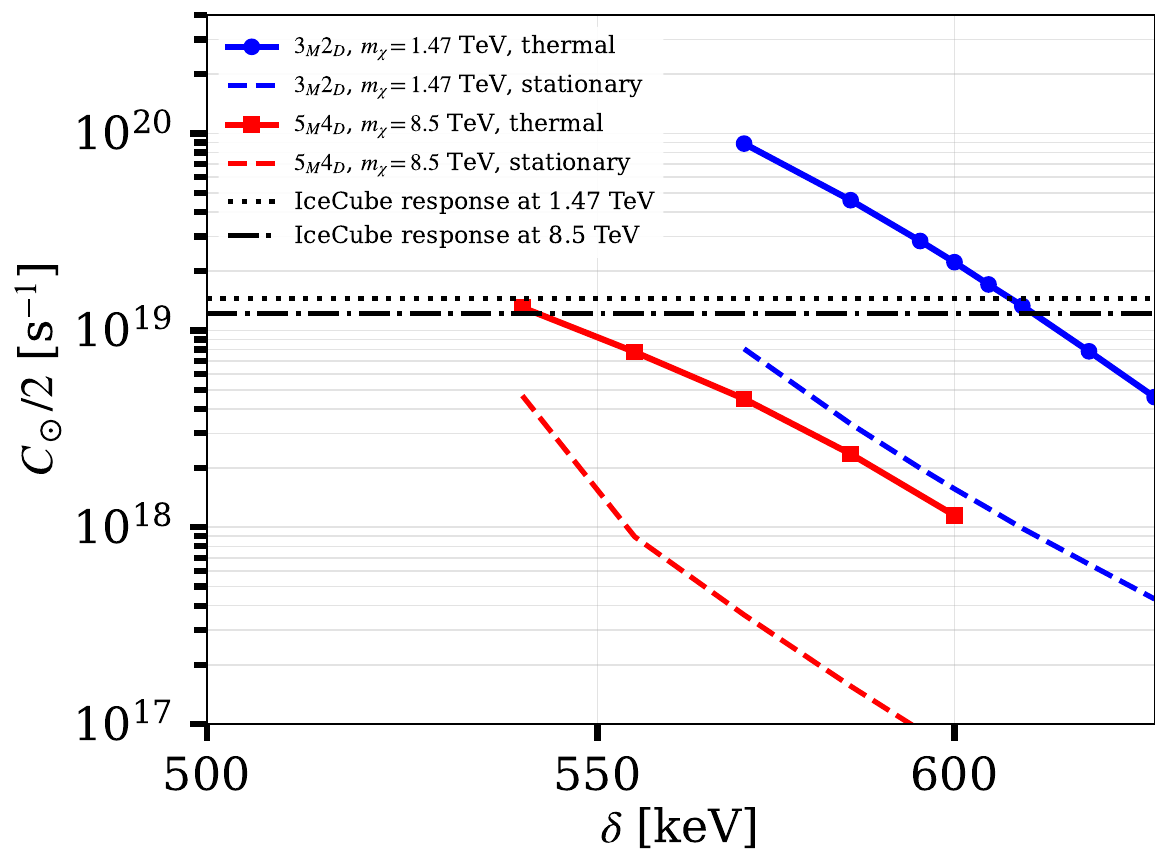}
\caption{Equilibrium Solar annihilation normalization $C_\odot/2$ as a function of the splitting for the $1.47\,\TeV$ and $8.5\,\TeV$ high-velocity trajectories. Solid curves include finite-temperature nuclear motion and dashed curves use stationary nuclei. The horizontal lines show the mass-matched IceCube $W^+W^-$ response. Thermal motion substantially reopens Solar capture near kinematic closure.}
\label{fig:highdelta}
\end{figure}

Fig.~\ref{fig:highdelta} shows how this thermal reopening develops with increasing splitting. The stationary-target rate falls rapidly as the endothermic channel approaches closure, while the thermal velocity tails of the Solar nuclei keep part of the Fe and Ni phase space accessible. For the $1.47\,\TeV$ trajectory, the finite-temperature prediction remains above the mass-matched IceCube $W^+W^-$ response up to approximately $\delta\simeq608\,\keV$.

Using the mass-dependent IceCube response defined in Sec.~\ref{sec:icecube}, we find
\begin{align}
\mathcal R_{\rm IC,resp}^{WW}
(1.47\,\TeV,570\,\keV)
&\simeq
6.07,
\\
\mathcal R_{\rm IC,resp}^{WW}
(8.5\,\TeV,570\,\keV)
&\simeq
0.36.
\label{eq:icratio570}
\end{align}
Thus the $1.47\,\TeV$ trajectory exceeds the $W^+W^-$ response normalization if more than approximately $16.5\%$ of its annihilation signal has a comparable neutrino response. The $8.5\,\TeV$ trajectory instead remains below the corresponding IceCube response even for a fully $W^+W^-$-equivalent spectrum.

It is useful to understand why these results differ substantially from the Solar estimates quoted in Ref.~\cite{SmirnovGriffithBeacom2026}. For the ordinary SHM benchmarks, our stationary-target capture calculation agrees closely with that work once the different weak-cross-section normalization is taken into account. The discrepancy therefore arises specifically in the treatment of the $\delta=570\,\keV$ high-velocity solutions.

Ref.~\cite{SmirnovGriffithBeacom2026} assumes a fast component with fractional local density $f_s\sim10^{-4}$--$10^{-3}$. Comparing the suppression of its quoted high-velocity Solar rates with the suppression produced by endothermic kinematics alone in our stationary-target calculation gives additional factors
\begin{equation}
\epsilon_{\rm hv}^{3_M2_D}
\simeq4.6\times10^{-4},
\qquad
\epsilon_{\rm hv}^{5_M4_D}
\simeq5.6\times10^{-4}.
\end{equation}
These values fall directly within the quoted range of $f_s$. The published high-velocity Solar rates are therefore numerically consistent with an additional suppression of order the fast-component density fraction.

We emphasize that this is a normalization diagnostic rather than a reconstruction of the unpublished numerical implementation of Ref.~\cite{SmirnovGriffithBeacom2026}. Physically, however, a subdominant fast component does not remove the ordinary Galactic population from the Solar capture integral. The capture windows in Fig.~\ref{fig:highdeltawindows} show that this slower population is precisely the one efficiently captured by Fe and Ni, while finite-temperature nuclear motion further enhances the rate close to threshold. We therefore identify the treatment of the halo components, together with the finite-temperature reopening of heavy-element capture, as the main origin of the difference between the two Solar predictions.

\begin{table*}[t]
\caption{Normalization diagnostic for the two $\delta=570\,\keV$ high-velocity trajectories. The second column gives the suppression of the $WW$ annihilation rate between the high-velocity and SHM entries in the Solar table of Ref.~\cite{SmirnovGriffithBeacom2026}. The third column gives the corresponding suppression in our stationary-nucleus capture calculation. Their ratio $\epsilon_{\rm hv}$ is compared with the quoted fast-component density fraction.}
\label{tab:endmatterhv}
\begin{ruledtabular}
\begin{tabular}{lcccc}
Model &
$(\Gamma_{WW}^{\rm hv}/\Gamma_{WW}^{\rm SHM})_{\rm Ref.}$ &
$[(C_\odot/2)_{\rm hv}/(C_\odot/2)_{\rm SHM}]_{\rm ours}$ &
$\epsilon_{\rm hv}$ &
quoted $f_s$
\\
\hline
$3_M2_D$ &
$1.11\times10^{-7}$ &
$2.43\times10^{-4}$ &
$4.57\times10^{-4}$ &
$10^{-4}$--$10^{-3}$
\\
$5_M4_D$ &
$1.67\times10^{-7}$ &
$2.98\times10^{-4}$ &
$5.60\times10^{-4}$ &
$10^{-4}$--$10^{-3}$
\end{tabular}
\end{ruledtabular}
\end{table*}

\section{Summary and conclusions}
\label{sec:conclusions}

The high-energy LZ recoil provides a particularly powerful test of endothermic electroweak DM because the interaction responsible for the terrestrial signal also controls capture in the Sun. In this work we have performed a substantially improved reassessment of the HC-MDM interpretation of Ref.~\cite{SmirnovGriffithBeacom2026}, combining an independent determination of the splittings required to reproduce the LZ event with a detailed calculation of Solar capture and of the subsequent evolution of the captured population.

Compared with the original Solar estimate, our treatment incorporates natural isotope mixtures, exact endothermic kinematics, finite-momentum nuclear form factors, finite-temperature nuclear motion close to kinematic closure, radiative and collisional de-excitation, loop-induced elastic cooling of the captured orbits, and an explicit test of capture--annihilation equilibrium. The model derivation, finite-temperature kernel, and first-orbit cooling treatment are given in Appendices~\ref{app:theory}--\ref{app:orbitcooling}. We have also verified the single-scatter regime and performed independent numerical and systematic checks, collected in Appendix~\ref{app:validation}. In particular, the finite-temperature calculation continuously reproduces the independent stationary-target result as $T_A\rightarrow0$, while the numerical integration errors are much smaller than the relevant physical uncertainties. These improvements make the Solar prediction considerably more robust, especially in the large-splitting region where isotope resolution and thermal nuclear motion become essential.

For the nominal thermal HC-MDM benchmarks, the captured population efficiently loses orbital energy through the loop-induced elastic interaction and reaches capture--annihilation equilibrium well within the Solar age. Post-capture thermalization therefore does not provide a viable mechanism for suppressing the Solar-neutrino signal. The resulting status of the six thermal representations is summarized in Tab.~\ref{tab:status}.

\begin{table*}[t]
\caption{Summary of the Solar-neutrino results for the six HC-MDM thermal benchmarks using the independently rate-matched splittings $\delta_\star$. The ratio $\mathcal R_{\rm IC,resp}^{WW}$ compares the predicted equilibrium Solar annihilation rate with the mass-dependent IceCube $W^+W^-$ response. For the four benchmarks above $10\,\TeV$, the values in parentheses use the conservative response-loss stress prescription of Eq.~\eqref{eq:highmassstress}. The two lighter benchmarks lie within the released IceCube DM grid, while the high-mass entries are response-based recast diagnostics rather than official IceCube confidence-level limits.}
\label{tab:status}
\begin{ruledtabular}
\begin{tabular}{lcccccc}
Model &
$m_\chi$ [TeV] &
$\delta_\star$ [keV] &
$C_\odot/2$ [$\mathrm{s}^{-1}$] &
$\mathcal R_{\rm IC,resp}^{WW}$ &
IceCube treatment &
Solar-neutrino status
\\
\hline
$3_M2_D$ & $1.4$ & $379.3$ &
$3.23\times10^{22}$ &
$2.21\times10^{3}$ &
released grid &
robustly constrained
\\
$5_M4_D$ & $7.7$ & $389.1$ &
$1.12\times10^{21}$ &
$9.09\times10^{1}$ &
released grid &
robustly constrained
\\
$7_M6_D$ & $22$ & $386.2$ &
$1.50\times10^{20}$ &
$13.5\ (8.43)$ &
response recast &
strongly above response
\\
$9_M8_D$ & $48$ & $382.3$ &
$3.42\times10^{19}$ &
$3.34\ (1.30)$ &
response recast &
above response
\\
$11_M10_D$ & $82$ & $378.5$ &
$1.27\times10^{19}$ &
$1.31\ (0.370)$ &
response recast &
response dependent
\\
$13_M12_D$ & $130$ & $375.4$ &
$5.31\times10^{18}$ &
$0.57\ (0.123)$ &
response recast &
below response
\end{tabular}
\end{ruledtabular}
\end{table*}

The clearest result concerns the two representations lying within the released IceCube DM mass grid. For $3_M2_D$ and $5_M4_D$ we obtain
\begin{equation}
\mathcal R_{\rm IC,resp}^{WW}
\simeq
2.21\times10^{3},
\qquad
90.9,
\end{equation}
respectively. Their Solar annihilation rates therefore exceed the corresponding IceCube $W^+W^-$ response by approximately three and two orders of magnitude. The margins are much larger than the astrophysical, nuclear, and numerical variations explored in our robustness tests. Within the $W^+W^-$ response mapping used here, the two light SHM thermal solutions are therefore very strongly constrained. A dedicated HC-MDM likelihood using the complete representation-dependent annihilation spectrum would sharpen the statistical interpretation, but is unlikely to change this qualitative conclusion.

The heavier representations provide an important target for future high-energy Solar-neutrino searches. Their masses, between $22$ and $130\,\TeV$, lie above the $10\,\TeV$ endpoint of the released IceCube DM templates. Our response-matched continuation nevertheless shows a clear hierarchy. The $7_M6_D$ benchmark remains well above the extrapolated response even under our deliberately conservative response-loss prescription, while $9_M8_D$ also remains above it throughout the adopted range. The $11_M10_D$ representation lies in the transition region and becomes sensitive to the assumed high-mass detector response, whereas $13_M12_D$ remains below the extrapolated response. These results make $7_M6_D$ and $9_M8_D$ particularly compelling targets for a dedicated extension of the IceCube analysis beyond $10\,\TeV$.

The $\delta=570\,\keV$ high-velocity solutions provide a second important result. Such a large splitting requires DM speeds of order $10^3\,\kms$ to reproduce the LZ recoil on Earth, but the corresponding Solar capture problem is qualitatively different. A subdominant fast component does not remove the ordinary Galactic halo, and DM falling into the Solar potential is strongly accelerated. Consequently, particles with ordinary Galactic velocities can cross the endothermic threshold and be efficiently captured by the abundant Fe and Ni isotopes.

This effect becomes even stronger once the finite-temperature velocity distribution of the Solar nuclei is included. Near kinematic closure, thermal motion reopens a substantial fraction of the Fe and Ni scattering phase space and enhances the capture rate by approximately an order of magnitude relative to the stationary-target calculation. We obtain
\begin{equation}
\mathcal R_{\rm IC,resp}^{WW}
\simeq
6.07
\,\,
(m_\chi=1.47\,\TeV,\ \delta=570\,\keV),
\end{equation}
whereas
\begin{equation}
\mathcal R_{\rm IC,resp}^{WW}
\simeq
0.36
\,\,
(m_\chi=8.5\,\TeV,\ \delta=570\,\keV).
\end{equation}
The light high-splitting trajectory therefore remains significantly constrained by Solar neutrinos, while the $8.5\,\TeV$ solution lies below the mass-matched IceCube response.

Our analysis also clarifies why the high-splitting Solar prediction differs from the much smaller rate quoted in Ref.~\cite{SmirnovGriffithBeacom2026}. The additional suppression in that estimate is numerically of the same order as the assumed $10^{-4}$--$10^{-3}$ density fraction of the fast component. In the Sun, however, the fast population is only an additional component: the much larger ordinary Galactic population remains available for capture and, because of Solar gravitational acceleration, is precisely the population efficiently captured by Fe and Ni. The finite-temperature reopening of these heavy-element channels then further strengthens the effect.

The central conclusion is therefore that Solar neutrinos provide a remarkably powerful and robust discriminator of the HC-MDM interpretations of the LZ event. They strongly constrain the two light SHM thermal solutions and continue to place substantial pressure on the lightest high-splitting alternative. At the same time, the $8.5\,\TeV$ high-splitting trajectory and part of the heavier thermal sequence remain viable with present public IceCube information. Our improved Solar-capture treatment makes this separation between constrained and surviving regions considerably sharper than was possible from the original stationary-target estimate.

A dedicated IceCube analysis extending the DM templates beyond $10\,\TeV$ and using the full HC-MDM electroweak annihilation spectra would now provide the decisive next test. In particular, the large predicted Solar rates of $7_M6_D$ and $9_M8_D$ make the high-mass HC-MDM sequence a well-motivated target for such an analysis.

\begin{acknowledgments}
MDM acknowledges support from the research grant {\sl TAsP (Theoretical Astroparticle Physics)} funded by Istituto Nazionale di Fisica Nucleare (INFN).
\end{acknowledgments}

\newpage

\appendix

\section{Detailed HC-MDM theoretical framework}
\label{app:theory}

This appendix gives a self-contained derivation of the HC-MDM ingredients used in Sec.~\ref{sec:model}. The purpose is twofold. First, it makes explicit which properties of the neutral sector are universal along the sequence $3_M2_D$ through $13_M12_D$. Second, it separates those tree-level results from the representation-dependent thermal and loop calculations imported from Refs.~\cite{LopezHonorez2018,GriffithSmirnov2026,SmirnovGriffithBeacom2026}. The construction is the generalization of the mixed electroweak-fermion models discussed in Refs.~\cite{TaitYu2016,LopezHonorez2018}.

\subsection{Gauge representations, field content, and renormalizable Lagrangian}

Let the Majorana multiplet have weak isospin $j_M=n$ and hypercharge $Y_M=0$. In two-component notation we denote it by $M_m$, with $m=-n,\ldots,n$. A vectorlike Dirac multiplet of isospin $j_D=n-1/2$ is represented by two left-handed Weyl multiplets,
\begin{equation}
D_1\sim(2n,-1/2),
\qquad
D_2\sim(2n,+1/2),
\end{equation}
which together form the four-component field $D$ used in Eq.~\eqref{eq:hcmdm_lagrangian}. The SM Higgs doublet and its conjugate are
\begin{equation}
H=\begin{pmatrix}H^+\\ H^0\end{pmatrix},
\qquad
\widetilde H=i\sigma_2 H^\ast,
\qquad
\langle H^0\rangle=\frac{v}{\sqrt{2}}.
\end{equation}
For definiteness, we impose a $Z_2$ parity under which all new fermions are odd and the SM fields are even. This makes the stability assumption uniform across the complete sequence considered here. For multiplets of sufficiently large dimension, stability against renormalizable decays can instead be accidental; however, when either representation is smaller than a quintuplet an additional discrete symmetry is required to forbid decays into SM fields \cite{CirelliFornengoStrumia2006,GriffithSmirnov2026}. Higher-dimensional operators can in general break accidental stability, so the $Z_2$ formulation is also a convenient statement of the effective theory assumed in this work.

With Lorentz contractions and the invariant $SU(2)_L$ contractions understood, the complete renormalizable dark-sector Lagrangian relevant here is
\begin{align}
\mathcal L_{\rm dark}={}&
 iM^\dagger\bar\sigma^\mu D_\mu M
\nonumber\\
&+iD_1^\dagger\bar\sigma^\mu D_\mu D_1
+iD_2^\dagger\bar\sigma^\mu D_\mu D_2
\nonumber\\
&-\left[
\frac{1}{2}m_M MM
+m_DD_1D_2
+y_1\,\mathcal I(M,D_1,H)
\right.\nonumber\\
&\left.\hspace{1.5cm}
+y_2\,\mathcal I(M,D_2,\widetilde H)
+\mathrm{h.c.}
\right],
\label{eq:app_full_lagrangian}
\end{align}
where
\begin{equation}
D_\mu=\partial_\mu-igW_\mu^aT^a-ig'YB_\mu
\end{equation}
is the electroweak covariant derivative. To make the $SU(2)_L$ singlet structure explicit, first couple $D_1\otimes H$ to isospin $j_M$ and then contract the result with the real $j_M$ representation carried by $M$. In a standard spherical basis one convenient convention is
\begin{align}
\mathcal I(M,D_1,H)
={}&
\sum_{m_M,m_D,m_H}(-1)^{j_M-m_M}
\nonumber\\
&\times
\left\langle
j_D,m_D;\frac12,m_H
\middle|j_M,-m_M
\right\rangle
\nonumber\\
&\times M_{m_M}D_{1,m_D}H_{m_H},
\label{eq:app_yukawa_cg}
\end{align}
where the Clebsch--Gordan coefficient enforces $m_D+m_H=-m_M$. The analogous expression with $D_2\widetilde H$ defines the second Yukawa invariant. Overall phase choices for the component fields can be absorbed into $y_1$ and $y_2$ and do not affect the spectrum or rates. Eq.~\eqref{eq:app_full_lagrangian} contains the four independent renormalizable dark-sector parameters $m_M$, $m_D$, $y_1$, and $y_2$ before the custodial conditions are imposed.

The electric charges follow from $Q=T_3+Y$. The Majorana multiplet contains charges $Q=-n,\ldots,n$, while the two Weyl multiplets forming $D$ supply the corresponding vectorlike tower. The neutral fields are $M^0$, the $T_3=+1/2$ component of $D_1$, and the $T_3=-1/2$ component of $D_2$. Charged partners are not populated as present-day DM, but they are essential for coannihilation, Sommerfeld enhancement, bound-state formation, and the broken-phase electroweak annihilation channels.

\subsection{Universal neutral Clebsch--Gordan coefficient and mass matrix}

The universality of the neutral sector follows from a simple group-theory identity. The neutral Dirac component carries $|T_3|=1/2$, whereas the neutral Majorana component has $T_3=0$. The neutral Higgs insertion therefore uses
\begin{equation}
\left|\left\langle
j_D,+\frac12;\frac12,-\frac12
\middle|
 j_D+\frac12,0
\right\rangle\right|
=
\sqrt{\frac{j_D+1/2}{2j_D+1}}
=
\frac{1}{\sqrt{2}},
\label{eq:app_neutral_cg}
\end{equation}
independently of $j_D$ \cite{SmirnovGriffithBeacom2026}. For the $D_1H$ invariant this is the neutral pair $m_D=+1/2$ and $m_H=-1/2$; the $D_2\widetilde H$ invariant gives the equivalent result with the signs reversed. Combining the magnitude of this coefficient with $\langle H^0\rangle=v/\sqrt{2}$ gives an off-diagonal neutral mass insertion $y_i v/2$, up to field phases that are absorbed into the definitions of $y_i$.

Denoting the two neutral Dirac-sector Weyl fields by $\psi^0$ and $\widetilde\psi^0$, the general neutral mass matrix is therefore
\begin{equation}
\mathcal M_0^{\rm gen}
=
\begin{pmatrix}
 m_M & y_1v/2 & y_2v/2\\
 y_1v/2 & 0 & m_D\\
 y_2v/2 & m_D & 0
\end{pmatrix}_{(M,\psi^0,\widetilde\psi^0)}.
\label{eq:app_general_massmatrix}
\end{equation}
This form is representation independent throughout the $(2n+1)_M(2n)_D$ sequence; representation dependence remains in the charged sector and in the multiplicity and group factors of the electroweak interactions.

At the custodial point of Eq.~\eqref{eq:custodial}, define
\begin{equation}
D_\pm=\frac{\psi^0\pm\widetilde\psi^0}{\sqrt{2}}.
\end{equation}
Eq.~\eqref{eq:app_general_massmatrix} then reduces to Eq.~\eqref{eq:massmatrix}. The crucial structural result is that $D_+$ decouples exactly from the $(M,D_-)$ block. Hence
\begin{equation}
\chi_1=D_+,
\qquad
m_{\chi_1}=m_\chi,
\end{equation}
while the remaining signed eigenvalues are $\pm M_{\ast}$ with
\begin{equation}
M_{\ast}=\sqrt{m_\chi^2+\frac{y^2v^2}{2}}.
\end{equation}
It is useful to parameterize the diagonalization of the heavy block by
\begin{equation}
\sin2\theta=\frac{yv}{\sqrt{2}M_{\ast}},
\qquad
\cos2\theta=\frac{m_\chi}{M_{\ast}}.
\label{eq:app_mixing_angle}
\end{equation}
Before the Majorana phase redefinition, one may choose the signed-mass eigenstates
\begin{align}
\eta_2&=\cos\theta\,M+\sin\theta\,D_-,\\
\eta_3&=-\sin\theta\,M+\cos\theta\,D_-,
\end{align}
with eigenvalues $+M_{\ast}$ and $-M_{\ast}$, respectively. Defining $\chi_2=\eta_2$ and $\chi_3=i\eta_3$ makes both physical masses positive. The two heavy states are exactly degenerate at this point,
\begin{equation}
m_{\chi_2}=m_{\chi_3}=M_{\ast}=m_\chi+\delta.
\end{equation}
The exact splitting and its small-mixing limit are given in Eq.~\eqref{eq:splitting}. Conversely, for a fixed $(m_\chi,\delta)$ the Yukawa coupling is
\begin{equation}
y
=
\frac{\sqrt{2}}{v}
\sqrt{(m_\chi+\delta)^2-m_\chi^2}
=
\frac{2}{v}
\sqrt{m_\chi\delta+\frac{\delta^2}{2}}.
\label{eq:app_y_from_delta}
\end{equation}
Eq.~\eqref{eq:app_y_from_delta} is the exact relation used to obtain the $y_\star$ values in Tab.~\ref{tab:benchmarks}; no $\delta/m_\chi$ expansion is needed numerically.

\subsection{Gauge interactions, charged sector, and stability of the DM state}

The neutral mass matrix is universal, but the full electroweak spectrum is representation dependent. For a Weyl multiplet $X$ with weak isospin $j$, hypercharge $Y$, and components $X_m$ with $m=-j,\ldots,j$, the covariant derivative fixes the gauge interactions without introducing additional free parameters. In particular, the charged-current interaction can be written as
\begin{align}
\mathcal L_W={}&
\frac{g}{\sqrt{2}}W^+_\mu
\sum_{m=-j}^{j-1}
\sqrt{(j-m)(j+m+1)}
\nonumber\\
&\times X_{m+1}^\dagger\bar\sigma^\mu X_m
+\mathrm{h.c.},
\label{eq:app_charged_current}
\end{align}
while the diagonal photon and $Z$ interactions are
\begin{align}
\mathcal L_{A,Z}={}&
e A_\mu\sum_m Q_m X_m^\dagger\bar\sigma^\mu X_m
\nonumber\\
&+\frac{g}{c_W}Z_\mu\sum_m
\left(m-s_W^2Q_m\right)
X_m^\dagger\bar\sigma^\mu X_m,
\label{eq:app_neutral_gauge_current}
\end{align}
where $Q_m=m+Y$ and $s_W\equiv\sin\theta_W$. Eqs.~\eqref{eq:app_charged_current} and \eqref{eq:app_neutral_gauge_current} apply separately to $M$, $D_1$, and $D_2$. They make explicit why the representation dimension matters strongly for coannihilation and for nonperturbative electroweak effects even though the neutral endothermic $Z$ transition is universal.

After electroweak symmetry breaking, fields with the same electric charge can also mix through the Yukawa terms in Eq.~\eqref{eq:app_full_lagrangian}. The relevant entries are obtained directly from the Clebsch--Gordan contraction in Eq.~\eqref{eq:app_yukawa_cg}; unlike the neutral coefficient in Eq.~\eqref{eq:app_neutral_cg}, the charged-sector coefficients depend on $j_M$ and on the charge of the component. Consequently, there is no representation-independent charged mass matrix analogous to Eq.~\eqref{eq:app_general_massmatrix}. This representation dependence is precisely what enters the coupled-channel potentials, coannihilation rates, Sommerfeld enhancement, and electroweak bound-state spectrum calculated in Ref.~\cite{GriffithSmirnov2026}.

The present analysis requires only that the lightest state on the adopted benchmark trajectory is the neutral state $\chi_1$. The charged partners are then unstable against electroweak transitions into lighter members of the dark multiplet and do not survive as a cosmological charged relic. We do not recompute the complete radiatively corrected charged spectrum here; instead, we use the benchmark thermal trajectories for which the neutral state is the DM candidate, as established in Refs.~\cite{LopezHonorez2018,GriffithSmirnov2026,SmirnovGriffithBeacom2026}. The charged states nevertheless remain essential virtual or coannihilating degrees of freedom and are therefore part of the theoretical model even though they do not enter the Solar capture calculation as incoming states.

\subsection{Custodial cancellation of the diagonal Higgs interaction}

The Higgs interaction in the neutral sector is obtained by replacing $v\rightarrow v+h$ in the Yukawa entries of the mass matrix. At the custodial point the light state $\chi_1=D_+$ is exactly orthogonal to the combination $D_-$ that mixes with $M$. Therefore
\begin{equation}
\frac{\partial m_{\chi_1}}{\partial v}=0,
\end{equation}
and the tree-level diagonal coupling $h\chi_1\chi_1$ vanishes. This is the origin of the tree-level SI blind direction used in the main text. The cancellation is a statement about the exact custodial point; departures from $m_M=m_D$ or from the custodial Yukawa relation generally restore a diagonal Higgs coupling \cite{TaitYu2016,LopezHonorez2018,GriffithSmirnov2026}.

The absence of the tree-level diagonal Higgs amplitude does not make the model inert. Gauge interactions remain, and electroweak loops generate the elastic SI amplitudes discussed in Sec.~\ref{sec:model}. Moreover, the same Yukawa coupling $y$ that disappears from the diagonal tree-level Higgs interaction of $\chi_1$ still determines the heavy-neutral splitting and modifies the early-Universe coupled-channel problem.

\subsection{Neutral-current interaction and inclusive inelastic strength}

The $Z$ current of the two neutral hypercharged Weyl fields is fixed by $T_3=\mp1/2$ and is independent of the total multiplet isospin,
\begin{equation}
\mathcal L_Z
=
\frac{g}{2c_W}Z_\mu
\left(
\widetilde\psi^{0\dagger}\bar\sigma^\mu\widetilde\psi^0
-
\psi^{0\dagger}\bar\sigma^\mu\psi^0
\right).
\label{eq:app_z_gauge_basis}
\end{equation}
Rotating to $D_\pm$ gives Eq.~\eqref{eq:zcurrent}. Thus there is no diagonal $Z\chi_1\chi_1$ interaction: $Z$ exchange connects $D_+$ to the $D_-$ direction in the heavy neutral subspace.

Using the phase convention above,
\begin{equation}
D_-
=
\sin\theta\,\chi_2-i\cos\theta\,\chi_3,
\end{equation}
up to an irrelevant overall phase. Hence the individual off-diagonal couplings depend on the mixing angle, but their inclusive strength does not,
\begin{align}
|g_{\chi_1\chi_2Z}|^2+|g_{\chi_1\chi_3Z}|^2
={}&\left(\frac{g}{2c_W}\right)^2
\left(\sin^2\theta+\cos^2\theta\right)\nonumber\\
={}&\left(\frac{g}{2c_W}\right)^2.
\label{eq:app_inclusive_z}
\end{align}
Because $\chi_2$ and $\chi_3$ are degenerate at the custodial point, both channels have the same kinematic threshold. This is why the inclusive tree-level endothermic interaction is universal at fixed $(m_\chi,y)$ even though the thermal relic trajectory is representation dependent.

\subsection{Low-energy weak interaction and nuclear recoil normalization}

At momentum transfers well below $m_Z$, integrating out the $Z$ gives the vector interaction
\begin{equation}
\mathcal L_{\rm eff}
=
\frac{G_F}{\sqrt{2}}J_\chi^\mu
\left[
\bar n\gamma_\mu n
-
\left(1-4s_W^2\right)\bar p\gamma_\mu p
\right],
\label{eq:app_fermi_operator}
\end{equation}
where $J_\chi^\mu$ is the off-diagonal transition current and $s_W\equiv\sin\theta_W$. Coherent scattering on a nucleus therefore carries the weak charge in Eq.~\eqref{eq:weakcharge}. The zero-momentum cross section is
\begin{equation}
\sigma_A^0
=
\frac{G_F^2\mu_{\chi A}^2}{2\pi}Q_V^2,
\label{eq:app_sigmaA_direct}
\end{equation}
which is equivalent to Eqs.~\eqref{eq:sigmaninel} and \eqref{eq:sigmaA0}. In particular, for a neutron in the heavy-DM limit,
\begin{equation}
\sigma_n^{\rm inel}
=
\frac{G_F^2m_n^2}{2\pi}
\simeq
7.4\times10^{-39}\,\mathrm{cm}^2,
\end{equation}
where the small difference from the rounded value in Eq.~\eqref{eq:sigmaninel} is only numerical rounding.

For a stationary target, the leading nonrelativistic differential cross section is Eq.~\eqref{eq:dsigma}. A useful normalization check follows by setting $\delta=0$ and $F_A(q)=1$. The elastic recoil interval is then
\begin{equation}
0<E_R<\frac{2\mu_{\chi A}^2w^2}{m_A},
\end{equation}
and direct integration gives
\begin{equation}
\int dE_R\,\frac{d\sigma_A}{dE_R}
=
\frac{G_F^2\mu_{\chi A}^2}{2\pi}Q_V^2
=
\sigma_A^0.
\label{eq:app_elastic_norm_check}
\end{equation}
For $\delta>0$ but $F_A(q)=1$, the interval in Eq.~\eqref{eq:erpm} instead gives
\begin{equation}
\sigma_A^{\rm inel}(w)
=
\sigma_A^0\frac{w'}{w},
\label{eq:app_inelastic_phase_space}
\end{equation}
so the expected final-state phase-space suppression is generated automatically by the shrinking recoil interval. Eqs.~\eqref{eq:app_elastic_norm_check} and \eqref{eq:app_inelastic_phase_space} are useful cross-checks of both the stationary and finite-temperature implementations.

The $1/(2\pi)$ normalization in Eq.~\eqref{eq:app_sigmaA_direct} is the vectorlike pseudo-Dirac normalization appropriate to the inclusive HC-MDM transition. As noted in Sec.~\ref{sec:model}, a factor-four smaller expression has been propagated in parts of the split-electroweak literature. The normalization adopted here agrees with the direct low-energy reduction above and with the Solar-capture normalization emphasized in Ref.~\cite{PospelovRamani2026}.

\subsection{Thermal relic trajectory, loop-induced SI scattering, and present-day annihilation}

At temperatures above electroweak symmetry breaking, all components of the two multiplets participate in the thermal system. Schematically, if chemical equilibrium among the coannihilating states holds, the total yield satisfies
\begin{equation}
\frac{dY}{dx}
=
-\frac{s}{Hx}
\left\langle\sigma_{\rm eff}v\right\rangle
\left(Y^2-Y_{\rm eq}^2\right),
\qquad
x\equiv\frac{m_\chi}{T},
\label{eq:app_boltzmann}
\end{equation}
with
\begin{equation}
\left\langle\sigma_{\rm eff}v\right\rangle
=
\sum_{i,j}
\frac{n_i^{\rm eq}n_j^{\rm eq}}{(n_{\rm tot}^{\rm eq})^2}
\left\langle\sigma_{ij}v\right\rangle
+
\hbox{bound-state contributions}.
\label{eq:app_sigmaeff}
\end{equation}
For heavy electroweak multiplets, the quantities in Eq.~\eqref{eq:app_sigmaeff} are not adequately described by tree-level perturbation theory alone. Multiple electroweak-boson exchange generates a coupled-channel Sommerfeld problem, and radiative formation and decay of electroweak bound states can also modify freeze-out \cite{CirelliStrumiaTamburini2007,Bottaro2022,GriffithSmirnov2026}. The Higgs Yukawa interaction changes both the short-distance annihilation matrix and the long-range potential in the unbroken phase. Consequently, each representation pair follows a distinct trajectory $m_\chi=m_{\rm th}^{(R_M,R_D)}(y)$ even though the late-time neutral $Z$ transition is universal.

We do not recompute this coupled-channel freeze-out problem in the present work. The thermal trajectories and the benchmark loop-induced SI cross sections are taken from Ref.~\cite{GriffithSmirnov2026} and from the LZ application in Ref.~\cite{SmirnovGriffithBeacom2026}. The loop-induced SI amplitude arises from electroweak matching onto scalar and spin-two quark and gluon operators and can exhibit substantial cancellations, as discussed in Sec.~\ref{sec:model} and Refs.~\cite{Hisano2011,Hisano2015,ChenHill2020,ChenDingHill2023}. This is why our Solar analysis stress-tests the elastic cooling rate over orders of magnitude rather than assigning a single Gaussian fractional uncertainty.

Present-day annihilation of the ground state does not require a thermally populated excited neutral state. The covariant derivatives in Eq.~\eqref{eq:app_full_lagrangian} provide charged-current and neutral-current interactions connecting the neutral and charged components, generating electroweak final states and the representation-dependent Sommerfeld structure relevant at Solar velocities. In this paper these annihilation dynamics enter only through the requirement of capture--annihilation equilibrium and through the comparison with an IceCube $W^+W^-$ response template. A full representation-dependent IceCube likelihood would require the complete annihilation branching fractions and propagated neutrino spectra, which are not presently available for all six HC-MDM representations.

\section{Finite-temperature scattering kernel}
\label{app:finiteT}

This appendix gives the finite-temperature generalization used in Sec.~\ref{subsec:finiteT}. Let the incoming local DM velocity at radius $r$ be $\bm w$ and the target-nucleus velocity be $\bm v_A$. The initial relative velocity is
\begin{equation}
\bm g=\bm w-\bm v_A,
\qquad
g=|\bm g|.
\end{equation}
The endothermic transition is possible only if
\begin{equation}
g^2>\frac{2\delta}{\mu_{\chi A}},
\end{equation}
and the final relative speed is
\begin{equation}
g'=\sqrt{g^2-\frac{2\delta}{\mu_{\chi A}}}.
\label{eq:app_gprime}
\end{equation}
The center-of-mass velocity is
\begin{equation}
\bm V_{\rm cm}
=
\frac{m_\chi\bm w+m_A\bm v_A}{m_\chi+m_A}.
\end{equation}
For a final relative direction $\hat{\bm n}$, the outgoing DM velocity is
\begin{equation}
\bm v_\chi'
=
\bm V_{\rm cm}
+
\frac{m_A}{m_\chi+m_A}g'\hat{\bm n}.
\label{eq:app_vout}
\end{equation}
The momentum transferred to the nucleus is
\begin{equation}
\bm q
=
\mu_{\chi A}\left(\bm g-g'\hat{\bm n}\right),
\qquad
E_R=\frac{q^2}{2m_A}.
\label{eq:app_qthermal}
\end{equation}
These expressions retain the full relative orientation of the halo particle, the thermally moving target, and the final-state scattering direction. They are exact at leading nonrelativistic order in the small ratios $v/c$ and $\delta/m_\chi$; corrections from using $m_{\chi_{2,3}}=m_\chi+\delta$ rather than $m_\chi$ in the center-of-mass mass factors are of relative order $\delta/m_\chi\lesssim10^{-6}$ for the benchmarks considered here.

At leading nonrelativistic order the coherent vector interaction is isotropic in the center-of-mass scattering angle apart from the momentum dependence of the nuclear form factor. A convenient form of the differential cross section is
\begin{equation}
\frac{d\sigma_A}{d\Omega_{\rm cm}}
=
\frac{\sigma_A^0}{4\pi}
\frac{g'}{g}
F_A^2(q),
\label{eq:app_dsigma_domega}
\end{equation}
with $\sigma_A^0$ from Eq.~\eqref{eq:app_sigmaA_direct}. The factor $g'/g$ is the two-body final-state phase-space factor. For $F_A=1$, integrating Eq.~\eqref{eq:app_dsigma_domega} gives Eq.~\eqref{eq:app_inelastic_phase_space}.

The normalized target velocity distribution is the local Maxwell--Boltzmann distribution in Eq.~\eqref{eq:targetMB}. The capture kernel for isotope $A$ can then be written as
\begin{align}
\Omega_A^-(w,r)={}&n_A(r)
\int d^3v_A\,f_A(\bm v_A;r)
\int d\Omega_{\rm cm}\,
 g\frac{d\sigma_A}{d\Omega_{\rm cm}}
\nonumber\\
&\times
\Theta\!\left[v_{\rm esc}(r)-|\bm v_\chi'|\right],
\label{eq:app_omega_thermal}
\end{align}
and the total finite-temperature capture rate is
\begin{equation}
C_\odot
=
\frac{\rho_\chi}{m_\chi}
\sum_A
\int_0^{R_\odot}4\pi r^2dr
\int_0^\infty du\,
\frac{f_\odot(u)}{u}\,w\,
\Omega_A^-(w,r).
\label{eq:app_capture_thermal_full}
\end{equation}
The factor $w/u$ is the usual gravitational-focusing Jacobian, while the collision rate inside $\Omega_A^-$ is controlled by the true relative speed $g$.

The stationary-target result follows continuously by taking $f_A(\bm v_A;r)\rightarrow\delta^{(3)}(\bm v_A)$. In that limit $g=w$, Eqs.~\eqref{eq:app_vout} and \eqref{eq:app_qthermal} map the angular integral onto the recoil interval in Eq.~\eqref{eq:erpm}, and Eq.~\eqref{eq:app_dsigma_domega} reduces to the differential-recoil formulation of Eq.~\eqref{eq:dsigma}. This analytical correspondence is supplemented by the numerical $T_A\rightarrow0$ tests in Appendix~\ref{app:validation}.

For the finite-temperature calculation we sample the radius, Solar-frame halo speed, three target-velocity components, and two center-of-mass scattering angles with scrambled Sobol sequences. The exact $q$ in Eq.~\eqref{eq:app_qthermal} is used in the Helm form factor for each natural isotope. This is essential near kinematic closure, where thermal motion reopens only a restricted part of the Fe and Ni phase space and where replacing $F_A(q)$ by its zero-momentum value would overestimate the rate.

\section{Capture-generated first orbits and loop-induced elastic cooling}
\label{app:orbitcooling}

The thermalization check in Sec.~\ref{sec:postcapture} follows the captured population from the first bound orbit generated by the capture event. It does not initialize the particle at $R_\odot$ or assume that capture immediately produces a compact orbit.

For a capture event with asymptotic speed $u$ and nuclear recoil $E_R$, the positive binding energy after the endothermic scatter is Eq.~\eqref{eq:bindingaftercapture}. We define the corresponding specific binding energy
\begin{equation}
b\equiv\frac{E_B}{m_\chi}>0.
\end{equation}
The capture-weighted distribution of $b$ is obtained from the same differential measure that appears in the stationary capture integral,
\begin{align}
dC_A\propto{}&
4\pi r^2dr\,
\frac{\rho_\chi}{m_\chi}
\frac{f_\odot(u)}{u}w\,du\,
 n_A(r)
\frac{d\sigma_A}{dE_R}dE_R,
\label{eq:app_capture_measure}
\end{align}
with the endothermic and capture conditions imposed event by event. Thus the very weakly bound tail is present with its physical capture weight rather than being added through an assumed orbital distribution.

If the first apocenter lies outside the Sun, the exterior part of the orbit is Keplerian. For a nearly radial orbit,
\begin{equation}
r_{\rm max}=\frac{GM_\odot}{b},
\qquad
P(b)=\frac{\pi GM_\odot}{\sqrt{2}\,b^{3/2}},
\label{eq:keplerouterapp}
\end{equation}
where $P$ is the full radial period of the degenerate Kepler orbit. The particle loses energy only while crossing the Solar interior. In the exterior stage we therefore evaluate the elastic energy loss per Solar transit and divide by the orbital period. The slowing column varies by only about $1$--$2\%$ across the capture-weighted weak-binding range tested in the package, validating the use of the weak-binding column until $r_{\rm max}=R_\odot$.

For the subsequent elastic interaction we define the energy-weighted slowing cross section
\begin{align}
\sigma_A^{\rm slow}(v)
={}&\int_0^{4\mu_{\chi A}^2v^2}dq^2\,
\frac{E_R}{E_\chi}
\frac{d\sigma_A}{dq^2},
\label{eq:sigmaslowapp}\\
E_R={}&\frac{q^2}{2m_A},
\qquad
E_\chi=\frac{m_\chi v^2}{2}.
\end{align}
For an orbit entirely inside the Sun, with turning point $r_{\rm max}$, the radial period is
\begin{equation}
T(r_{\rm max})
=
4\int_0^{r_{\rm max}}
\frac{dr}{v(r;r_{\rm max})}.
\label{eq:orbitperiodapp}
\end{equation}
The factor $4$ accounts for the two radial legs on both sides of the Solar center in the radial-orbit convention. The local specific-energy loss rate from isotope $A$ is
\begin{equation}
\dot\varepsilon_A(r)
=
-\frac{1}{2}v^2(r)\,
 n_A(r)v(r)\,
\sigma_A^{\rm slow}[v(r)],
\label{eq:localcoolapp}
\end{equation}
and the orbit average is
\begin{equation}
\left\langle\dot\varepsilon\right\rangle
=
\frac{4}{T(r_{\rm max})}
\int_0^{r_{\rm max}}
\frac{dr}{v(r;r_{\rm max})}
\sum_A\dot\varepsilon_A(r).
\label{eq:orbitcoolapp}
\end{equation}
We integrate Eq.~\eqref{eq:orbitcoolapp} until the characteristic thermal radius in Eq.~\eqref{eq:rthermal} is reached.

The cooling calculation uses the loop-induced SI nucleon cross sections in Eq.~\eqref{eq:loopsix}, coherent isotope-dependent SI enhancement, and the same Helm form factor used in the capture calculation. Because the initial first-orbit distribution is generated from Eq.~\eqref{eq:app_capture_measure}, the quoted median and $99$th-percentile cooling times include both ordinary compact captures and the long-period weak-binding tail. The stress tests discussed in Sec.~\ref{sec:postcapture} then rescale the elastic loop cross section while leaving the capture-generated orbital distribution fixed. This cleanly separates uncertainty in the post-capture cooling interaction from uncertainty in the initial tree-level inelastic capture process.

\section{Numerical validation and systematic uncertainties}
\label{app:validation}

The Solar-capture calculation combines nested quadratures, natural isotope mixtures, a quasi-Monte-Carlo finite-temperature kernel, and several astrophysical and nuclear inputs. We therefore collect the numerical and physical validation in one appendix.

\subsection{Stationary-target convergence and natural-isotope resolution}
\label{app:validation:numerics}

Tab.~\ref{tab:conv} shows the convergence of the stationary-target integral at a representative $3_M2_D$ near-benchmark point. We intentionally retain $\delta=378\,\keV$ for this code-validation test because it is the original reference point used to validate the stationary integration; the physics results in the main text use the independently rate-matched splitting in Tab.~\ref{tab:benchmarks}. The result is stable at the sub-per-mille level at the resolution used for the quoted rates.

\begin{table}[t]
\caption{Quadrature convergence for $m_\chi=1.4\,\TeV$ and $\delta=378\,\keV$. The columns give the numbers of radial, asymptotic-speed, and recoil-energy nodes.}
\label{tab:conv}
\begin{ruledtabular}
\begin{tabular}{cccc}
$n_r$ & $n_u$ & $n_E$ & $C_\odot$ [$\mathrm{s}^{-1}$]\\
\hline
50 & 70 & 24 & $6.638\times10^{22}$\\
70 & 100 & 32 & $6.646\times10^{22}$\\
100 & 140 & 40 & $6.642\times10^{22}$\\
130 & 180 & 48 & $6.643\times10^{22}$\\
\end{tabular}
\end{ruledtabular}
\end{table}

Natural isotope resolution has little impact in the ordinary benchmark range but becomes important close to Solar kinematic closure. As an external check, we reproduce the thermal-Higgsino capture setup of Ref.~\cite{PospelovRamani2026}. The ratios of the natural-isotope rate to a representative-single-isotope calculation are $0.997$, $0.997$, and $1.006$ at $\delta=350$, $377$, and $506\,\keV$, respectively, but rise to $1.47$ at $566\,\keV$. The enhancement is dominated by neutron-richer Ni isotopes that remain accessible after lighter isotopes have approached closure. We therefore retain natural isotope mixtures in all quoted HC-MDM results.

\subsection{Independent finite-temperature zero-temperature limit}

The finite-temperature implementation is numerically independent of the stationary Gauss--Legendre code. Sending the target temperature to zero must nevertheless reproduce the stationary result. For
\begin{equation}
(m_\chi/\TeV,\delta/\keV)
=
(1.4,378),\,
(1.47,570),\,
(8.5,570),
\end{equation}
we obtain
\begin{equation}
\frac{C_\odot^{\rm QMC}(T_A=0)}{C_\odot^{\rm GL}}
=
0.99984,\,
1.00282,\,
1.00340,
\label{eq:app_qmc_zeroT}
\end{equation}
respectively. Thus the two independent implementations agree to better than $0.4\%$, including the near-threshold configurations for which finite-temperature effects are largest. This test also verifies the normalization relation between Eqs.~\eqref{eq:dsigma} and \eqref{eq:app_dsigma_domega}.

\subsection{Astrophysical and nuclear systematic variations}

We assess representative one-at-a-time variations for the rate-matched $3_M2_D$ and $5_M4_D$ benchmarks, the two candidates lying within the released IceCube mass grid. The reference SHM uses $(v_0,v_\odot)=(220,232)\,\kms$. We vary these jointly to $(200,220)\,\kms$ and $(240,250)\,\kms$, vary $v_{\rm esc}^{\rm Gal}$ from $520$ to $580\,\kms$, rescale the effective Helm radius by $\pm6\%$, rescale the heavy-element Solar abundances coherently by factors $0.8$ and $1.2$, and vary the local DM density between $0.24$ and $0.36\,\GeV\,\mathrm{cm}^{-3}$.

\begin{figure}[t]
\centering
\includegraphics[width=\columnwidth]{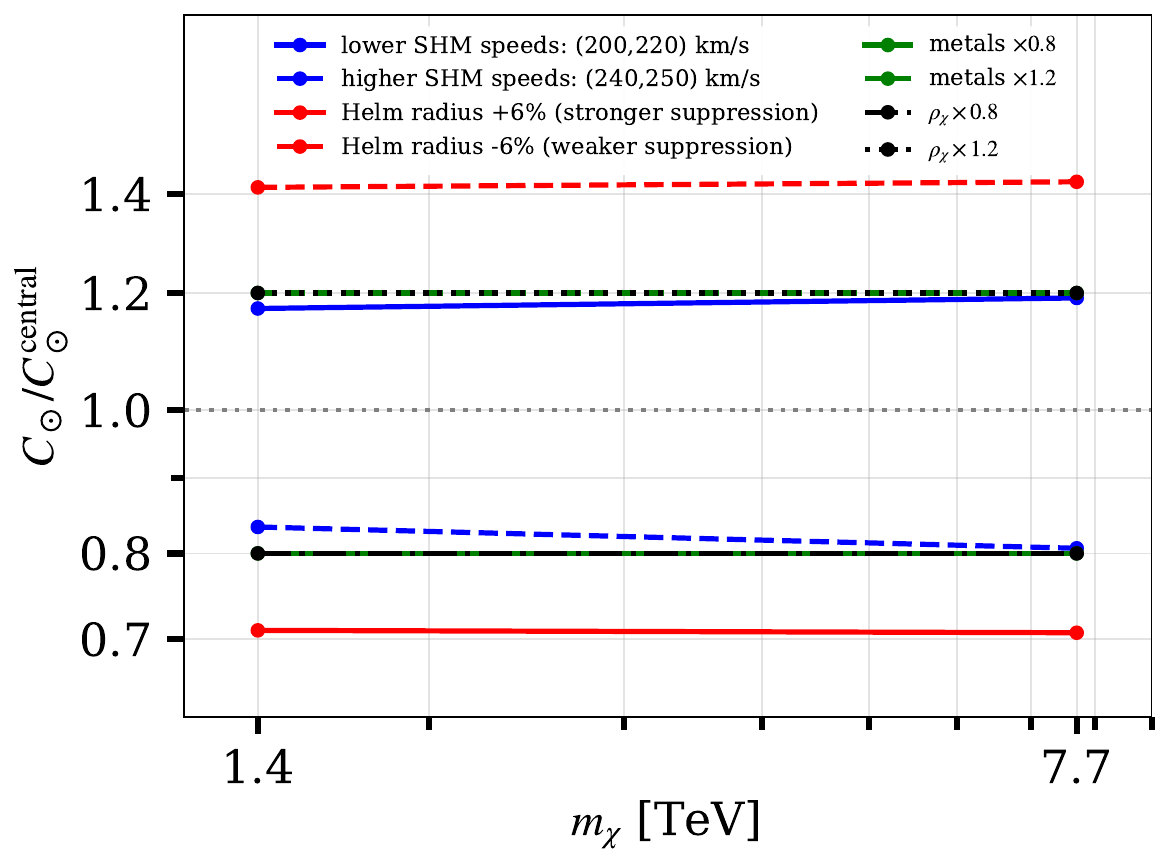}
\caption{One-at-a-time variations of the Solar capture rate relative to the central prediction for the $3_M2_D$ and $5_M4_D$ benchmarks. We vary the SHM velocity parameters, the effective Helm radius, the heavy-element Solar abundances, and the local DM density. The horizontal gray line denotes the reference calculation. The largest individual variation in this test arises from the finite-momentum nuclear form factor.}
\label{fig:systematics}
\end{figure}

The corresponding rate ratios are given in Tab.~\ref{tab:syst} and shown in Fig.~\ref{fig:systematics}. The SHM-speed variations modify the capture rate by about $17$--$20\%$, while the adopted Galactic escape-speed variation changes it by less than $1\%$. The local density and coherent heavy-element rescaling act nearly linearly. The largest individual variation in this stress test comes from the nuclear form factor: the adopted Helm-radius change moves the rate by roughly $-29\%$ to $+43\%$.

\begin{table}[t]
\caption{Capture-rate ratios relative to the central calculation for the two HC-MDM candidates inside the released IceCube mass grid.}
\label{tab:syst}
\begin{ruledtabular}
\begin{tabular}{lcc}
Variation & $3_M2_D$ & $5_M4_D$\\
\hline
lower SHM speeds & 1.171 & 1.191\\
higher SHM speeds & 0.834 & 0.806\\
low $v_{\rm esc}^{\rm Gal}$ & 1.003 & 0.990\\
high $v_{\rm esc}^{\rm Gal}$ & 0.995 & 0.991\\
Helm radius $+6\%$ & 0.710 & 0.707\\
Helm radius $-6\%$ & 1.415 & 1.427\\
$0.8Z_\odot$ & 0.800 & 0.800\\
$1.2Z_\odot$ & 1.200 & 1.200\\
$0.8\rho_\chi$ & 0.800 & 0.800\\
$1.2\rho_\chi$ & 1.200 & 1.200\\
\end{tabular}
\end{ruledtabular}
\end{table}

These variations are robustness tests rather than statistically independent nuisance parameters and should not be combined in quadrature. In particular, the Solar heavy-element abundances are correlated, the local DM density is an overall normalization uncertainty, and the Helm-radius variation is a proxy for nuclear-model dependence. Their magnitude is nevertheless far smaller than the factors by which the two light thermal benchmarks exceed the mass-matched IceCube $W^+W^-$ response.

\subsection{Single-scatter optical-depth test}
\label{subsec:optical}

The initial capture calculation assumes that an incoming DM particle undergoes at most one relevant endothermic collision in a Solar transit. We test this by integrating the physical scattering probability along a central Solar diameter, including the endothermic threshold and finite-momentum Helm suppression. The required condition is
\begin{equation}
\tau_{\rm inel}\ll1.
\end{equation}
For a representative asymptotic speed $u=232\,\kms$, we find
\begin{align}
\tau_{\rm inel}(3_M2_D)&=5.12\times10^{-3},\\
\tau_{\rm inel}(5_M4_D)&=5.00\times10^{-3},
\end{align}
while the corresponding $\delta=570\,\keV$ trajectories give
\begin{align}
\tau_{\rm inel}(1.47\,\TeV)&=2.24\times10^{-6},\\
\tau_{\rm inel}(8.5\,\TeV)&=6.16\times10^{-6}.
\end{align}
All values are well below unity; even the largest optical depths are smaller than the multiple-scattering threshold by more than two orders of magnitude. The single-scatter capture calculation is therefore self-consistent.

\begin{figure}[t]
\centering
\includegraphics[width=\columnwidth]{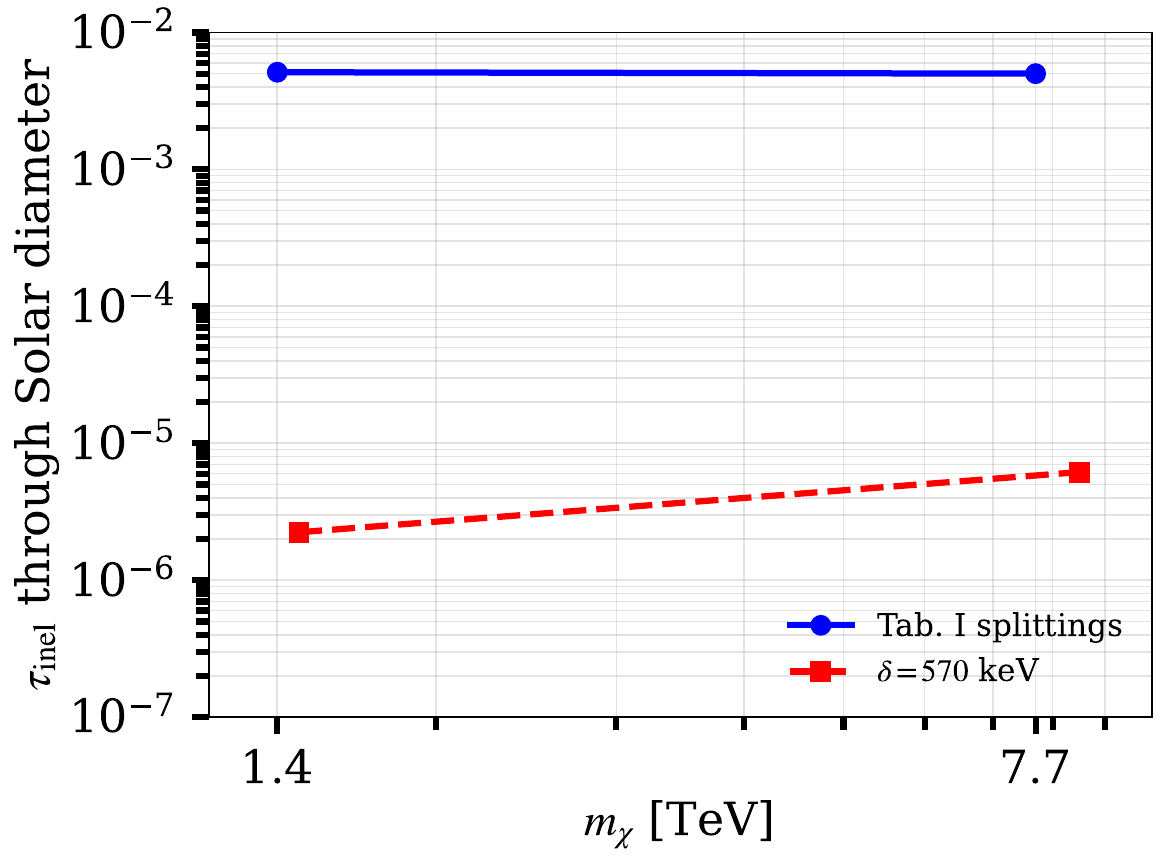}
\caption{Single-pass endothermic optical depth through a Solar diameter for $u=232\,\kms$. Blue circles show the $3_M2_D$ and $5_M4_D$ thermal benchmarks, while red squares correspond to their $\delta=570\,\keV$ high-splitting trajectories. All values satisfy $\tau_{\rm inel}\ll1$, validating the single-scatter approximation for the initial Solar capture.}
\label{fig:optical}
\end{figure}

This single-transit test should be distinguished from the post-capture cooling discussed in Sec.~\ref{sec:postcapture} and Appendix~\ref{app:orbitcooling}. Once a particle is gravitationally bound, it can cross the Solar interior repeatedly over many orbits, so a much smaller elastic scattering probability per passage can still remove sufficient orbital energy over the Solar lifetime.

\bibliography{main}

\begin{thebibliography}{62}%
\makeatletter
\providecommand \@ifxundefined [1]{%
 \@ifx{#1\undefined}
}%
\providecommand \@ifnum [1]{%
 \ifnum #1\expandafter \@firstoftwo
 \else \expandafter \@secondoftwo
 \fi
}%
\providecommand \@ifx [1]{%
 \ifx #1\expandafter \@firstoftwo
 \else \expandafter \@secondoftwo
 \fi
}%
\providecommand \natexlab [1]{#1}%
\providecommand \enquote  [1]{``#1''}%
\providecommand \bibnamefont  [1]{#1}%
\providecommand \bibfnamefont [1]{#1}%
\providecommand \citenamefont [1]{#1}%
\providecommand \href@noop [0]{\@secondoftwo}%
\providecommand \href [0]{\begingroup \@sanitize@url \@href}%
\providecommand \@href[1]{\@@startlink{#1}\@@href}%
\providecommand \@@href[1]{\endgroup#1\@@endlink}%
\providecommand \@sanitize@url [0]{\catcode `\\12\catcode `\$12\catcode
  `\&12\catcode `\#12\catcode `\^12\catcode `\_12\catcode `\%12\relax}%
\providecommand \@@startlink[1]{}%
\providecommand \@@endlink[0]{}%
\providecommand \url  [0]{\begingroup\@sanitize@url \@url }%
\providecommand \@url [1]{\endgroup\@href {#1}{\urlprefix }}%
\providecommand \urlprefix  [0]{URL }%
\providecommand \Eprint [0]{\href }%
\providecommand \doibase [0]{https://doi.org/}%
\providecommand \selectlanguage [0]{\@gobble}%
\providecommand \bibinfo  [0]{\@secondoftwo}%
\providecommand \bibfield  [0]{\@secondoftwo}%
\providecommand \translation [1]{[#1]}%
\providecommand \BibitemOpen [0]{}%
\providecommand \bibitemStop [0]{}%
\providecommand \bibitemNoStop [0]{.\EOS\space}%
\providecommand \EOS [0]{\spacefactor3000\relax}%
\providecommand \BibitemShut  [1]{\csname bibitem#1\endcsname}%
\let\auto@bib@innerbib\@empty
\bibitem [{\citenamefont {Bertone}\ \emph {et~al.}(2005)\citenamefont
  {Bertone}, \citenamefont {Hooper},\ and\ \citenamefont
  {Silk}}]{BertoneHooperSilk2005}%
  \BibitemOpen
  \bibfield  {author} {\bibinfo {author} {\bibfnamefont {G.}~\bibnamefont
  {Bertone}}, \bibinfo {author} {\bibfnamefont {D.}~\bibnamefont {Hooper}},\
  and\ \bibinfo {author} {\bibfnamefont {J.}~\bibnamefont {Silk}},\ }\bibfield
  {title} {\bibinfo {title} {Particle dark matter: Evidence, candidates and
  constraints},\ }\href {https://doi.org/10.1016/j.physrep.2004.08.031}
  {\bibfield  {journal} {\bibinfo  {journal} {Phys. Rept.}\ }\textbf {\bibinfo
  {volume} {405}},\ \bibinfo {pages} {279} (\bibinfo {year} {2005})},\ \Eprint
  {https://arxiv.org/abs/hep-ph/0404175} {arXiv:hep-ph/0404175} \BibitemShut
  {NoStop}%
\bibitem [{\citenamefont {Bertone}\ and\ \citenamefont
  {Hooper}(2018)}]{BertoneHooper2018}%
  \BibitemOpen
  \bibfield  {author} {\bibinfo {author} {\bibfnamefont {G.}~\bibnamefont
  {Bertone}}\ and\ \bibinfo {author} {\bibfnamefont {D.}~\bibnamefont
  {Hooper}},\ }\bibfield  {title} {\bibinfo {title} {A history of dark
  matter},\ }\href {https://doi.org/10.1103/RevModPhys.90.045002} {\bibfield
  {journal} {\bibinfo  {journal} {Rev. Mod. Phys.}\ }\textbf {\bibinfo {volume}
  {90}},\ \bibinfo {pages} {045002} (\bibinfo {year} {2018})},\ \Eprint
  {https://arxiv.org/abs/1605.04909} {arXiv:1605.04909 [astro-ph.CO]}
  \BibitemShut {NoStop}%
\bibitem [{\citenamefont {Aghanim}\ \emph {et~al.}(2020)\citenamefont {Aghanim}
  \emph {et~al.}}]{Planck2020}%
  \BibitemOpen
  \bibfield  {author} {\bibinfo {author} {\bibfnamefont {N.}~\bibnamefont
  {Aghanim}} \emph {et~al.} (\bibinfo {collaboration} {Planck Collaboration}),\
  }\bibfield  {title} {\bibinfo {title} {Planck 2018 results. vi. cosmological
  parameters},\ }\href {https://doi.org/10.1051/0004-6361/201833910} {\bibfield
   {journal} {\bibinfo  {journal} {Astron. Astrophys.}\ }\textbf {\bibinfo
  {volume} {641}},\ \bibinfo {pages} {A6} (\bibinfo {year} {2020})},\ \Eprint
  {https://arxiv.org/abs/1807.06209} {arXiv:1807.06209 [astro-ph.CO]}
  \BibitemShut {NoStop}%
\bibitem [{\citenamefont {Cirelli}\ \emph {et~al.}(2024)\citenamefont
  {Cirelli}, \citenamefont {Strumia},\ and\ \citenamefont
  {Zupan}}]{CirelliStrumiaZupan2024}%
  \BibitemOpen
  \bibfield  {author} {\bibinfo {author} {\bibfnamefont {M.}~\bibnamefont
  {Cirelli}}, \bibinfo {author} {\bibfnamefont {A.}~\bibnamefont {Strumia}},\
  and\ \bibinfo {author} {\bibfnamefont {J.}~\bibnamefont {Zupan}},\ }\bibfield
   {title} {\bibinfo {title} {Dark matter},\ }\href@noop {} {\bibfield
  {journal} {\bibinfo  {journal} {arXiv preprint}\ } (\bibinfo {year}
  {2024})},\ \Eprint {https://arxiv.org/abs/2406.01705} {arXiv:2406.01705
  [hep-ph]} \BibitemShut {NoStop}%
\bibitem [{\citenamefont {Jungman}\ \emph {et~al.}(1996)\citenamefont
  {Jungman}, \citenamefont {Kamionkowski},\ and\ \citenamefont
  {Griest}}]{Jungman1996}%
  \BibitemOpen
  \bibfield  {author} {\bibinfo {author} {\bibfnamefont {G.}~\bibnamefont
  {Jungman}}, \bibinfo {author} {\bibfnamefont {M.}~\bibnamefont
  {Kamionkowski}},\ and\ \bibinfo {author} {\bibfnamefont {K.}~\bibnamefont
  {Griest}},\ }\bibfield  {title} {\bibinfo {title} {Supersymmetric dark
  matter},\ }\href {https://doi.org/10.1016/0370-1573(95)00058-5} {\bibfield
  {journal} {\bibinfo  {journal} {Phys. Rept.}\ }\textbf {\bibinfo {volume}
  {267}},\ \bibinfo {pages} {195} (\bibinfo {year} {1996})},\ \Eprint
  {https://arxiv.org/abs/hep-ph/9506380} {arXiv:hep-ph/9506380} \BibitemShut
  {NoStop}%
\bibitem [{\citenamefont {Feng}(2010)}]{Feng2010}%
  \BibitemOpen
  \bibfield  {author} {\bibinfo {author} {\bibfnamefont {J.~L.}\ \bibnamefont
  {Feng}},\ }\bibfield  {title} {\bibinfo {title} {Dark matter candidates from
  particle physics and methods of detection},\ }\href
  {https://doi.org/10.1146/annurev-astro-082708-101659} {\bibfield  {journal}
  {\bibinfo  {journal} {Ann. Rev. Astron. Astrophys.}\ }\textbf {\bibinfo
  {volume} {48}},\ \bibinfo {pages} {495} (\bibinfo {year} {2010})},\ \Eprint
  {https://arxiv.org/abs/1003.0904} {arXiv:1003.0904 [astro-ph.CO]}
  \BibitemShut {NoStop}%
\bibitem [{\citenamefont {Arcadi}\ \emph {et~al.}(2018)\citenamefont {Arcadi},
  \citenamefont {Dutra}, \citenamefont {Ghosh}, \citenamefont {Lindner},
  \citenamefont {Mambrini}, \citenamefont {Pierre}, \citenamefont {Profumo},\
  and\ \citenamefont {Queiroz}}]{Arcadi2018}%
  \BibitemOpen
  \bibfield  {author} {\bibinfo {author} {\bibfnamefont {G.}~\bibnamefont
  {Arcadi}}, \bibinfo {author} {\bibfnamefont {M.}~\bibnamefont {Dutra}},
  \bibinfo {author} {\bibfnamefont {P.}~\bibnamefont {Ghosh}}, \bibinfo
  {author} {\bibfnamefont {M.}~\bibnamefont {Lindner}}, \bibinfo {author}
  {\bibfnamefont {Y.}~\bibnamefont {Mambrini}}, \bibinfo {author}
  {\bibfnamefont {M.}~\bibnamefont {Pierre}}, \bibinfo {author} {\bibfnamefont
  {S.}~\bibnamefont {Profumo}},\ and\ \bibinfo {author} {\bibfnamefont {F.~S.}\
  \bibnamefont {Queiroz}},\ }\bibfield  {title} {\bibinfo {title} {The waning
  of the wimp? a review of models, searches, and constraints},\ }\href
  {https://doi.org/10.1140/epjc/s10052-018-5662-y} {\bibfield  {journal}
  {\bibinfo  {journal} {Eur. Phys. J. C}\ }\textbf {\bibinfo {volume} {78}},\
  \bibinfo {pages} {203} (\bibinfo {year} {2018})},\ \Eprint
  {https://arxiv.org/abs/1703.07364} {arXiv:1703.07364 [hep-ph]} \BibitemShut
  {NoStop}%
\bibitem [{\citenamefont {Arcadi}\ \emph {et~al.}(2025)\citenamefont {Arcadi}
  \emph {et~al.}}]{Arcadi2025}%
  \BibitemOpen
  \bibfield  {author} {\bibinfo {author} {\bibfnamefont {G.}~\bibnamefont
  {Arcadi}} \emph {et~al.},\ }\bibfield  {title} {\bibinfo {title} {The waning
  of the wimp: Endgame?},\ }\href
  {https://doi.org/10.1140/epjc/s10052-024-13672-y} {\bibfield  {journal}
  {\bibinfo  {journal} {Eur. Phys. J. C}\ }\textbf {\bibinfo {volume} {85}},\
  \bibinfo {pages} {152} (\bibinfo {year} {2025})},\ \Eprint
  {https://arxiv.org/abs/2403.15860} {arXiv:2403.15860 [hep-ph]} \BibitemShut
  {NoStop}%
\bibitem [{\citenamefont {Pospelov}\ \emph {et~al.}(2008)\citenamefont
  {Pospelov}, \citenamefont {Ritz},\ and\ \citenamefont
  {Voloshin}}]{PospelovRitzVoloshin2008}%
  \BibitemOpen
  \bibfield  {author} {\bibinfo {author} {\bibfnamefont {M.}~\bibnamefont
  {Pospelov}}, \bibinfo {author} {\bibfnamefont {A.}~\bibnamefont {Ritz}},\
  and\ \bibinfo {author} {\bibfnamefont {M.~B.}\ \bibnamefont {Voloshin}},\
  }\bibfield  {title} {\bibinfo {title} {Secluded wimp dark matter},\ }\href
  {https://doi.org/10.1016/j.physletb.2008.02.052} {\bibfield  {journal}
  {\bibinfo  {journal} {Phys. Lett. B}\ }\textbf {\bibinfo {volume} {662}},\
  \bibinfo {pages} {53} (\bibinfo {year} {2008})},\ \Eprint
  {https://arxiv.org/abs/0711.4866} {arXiv:0711.4866 [hep-ph]} \BibitemShut
  {NoStop}%
\bibitem [{\citenamefont {Di~Mauro}\ \emph {et~al.}(2023)\citenamefont
  {Di~Mauro}, \citenamefont {Arina}, \citenamefont {Fornengo}, \citenamefont
  {Heisig},\ and\ \citenamefont {Massaro}}]{DiMauroArina2023}%
  \BibitemOpen
  \bibfield  {author} {\bibinfo {author} {\bibfnamefont {M.}~\bibnamefont
  {Di~Mauro}}, \bibinfo {author} {\bibfnamefont {C.}~\bibnamefont {Arina}},
  \bibinfo {author} {\bibfnamefont {N.}~\bibnamefont {Fornengo}}, \bibinfo
  {author} {\bibfnamefont {J.}~\bibnamefont {Heisig}},\ and\ \bibinfo {author}
  {\bibfnamefont {D.}~\bibnamefont {Massaro}},\ }\bibfield  {title} {\bibinfo
  {title} {Dark matter in the higgs resonance region},\ }\href
  {https://doi.org/10.1103/PhysRevD.108.095008} {\bibfield  {journal} {\bibinfo
   {journal} {Phys. Rev. D}\ }\textbf {\bibinfo {volume} {108}},\ \bibinfo
  {pages} {095008} (\bibinfo {year} {2023})},\ \Eprint
  {https://arxiv.org/abs/2305.11937} {arXiv:2305.11937 [hep-ph]} \BibitemShut
  {NoStop}%
\bibitem [{\citenamefont {Di~Mauro}\ and\ \citenamefont
  {Xie}(2025)}]{DiMauroXie2025}%
  \BibitemOpen
  \bibfield  {author} {\bibinfo {author} {\bibfnamefont {M.}~\bibnamefont
  {Di~Mauro}}\ and\ \bibinfo {author} {\bibfnamefont {B.}~\bibnamefont {Xie}},\
  }\bibfield  {title} {\bibinfo {title} {Dark matter simplified models in the
  resonance region},\ }\href@noop {} {\bibfield  {journal} {\bibinfo  {journal}
  {arXiv preprint}\ } (\bibinfo {year} {2025})},\ \Eprint
  {https://arxiv.org/abs/2510.08677} {arXiv:2510.08677 [hep-ph]} \BibitemShut
  {NoStop}%
\bibitem [{\citenamefont {Di~Mauro}\ and\ \citenamefont
  {Wang}(2025)}]{DiMauroWang2025}%
  \BibitemOpen
  \bibfield  {author} {\bibinfo {author} {\bibfnamefont {M.}~\bibnamefont
  {Di~Mauro}}\ and\ \bibinfo {author} {\bibfnamefont {Y.}~\bibnamefont
  {Wang}},\ }\bibfield  {title} {\bibinfo {title} {Wimp shadows: Phenomenology
  of secluded dark matter in three minimal bsm scenarios},\ }\href@noop {}
  {\bibfield  {journal} {\bibinfo  {journal} {arXiv preprint}\ } (\bibinfo
  {year} {2025})},\ \Eprint {https://arxiv.org/abs/2510.23771}
  {arXiv:2510.23771 [hep-ph]} \BibitemShut {NoStop}%
\bibitem [{\citenamefont {Shaikh}\ and\ \citenamefont
  {Di~Mauro}(2026)}]{ShaikhDiMauro2026}%
  \BibitemOpen
  \bibfield  {author} {\bibinfo {author} {\bibfnamefont {H.}~\bibnamefont
  {Shaikh}}\ and\ \bibinfo {author} {\bibfnamefont {M.}~\bibnamefont
  {Di~Mauro}},\ }\bibfield  {title} {\bibinfo {title} {Vector higgs-portal dark
  matter: How uv completion reopens viable parameter space},\ }\href@noop {}
  {\bibfield  {journal} {\bibinfo  {journal} {arXiv preprint}\ } (\bibinfo
  {year} {2026})},\ \Eprint {https://arxiv.org/abs/2603.11233}
  {arXiv:2603.11233 [hep-ph]} \BibitemShut {NoStop}%
\bibitem [{\citenamefont {Tucker-Smith}\ and\ \citenamefont
  {Weiner}(2001)}]{TuckerSmith2001}%
  \BibitemOpen
  \bibfield  {author} {\bibinfo {author} {\bibfnamefont {D.}~\bibnamefont
  {Tucker-Smith}}\ and\ \bibinfo {author} {\bibfnamefont {N.}~\bibnamefont
  {Weiner}},\ }\bibfield  {title} {\bibinfo {title} {Inelastic dark matter},\
  }\href {https://doi.org/10.1103/PhysRevD.64.043502} {\bibfield  {journal}
  {\bibinfo  {journal} {Phys. Rev. D}\ }\textbf {\bibinfo {volume} {64}},\
  \bibinfo {pages} {043502} (\bibinfo {year} {2001})},\ \Eprint
  {https://arxiv.org/abs/hep-ph/0101138} {arXiv:hep-ph/0101138} \BibitemShut
  {NoStop}%
\bibitem [{\citenamefont {Tucker-Smith}\ and\ \citenamefont
  {Weiner}(2005)}]{TuckerSmith2005}%
  \BibitemOpen
  \bibfield  {author} {\bibinfo {author} {\bibfnamefont {D.}~\bibnamefont
  {Tucker-Smith}}\ and\ \bibinfo {author} {\bibfnamefont {N.}~\bibnamefont
  {Weiner}},\ }\bibfield  {title} {\bibinfo {title} {The status of inelastic
  dark matter},\ }\href {https://doi.org/10.1103/PhysRevD.72.063509} {\bibfield
   {journal} {\bibinfo  {journal} {Phys. Rev. D}\ }\textbf {\bibinfo {volume}
  {72}},\ \bibinfo {pages} {063509} (\bibinfo {year} {2005})},\ \Eprint
  {https://arxiv.org/abs/hep-ph/0402065} {arXiv:hep-ph/0402065} \BibitemShut
  {NoStop}%
\bibitem [{\citenamefont {Bramante}\ \emph {et~al.}(2016)\citenamefont
  {Bramante}, \citenamefont {Fox}, \citenamefont {Kribs},\ and\ \citenamefont
  {Martin}}]{Bramante2016}%
  \BibitemOpen
  \bibfield  {author} {\bibinfo {author} {\bibfnamefont {J.}~\bibnamefont
  {Bramante}}, \bibinfo {author} {\bibfnamefont {P.~J.}\ \bibnamefont {Fox}},
  \bibinfo {author} {\bibfnamefont {G.~D.}\ \bibnamefont {Kribs}},\ and\
  \bibinfo {author} {\bibfnamefont {A.}~\bibnamefont {Martin}},\ }\bibfield
  {title} {\bibinfo {title} {Inelastic frontier: Discovering dark matter at
  high recoil energy},\ }\href {https://doi.org/10.1103/PhysRevD.94.115026}
  {\bibfield  {journal} {\bibinfo  {journal} {Phys. Rev. D}\ }\textbf {\bibinfo
  {volume} {94}},\ \bibinfo {pages} {115026} (\bibinfo {year} {2016})},\
  \Eprint {https://arxiv.org/abs/1608.02662} {arXiv:1608.02662 [hep-ph]}
  \BibitemShut {NoStop}%
\bibitem [{\citenamefont {Dalla Valle~Garcia}\ \emph
  {et~al.}(2025)\citenamefont {Dalla Valle~Garcia}, \citenamefont {Kahlhoefer},
  \citenamefont {Ovchynnikov},\ and\ \citenamefont
  {Schwetz}}]{DallaValleGarcia2025}%
  \BibitemOpen
  \bibfield  {author} {\bibinfo {author} {\bibfnamefont {G.}~\bibnamefont
  {Dalla Valle~Garcia}}, \bibinfo {author} {\bibfnamefont {F.}~\bibnamefont
  {Kahlhoefer}}, \bibinfo {author} {\bibfnamefont {M.}~\bibnamefont
  {Ovchynnikov}},\ and\ \bibinfo {author} {\bibfnamefont {T.}~\bibnamefont
  {Schwetz}},\ }\bibfield  {title} {\bibinfo {title} {Not-so-inelastic dark
  matter},\ }\href {https://doi.org/10.1007/JHEP02(2025)127} {\bibfield
  {journal} {\bibinfo  {journal} {JHEP}\ }\textbf {\bibinfo {volume} {02}},\
  \bibinfo {pages} {127}},\ \Eprint {https://arxiv.org/abs/2405.08081}
  {arXiv:2405.08081 [hep-ph]} \BibitemShut {NoStop}%
\bibitem [{\citenamefont {Foguel}\ \emph {et~al.}(2025)\citenamefont {Foguel},
  \citenamefont {Reimitz},\ and\ \citenamefont
  {Zukanovich~Funchal}}]{Foguel2025}%
  \BibitemOpen
  \bibfield  {author} {\bibinfo {author} {\bibfnamefont {A.~L.}\ \bibnamefont
  {Foguel}}, \bibinfo {author} {\bibfnamefont {P.}~\bibnamefont {Reimitz}},\
  and\ \bibinfo {author} {\bibfnamefont {R.}~\bibnamefont
  {Zukanovich~Funchal}},\ }\bibfield  {title} {\bibinfo {title} {Unlocking the
  inelastic dark matter window with vector mediators},\ }\href
  {https://doi.org/10.1007/JHEP05(2025)001} {\bibfield  {journal} {\bibinfo
  {journal} {JHEP}\ }\textbf {\bibinfo {volume} {05}},\ \bibinfo {pages}
  {001}},\ \Eprint {https://arxiv.org/abs/2410.00881} {arXiv:2410.00881
  [hep-ph]} \BibitemShut {NoStop}%
\bibitem [{\citenamefont {Akerib}\ \emph {et~al.}(2026)\citenamefont {Akerib}
  \emph {et~al.}}]{LZ2026}%
  \BibitemOpen
  \bibfield  {author} {\bibinfo {author} {\bibfnamefont {D.~S.}\ \bibnamefont
  {Akerib}} \emph {et~al.} (\bibinfo {collaboration} {LUX-ZEPLIN
  Collaboration}),\ }\bibfield  {title} {\bibinfo {title} {Search for dark
  matter particle interactions in an extended nuclear recoil energy window with
  the lux-zeplin (lz) experiment},\ }\href@noop {} {\bibfield  {journal}
  {\bibinfo  {journal} {arXiv preprint}\ } (\bibinfo {year} {2026})},\ \Eprint
  {https://arxiv.org/abs/2609.02823} {arXiv:2609.02823 [hep-ex]} \BibitemShut
  {NoStop}%
\bibitem [{\citenamefont {Di~Mauro}(2026)}]{DiMauro2026}%
  \BibitemOpen
  \bibfield  {author} {\bibinfo {author} {\bibfnamefont {M.}~\bibnamefont
  {Di~Mauro}},\ }\bibfield  {title} {\bibinfo {title} {Dark matter at the
  kinematic edge: Interpreting the 248 kev lz nuclear-recoil candidate},\
  }\href@noop {} {\bibfield  {journal} {\bibinfo  {journal} {arXiv preprint}\ }
  (\bibinfo {year} {2026})},\ \Eprint {https://arxiv.org/abs/2609.02608}
  {arXiv:2609.02608 [hep-ph]} \BibitemShut {NoStop}%
\bibitem [{\citenamefont {Fan}\ and\ \citenamefont
  {Reece}(2026)}]{FanReece2026}%
  \BibitemOpen
  \bibfield  {author} {\bibinfo {author} {\bibfnamefont {J.}~\bibnamefont
  {Fan}}\ and\ \bibinfo {author} {\bibfnamefont {M.}~\bibnamefont {Reece}},\
  }\bibfield  {title} {\bibinfo {title} {Higgsino above the sea of fog},\
  }\href@noop {} {\bibfield  {journal} {\bibinfo  {journal} {arXiv preprint}\ }
  (\bibinfo {year} {2026})},\ \Eprint {https://arxiv.org/abs/2609.01504}
  {arXiv:2609.01504 [hep-ph]} \BibitemShut {NoStop}%
\bibitem [{\citenamefont {Freese}\ and\ \citenamefont
  {Theodosopoulos}(2026)}]{FreeseTheodosopoulos2026}%
  \BibitemOpen
  \bibfield  {author} {\bibinfo {author} {\bibfnamefont {K.}~\bibnamefont
  {Freese}}\ and\ \bibinfo {author} {\bibfnamefont {D.~P.}\ \bibnamefont
  {Theodosopoulos}},\ }\bibfield  {title} {\bibinfo {title} {Higgsino dark
  matter interpretation of the lux-zeplin 248 kev nuclear-recoil event},\
  }\href@noop {} {\bibfield  {journal} {\bibinfo  {journal} {arXiv preprint}\ }
  (\bibinfo {year} {2026})},\ \Eprint {https://arxiv.org/abs/2609.01583}
  {arXiv:2609.01583 [hep-ph]} \BibitemShut {NoStop}%
\bibitem [{\citenamefont {Wu}\ \emph {et~al.}(2026)\citenamefont {Wu},
  \citenamefont {Zhang},\ and\ \citenamefont {Zhu}}]{WuZhangZhu2026}%
  \BibitemOpen
  \bibfield  {author} {\bibinfo {author} {\bibfnamefont {L.}~\bibnamefont
  {Wu}}, \bibinfo {author} {\bibfnamefont {Y.}~\bibnamefont {Zhang}},\ and\
  \bibinfo {author} {\bibfnamefont {B.}~\bibnamefont {Zhu}},\ }\bibfield
  {title} {\bibinfo {title} {Tev higgsino dark matter from lz nuclear recoil to
  fermi-lat gamma rays},\ }\href@noop {} {\bibfield  {journal} {\bibinfo
  {journal} {arXiv preprint}\ } (\bibinfo {year} {2026})},\ \Eprint
  {https://arxiv.org/abs/2609.01590} {arXiv:2609.01590 [hep-ph]} \BibitemShut
  {NoStop}%
\bibitem [{\citenamefont {Du}\ and\ \citenamefont {Wang}(2026)}]{DuWang2026}%
  \BibitemOpen
  \bibfield  {author} {\bibinfo {author} {\bibfnamefont {X.}~\bibnamefont
  {Du}}\ and\ \bibinfo {author} {\bibfnamefont {F.}~\bibnamefont {Wang}},\
  }\bibfield  {title} {\bibinfo {title} {Tev higgsino interpretation of the lz
  high-recoil event with intermediate-scale electroweak gauginos},\ }\href@noop
  {} {\bibfield  {journal} {\bibinfo  {journal} {arXiv preprint}\ } (\bibinfo
  {year} {2026})},\ \Eprint {https://arxiv.org/abs/2609.04163}
  {arXiv:2609.04163 [hep-ph]} \BibitemShut {NoStop}%
\bibitem [{\citenamefont {Smirnov}\ \emph {et~al.}(2026)\citenamefont
  {Smirnov}, \citenamefont {Griffith},\ and\ \citenamefont
  {Beacom}}]{SmirnovGriffithBeacom2026}%
  \BibitemOpen
  \bibfield  {author} {\bibinfo {author} {\bibfnamefont {J.}~\bibnamefont
  {Smirnov}}, \bibinfo {author} {\bibfnamefont {S.}~\bibnamefont {Griffith}},\
  and\ \bibinfo {author} {\bibfnamefont {J.~F.}\ \bibnamefont {Beacom}},\
  }\bibfield  {title} {\bibinfo {title} {Inelastic signatures of electroweak
  dark matter},\ }\href@noop {} {\bibfield  {journal} {\bibinfo  {journal}
  {arXiv preprint}\ } (\bibinfo {year} {2026})},\ \Eprint
  {https://arxiv.org/abs/2609.04144} {arXiv:2609.04144 [hep-ph]} \BibitemShut
  {NoStop}%
\bibitem [{\citenamefont {Su}\ \emph {et~al.}(2026)\citenamefont {Su},
  \citenamefont {Yang},\ and\ \citenamefont {Yang}}]{SuYangYang2026}%
  \BibitemOpen
  \bibfield  {author} {\bibinfo {author} {\bibfnamefont {L.}~\bibnamefont
  {Su}}, \bibinfo {author} {\bibfnamefont {J.~M.}\ \bibnamefont {Yang}},\ and\
  \bibinfo {author} {\bibfnamefont {W.-N.}\ \bibnamefont {Yang}},\ }\bibfield
  {title} {\bibinfo {title} {Inelastic dark matter signature at high recoil
  energy in lux-zeplin and cresst},\ }\href@noop {} {\bibfield  {journal}
  {\bibinfo  {journal} {arXiv preprint}\ } (\bibinfo {year} {2026})},\ \Eprint
  {https://arxiv.org/abs/2609.01475} {arXiv:2609.01475 [hep-ph]} \BibitemShut
  {NoStop}%
\bibitem [{\citenamefont {Yamashita}(2026)}]{Yamashita2026}%
  \BibitemOpen
  \bibfield  {author} {\bibinfo {author} {\bibfnamefont {K.}~\bibnamefont
  {Yamashita}},\ }\bibfield  {title} {\bibinfo {title} {Inelastic dark photon
  dark matter for the lux-zeplin high-recoil event and the galactic halo
  gamma-ray excess},\ }\href@noop {} {\bibfield  {journal} {\bibinfo  {journal}
  {arXiv preprint}\ } (\bibinfo {year} {2026})},\ \Eprint
  {https://arxiv.org/abs/2609.02868} {arXiv:2609.02868 [hep-ph]} \BibitemShut
  {NoStop}%
\bibitem [{\citenamefont {Lee}\ and\ \citenamefont {Youn}(2026)}]{LeeYoun2026}%
  \BibitemOpen
  \bibfield  {author} {\bibinfo {author} {\bibfnamefont {S.~J.}\ \bibnamefont
  {Lee}}\ and\ \bibinfo {author} {\bibfnamefont {T.}~\bibnamefont {Youn}},\
  }\bibfield  {title} {\bibinfo {title} {Mixing-suppressed inelastic dark
  matter: a minimal model for the lz 248 kev event},\ }\href@noop {} {\bibfield
   {journal} {\bibinfo  {journal} {arXiv preprint}\ } (\bibinfo {year}
  {2026})},\ \Eprint {https://arxiv.org/abs/2609.09138} {arXiv:2609.09138
  [hep-ph]} \BibitemShut {NoStop}%
\bibitem [{\citenamefont {McCabe}(2026)}]{McCabe2026}%
  \BibitemOpen
  \bibfield  {author} {\bibinfo {author} {\bibfnamefont {C.}~\bibnamefont
  {McCabe}},\ }\bibfield  {title} {\bibinfo {title} {Seasonal dark matter from
  the lux-zeplin high-energy event},\ }\href@noop {} {\bibfield  {journal}
  {\bibinfo  {journal} {arXiv preprint}\ } (\bibinfo {year} {2026})},\ \Eprint
  {https://arxiv.org/abs/2609.04181} {arXiv:2609.04181 [hep-ph]} \BibitemShut
  {NoStop}%
\bibitem [{\citenamefont {Rodd}\ \emph {et~al.}(2026)\citenamefont {Rodd},
  \citenamefont {Safdi}, \citenamefont {Slatyer},\ and\ \citenamefont
  {Xu}}]{RoddSafdiSlatyerXu2026}%
  \BibitemOpen
  \bibfield  {author} {\bibinfo {author} {\bibfnamefont {N.~L.}\ \bibnamefont
  {Rodd}}, \bibinfo {author} {\bibfnamefont {B.~R.}\ \bibnamefont {Safdi}},
  \bibinfo {author} {\bibfnamefont {T.~R.}\ \bibnamefont {Slatyer}},\ and\
  \bibinfo {author} {\bibfnamefont {W.~L.}\ \bibnamefont {Xu}},\ }\bibfield
  {title} {\bibinfo {title} {Confronting the higgsino interpretation of the lz
  event with the high-energy sideband},\ }\href@noop {} {\bibfield  {journal}
  {\bibinfo  {journal} {arXiv preprint}\ } (\bibinfo {year} {2026})},\ \Eprint
  {https://arxiv.org/abs/2609.04175} {arXiv:2609.04175 [hep-ph]} \BibitemShut
  {NoStop}%
\bibitem [{\citenamefont {Gould}(1987)}]{Gould1987a}%
  \BibitemOpen
  \bibfield  {author} {\bibinfo {author} {\bibfnamefont {A.}~\bibnamefont
  {Gould}},\ }\bibfield  {title} {\bibinfo {title} {Wimp distribution in and
  evaporation from the sun},\ }\href {https://doi.org/10.1086/165653}
  {\bibfield  {journal} {\bibinfo  {journal} {Astrophys. J.}\ }\textbf
  {\bibinfo {volume} {321}},\ \bibinfo {pages} {560} (\bibinfo {year}
  {1987})}\BibitemShut {NoStop}%
\bibitem [{\citenamefont {Nussinov}\ \emph {et~al.}(2009)\citenamefont
  {Nussinov}, \citenamefont {Wang},\ and\ \citenamefont
  {Yavin}}]{NussinovWangYavin2009}%
  \BibitemOpen
  \bibfield  {author} {\bibinfo {author} {\bibfnamefont {S.}~\bibnamefont
  {Nussinov}}, \bibinfo {author} {\bibfnamefont {L.-T.}\ \bibnamefont {Wang}},\
  and\ \bibinfo {author} {\bibfnamefont {I.}~\bibnamefont {Yavin}},\ }\bibfield
   {title} {\bibinfo {title} {Capture of inelastic dark matter in the sun},\
  }\href {https://doi.org/10.1088/1475-7516/2009/08/037} {\bibfield  {journal}
  {\bibinfo  {journal} {JCAP}\ }\textbf {\bibinfo {volume} {08}},\ \bibinfo
  {pages} {037}},\ \Eprint {https://arxiv.org/abs/0905.1333} {arXiv:0905.1333
  [hep-ph]} \BibitemShut {NoStop}%
\bibitem [{\citenamefont {Menon}\ \emph {et~al.}(2010)\citenamefont {Menon},
  \citenamefont {Morris}, \citenamefont {Pierce},\ and\ \citenamefont
  {Weiner}}]{MenonMorrisPierceWeiner2010}%
  \BibitemOpen
  \bibfield  {author} {\bibinfo {author} {\bibfnamefont {A.}~\bibnamefont
  {Menon}}, \bibinfo {author} {\bibfnamefont {R.}~\bibnamefont {Morris}},
  \bibinfo {author} {\bibfnamefont {A.}~\bibnamefont {Pierce}},\ and\ \bibinfo
  {author} {\bibfnamefont {N.}~\bibnamefont {Weiner}},\ }\bibfield  {title}
  {\bibinfo {title} {Capture and indirect detection of inelastic dark matter},\
  }\href {https://doi.org/10.1103/PhysRevD.82.015011} {\bibfield  {journal}
  {\bibinfo  {journal} {Phys. Rev. D}\ }\textbf {\bibinfo {volume} {82}},\
  \bibinfo {pages} {015011} (\bibinfo {year} {2010})},\ \Eprint
  {https://arxiv.org/abs/0905.1847} {arXiv:0905.1847 [hep-ph]} \BibitemShut
  {NoStop}%
\bibitem [{\citenamefont {Blennow}\ \emph {et~al.}(2016)\citenamefont
  {Blennow}, \citenamefont {Clementz},\ and\ \citenamefont
  {Herrero-Garcia}}]{BlennowClementzHerreroGarcia2016}%
  \BibitemOpen
  \bibfield  {author} {\bibinfo {author} {\bibfnamefont {M.}~\bibnamefont
  {Blennow}}, \bibinfo {author} {\bibfnamefont {S.}~\bibnamefont {Clementz}},\
  and\ \bibinfo {author} {\bibfnamefont {J.}~\bibnamefont {Herrero-Garcia}},\
  }\bibfield  {title} {\bibinfo {title} {Pinning down inelastic dark matter in
  the sun and in direct detection},\ }\href
  {https://doi.org/10.1088/1475-7516/2016/04/004} {\bibfield  {journal}
  {\bibinfo  {journal} {JCAP}\ }\textbf {\bibinfo {volume} {04}},\ \bibinfo
  {pages} {004}},\ \Eprint {https://arxiv.org/abs/1512.03317} {arXiv:1512.03317
  [hep-ph]} \BibitemShut {NoStop}%
\bibitem [{\citenamefont {Catena}\ and\ \citenamefont
  {Hellstrom}(2018)}]{Catena2018}%
  \BibitemOpen
  \bibfield  {author} {\bibinfo {author} {\bibfnamefont {R.}~\bibnamefont
  {Catena}}\ and\ \bibinfo {author} {\bibfnamefont {F.}~\bibnamefont
  {Hellstrom}},\ }\bibfield  {title} {\bibinfo {title} {New constraints on
  inelastic dark matter from icecube},\ }\href
  {https://doi.org/10.1088/1475-7516/2018/10/039} {\bibfield  {journal}
  {\bibinfo  {journal} {JCAP}\ }\textbf {\bibinfo {volume} {10}},\ \bibinfo
  {pages} {039}},\ \Eprint {https://arxiv.org/abs/1808.08082} {arXiv:1808.08082
  [hep-ph]} \BibitemShut {NoStop}%
\bibitem [{\citenamefont {Abbasi}\ \emph {et~al.}(2025)\citenamefont {Abbasi}
  \emph {et~al.}}]{IceCubeSolar2025}%
  \BibitemOpen
  \bibfield  {author} {\bibinfo {author} {\bibfnamefont {R.}~\bibnamefont
  {Abbasi}} \emph {et~al.} (\bibinfo {collaboration} {IceCube Collaboration}),\
  }\bibfield  {title} {\bibinfo {title} {Search for high-energy neutrinos from
  the sun using ten years of icecube data},\ }\href@noop {} {\bibfield
  {journal} {\bibinfo  {journal} {arXiv preprint}\ } (\bibinfo {year}
  {2025})},\ \Eprint {https://arxiv.org/abs/2507.08457} {arXiv:2507.08457
  [hep-ex]} \BibitemShut {NoStop}%
\bibitem [{\citenamefont {Pospelov}\ and\ \citenamefont
  {Ramani}(2026)}]{PospelovRamani2026}%
  \BibitemOpen
  \bibfield  {author} {\bibinfo {author} {\bibfnamefont {M.}~\bibnamefont
  {Pospelov}}\ and\ \bibinfo {author} {\bibfnamefont {H.}~\bibnamefont
  {Ramani}},\ }\bibfield  {title} {\bibinfo {title} {Strong constraints on
  higgsino dark matter from solar capture},\ }\href@noop {} {\bibfield
  {journal} {\bibinfo  {journal} {arXiv preprint}\ } (\bibinfo {year}
  {2026})},\ \Eprint {https://arxiv.org/abs/2609.02775} {arXiv:2609.02775
  [hep-ph]} \BibitemShut {NoStop}%
\bibitem [{\citenamefont {Di~Mauro}\ and\ \citenamefont
  {Shaikh}(2026)}]{DiMauroShaikh2026Solar}%
  \BibitemOpen
  \bibfield  {author} {\bibinfo {author} {\bibfnamefont {M.}~\bibnamefont
  {Di~Mauro}}\ and\ \bibinfo {author} {\bibfnamefont {H.}~\bibnamefont
  {Shaikh}},\ }\bibfield  {title} {\bibinfo {title} {Solar capture tests of
  inelastic dark matter after the lz high-recoil event},\ }\href@noop {}
  {\bibfield  {journal} {\bibinfo  {journal} {arXiv preprint}\ } (\bibinfo
  {year} {2026})},\ \Eprint {https://arxiv.org/abs/2609.06760}
  {arXiv:2609.06760 [hep-ph]} \BibitemShut {NoStop}%
\bibitem [{\citenamefont {Bose}\ \emph {et~al.}(2026)\citenamefont {Bose} \emph
  {et~al.}}]{BoseEtAl2026}%
  \BibitemOpen
  \bibfield  {author} {\bibinfo {author} {\bibfnamefont {D.}~\bibnamefont
  {Bose}} \emph {et~al.},\ }\bibfield  {title} {\bibinfo {title} {Not so good
  neutrinos for higgsino dark matter as lz excess: stringent limits from
  super-kamiokande and icecube},\ }\href@noop {} {\bibfield  {journal}
  {\bibinfo  {journal} {arXiv preprint}\ } (\bibinfo {year} {2026})},\ \Eprint
  {https://arxiv.org/abs/2609.07807} {arXiv:2609.07807 [hep-ph]} \BibitemShut
  {NoStop}%
\bibitem [{\citenamefont {Griffith}\ \emph {et~al.}(2026)\citenamefont
  {Griffith}, \citenamefont {Smirnov}, \citenamefont {Lopez-Honorez},\ and\
  \citenamefont {Beacom}}]{GriffithSmirnov2026}%
  \BibitemOpen
  \bibfield  {author} {\bibinfo {author} {\bibfnamefont {S.}~\bibnamefont
  {Griffith}}, \bibinfo {author} {\bibfnamefont {J.}~\bibnamefont {Smirnov}},
  \bibinfo {author} {\bibfnamefont {L.}~\bibnamefont {Lopez-Honorez}},\ and\
  \bibinfo {author} {\bibfnamefont {J.~F.}\ \bibnamefont {Beacom}},\ }\bibfield
   {title} {\bibinfo {title} {Minimal dark matter: Generalized framework and
  direct-detection sensitivity},\ }\href {https://doi.org/10.1103/jv6s-76s9}
  {\bibfield  {journal} {\bibinfo  {journal} {Phys. Rev. D}\ }\textbf {\bibinfo
  {volume} {114}},\ \bibinfo {pages} {035016} (\bibinfo {year} {2026})},\
  \Eprint {https://arxiv.org/abs/2602.17764} {arXiv:2602.17764 [hep-ph]}
  \BibitemShut {NoStop}%
\bibitem [{\citenamefont {Cirelli}\ \emph
  {et~al.}(2006{\natexlab{a}})\citenamefont {Cirelli}, \citenamefont
  {Fornengo},\ and\ \citenamefont {Strumia}}]{Cirelli:2005uq}%
  \BibitemOpen
  \bibfield  {author} {\bibinfo {author} {\bibfnamefont {M.}~\bibnamefont
  {Cirelli}}, \bibinfo {author} {\bibfnamefont {N.}~\bibnamefont {Fornengo}},\
  and\ \bibinfo {author} {\bibfnamefont {A.}~\bibnamefont {Strumia}},\
  }\bibfield  {title} {\bibinfo {title} {{Minimal Dark Matter}},\ }\href
  {https://doi.org/10.1016/j.nuclphysb.2006.07.012} {\bibfield  {journal}
  {\bibinfo  {journal} {Nucl. Phys. B}\ }\textbf {\bibinfo {volume} {753}},\
  \bibinfo {pages} {178} (\bibinfo {year} {2006}{\natexlab{a}})},\ \Eprint
  {https://arxiv.org/abs/hep-ph/0512090} {arXiv:hep-ph/0512090} \BibitemShut
  {NoStop}%
\bibitem [{\citenamefont {Cirelli}\ \emph
  {et~al.}(2007{\natexlab{a}})\citenamefont {Cirelli}, \citenamefont
  {Strumia},\ and\ \citenamefont {Tamburini}}]{Cirelli:2007xd}%
  \BibitemOpen
  \bibfield  {author} {\bibinfo {author} {\bibfnamefont {M.}~\bibnamefont
  {Cirelli}}, \bibinfo {author} {\bibfnamefont {A.}~\bibnamefont {Strumia}},\
  and\ \bibinfo {author} {\bibfnamefont {M.}~\bibnamefont {Tamburini}},\
  }\bibfield  {title} {\bibinfo {title} {{Cosmology and Astrophysics of Minimal
  Dark Matter}},\ }\href {https://doi.org/10.1016/j.nuclphysb.2007.07.023}
  {\bibfield  {journal} {\bibinfo  {journal} {Nucl. Phys. B}\ }\textbf
  {\bibinfo {volume} {787}},\ \bibinfo {pages} {152} (\bibinfo {year}
  {2007}{\natexlab{a}})},\ \Eprint {https://arxiv.org/abs/0706.4071}
  {arXiv:0706.4071 [hep-ph]} \BibitemShut {NoStop}%
\bibitem [{\citenamefont {Hisano}\ \emph
  {et~al.}(2011{\natexlab{a}})\citenamefont {Hisano}, \citenamefont {Ishiwata},
  \citenamefont {Nagata},\ and\ \citenamefont {Takesako}}]{Hisano:2011cs}%
  \BibitemOpen
  \bibfield  {author} {\bibinfo {author} {\bibfnamefont {J.}~\bibnamefont
  {Hisano}}, \bibinfo {author} {\bibfnamefont {K.}~\bibnamefont {Ishiwata}},
  \bibinfo {author} {\bibfnamefont {N.}~\bibnamefont {Nagata}},\ and\ \bibinfo
  {author} {\bibfnamefont {T.}~\bibnamefont {Takesako}},\ }\bibfield  {title}
  {\bibinfo {title} {{Direct Detection of Electroweak-Interacting Dark
  Matter}},\ }\href {https://doi.org/10.1007/JHEP07(2011)005} {\bibfield
  {journal} {\bibinfo  {journal} {JHEP}\ }\textbf {\bibinfo {volume} {07}},\
  \bibinfo {pages} {005}},\ \Eprint {https://arxiv.org/abs/1104.0228}
  {arXiv:1104.0228 [hep-ph]} \BibitemShut {NoStop}%
\bibitem [{\citenamefont {Bottaro}\ \emph
  {et~al.}(2022{\natexlab{a}})\citenamefont {Bottaro}, \citenamefont
  {Buttazzo}, \citenamefont {Costa}, \citenamefont {Franceschini},
  \citenamefont {Panci}, \citenamefont {Redigolo},\ and\ \citenamefont
  {Vittorio}}]{Bottaro:2021snn}%
  \BibitemOpen
  \bibfield  {author} {\bibinfo {author} {\bibfnamefont {S.}~\bibnamefont
  {Bottaro}}, \bibinfo {author} {\bibfnamefont {D.}~\bibnamefont {Buttazzo}},
  \bibinfo {author} {\bibfnamefont {M.}~\bibnamefont {Costa}}, \bibinfo
  {author} {\bibfnamefont {R.}~\bibnamefont {Franceschini}}, \bibinfo {author}
  {\bibfnamefont {P.}~\bibnamefont {Panci}}, \bibinfo {author} {\bibfnamefont
  {D.}~\bibnamefont {Redigolo}},\ and\ \bibinfo {author} {\bibfnamefont
  {L.}~\bibnamefont {Vittorio}},\ }\bibfield  {title} {\bibinfo {title}
  {{Closing the window on WIMP Dark Matter}},\ }\href
  {https://doi.org/10.1140/epjc/s10052-021-09917-9} {\bibfield  {journal}
  {\bibinfo  {journal} {Eur. Phys. J. C}\ }\textbf {\bibinfo {volume} {82}},\
  \bibinfo {pages} {31} (\bibinfo {year} {2022}{\natexlab{a}})},\ \Eprint
  {https://arxiv.org/abs/2107.09688} {arXiv:2107.09688 [hep-ph]} \BibitemShut
  {NoStop}%
\bibitem [{\citenamefont {Bloch}\ \emph {et~al.}(2025)\citenamefont {Bloch},
  \citenamefont {Bottaro}, \citenamefont {Redigolo},\ and\ \citenamefont
  {Vittorio}}]{Bloch:2024wimps}%
  \BibitemOpen
  \bibfield  {author} {\bibinfo {author} {\bibfnamefont {I.~M.}\ \bibnamefont
  {Bloch}}, \bibinfo {author} {\bibfnamefont {S.}~\bibnamefont {Bottaro}},
  \bibinfo {author} {\bibfnamefont {D.}~\bibnamefont {Redigolo}},\ and\
  \bibinfo {author} {\bibfnamefont {L.}~\bibnamefont {Vittorio}},\ }\bibfield
  {title} {\bibinfo {title} {{Looking for WIMPs through the neutrino fogs}},\
  }\href {https://doi.org/10.1007/JHEP08(2025)216} {\bibfield  {journal}
  {\bibinfo  {journal} {JHEP}\ }\textbf {\bibinfo {volume} {08}},\ \bibinfo
  {pages} {216}},\ \Eprint {https://arxiv.org/abs/2410.02723} {arXiv:2410.02723
  [hep-ph]} \BibitemShut {NoStop}%
\bibitem [{\citenamefont {Safdi}\ and\ \citenamefont
  {Xu}(2026)}]{Safdi:2025mdm}%
  \BibitemOpen
  \bibfield  {author} {\bibinfo {author} {\bibfnamefont {B.~R.}\ \bibnamefont
  {Safdi}}\ and\ \bibinfo {author} {\bibfnamefont {W.~L.}\ \bibnamefont {Xu}},\
  }\bibfield  {title} {\bibinfo {title} {{Wino and real minimal dark matter
  disfavored by Fermi gamma-ray observations}},\ }\bibfield  {journal}
  {\bibinfo  {journal} {Phys. Rev. D}\ }\href
  {https://doi.org/10.1103/pc9s-b56k} {10.1103/pc9s-b56k} (\bibinfo {year}
  {2026}),\ \Eprint {https://arxiv.org/abs/2507.15934} {arXiv:2507.15934
  [hep-ph]} \BibitemShut {NoStop}%
\bibitem [{\citenamefont {Cirelli}\ \emph
  {et~al.}(2006{\natexlab{b}})\citenamefont {Cirelli}, \citenamefont
  {Fornengo},\ and\ \citenamefont {Strumia}}]{CirelliFornengoStrumia2006}%
  \BibitemOpen
  \bibfield  {author} {\bibinfo {author} {\bibfnamefont {M.}~\bibnamefont
  {Cirelli}}, \bibinfo {author} {\bibfnamefont {N.}~\bibnamefont {Fornengo}},\
  and\ \bibinfo {author} {\bibfnamefont {A.}~\bibnamefont {Strumia}},\
  }\bibfield  {title} {\bibinfo {title} {Minimal dark matter},\ }\href
  {https://doi.org/10.1016/j.nuclphysb.2006.07.012} {\bibfield  {journal}
  {\bibinfo  {journal} {Nucl. Phys. B}\ }\textbf {\bibinfo {volume} {753}},\
  \bibinfo {pages} {178} (\bibinfo {year} {2006}{\natexlab{b}})},\ \Eprint
  {https://arxiv.org/abs/hep-ph/0512090} {arXiv:hep-ph/0512090} \BibitemShut
  {NoStop}%
\bibitem [{\citenamefont {Cirelli}\ \emph
  {et~al.}(2007{\natexlab{b}})\citenamefont {Cirelli}, \citenamefont
  {Strumia},\ and\ \citenamefont {Tamburini}}]{CirelliStrumiaTamburini2007}%
  \BibitemOpen
  \bibfield  {author} {\bibinfo {author} {\bibfnamefont {M.}~\bibnamefont
  {Cirelli}}, \bibinfo {author} {\bibfnamefont {A.}~\bibnamefont {Strumia}},\
  and\ \bibinfo {author} {\bibfnamefont {M.}~\bibnamefont {Tamburini}},\
  }\bibfield  {title} {\bibinfo {title} {Cosmology and astrophysics of minimal
  dark matter},\ }\href {https://doi.org/10.1016/j.nuclphysb.2007.07.023}
  {\bibfield  {journal} {\bibinfo  {journal} {Nucl. Phys. B}\ }\textbf
  {\bibinfo {volume} {787}},\ \bibinfo {pages} {152} (\bibinfo {year}
  {2007}{\natexlab{b}})},\ \Eprint {https://arxiv.org/abs/0706.4071}
  {arXiv:0706.4071 [hep-ph]} \BibitemShut {NoStop}%
\bibitem [{\citenamefont {Cirelli}\ \emph {et~al.}(2019)\citenamefont
  {Cirelli}, \citenamefont {Gouttenoire}, \citenamefont {Petraki},\ and\
  \citenamefont {Sala}}]{CirelliGouttenoirePetrakiSala2019}%
  \BibitemOpen
  \bibfield  {author} {\bibinfo {author} {\bibfnamefont {M.}~\bibnamefont
  {Cirelli}}, \bibinfo {author} {\bibfnamefont {Y.}~\bibnamefont
  {Gouttenoire}}, \bibinfo {author} {\bibfnamefont {K.}~\bibnamefont
  {Petraki}},\ and\ \bibinfo {author} {\bibfnamefont {F.}~\bibnamefont
  {Sala}},\ }\bibfield  {title} {\bibinfo {title} {Homeopathic dark matter, or
  how diluted heavy substances produce high energy cosmic rays},\ }\href
  {https://doi.org/10.1088/1475-7516/2019/02/014} {\bibfield  {journal}
  {\bibinfo  {journal} {JCAP}\ }\textbf {\bibinfo {volume} {02}},\ \bibinfo
  {pages} {014}},\ \Eprint {https://arxiv.org/abs/1811.03608} {arXiv:1811.03608
  [hep-ph]} \BibitemShut {NoStop}%
\bibitem [{\citenamefont {Bottaro}\ \emph
  {et~al.}(2022{\natexlab{b}})\citenamefont {Bottaro}, \citenamefont
  {Buttazzo}, \citenamefont {Costa}, \citenamefont {Franceschini},
  \citenamefont {Panci}, \citenamefont {Redigolo},\ and\ \citenamefont
  {Vittorio}}]{Bottaro2022}%
  \BibitemOpen
  \bibfield  {author} {\bibinfo {author} {\bibfnamefont {S.}~\bibnamefont
  {Bottaro}}, \bibinfo {author} {\bibfnamefont {D.}~\bibnamefont {Buttazzo}},
  \bibinfo {author} {\bibfnamefont {M.}~\bibnamefont {Costa}}, \bibinfo
  {author} {\bibfnamefont {R.}~\bibnamefont {Franceschini}}, \bibinfo {author}
  {\bibfnamefont {P.}~\bibnamefont {Panci}}, \bibinfo {author} {\bibfnamefont
  {D.}~\bibnamefont {Redigolo}},\ and\ \bibinfo {author} {\bibfnamefont
  {L.}~\bibnamefont {Vittorio}},\ }\bibfield  {title} {\bibinfo {title}
  {Closing the window on wimp dark matter},\ }\href
  {https://doi.org/10.1140/epjc/s10052-021-09917-9} {\bibfield  {journal}
  {\bibinfo  {journal} {Eur. Phys. J. C}\ }\textbf {\bibinfo {volume} {82}},\
  \bibinfo {pages} {31} (\bibinfo {year} {2022}{\natexlab{b}})},\ \Eprint
  {https://arxiv.org/abs/2107.09688} {arXiv:2107.09688 [hep-ph]} \BibitemShut
  {NoStop}%
\bibitem [{\citenamefont {Tait}\ and\ \citenamefont {Yu}(2016)}]{TaitYu2016}%
  \BibitemOpen
  \bibfield  {author} {\bibinfo {author} {\bibfnamefont {T.~M.~P.}\
  \bibnamefont {Tait}}\ and\ \bibinfo {author} {\bibfnamefont {Z.-H.}\
  \bibnamefont {Yu}},\ }\bibfield  {title} {\bibinfo {title}
  {Triplet-quadruplet dark matter},\ }\href
  {https://doi.org/10.1007/JHEP03(2016)204} {\bibfield  {journal} {\bibinfo
  {journal} {JHEP}\ }\textbf {\bibinfo {volume} {03}},\ \bibinfo {pages}
  {204}},\ \Eprint {https://arxiv.org/abs/1601.01354} {arXiv:1601.01354
  [hep-ph]} \BibitemShut {NoStop}%
\bibitem [{\citenamefont {Lopez~Honorez}\ \emph {et~al.}(2018)\citenamefont
  {Lopez~Honorez}, \citenamefont {Tytgat}, \citenamefont {Tziveloglou},\ and\
  \citenamefont {Zaldivar}}]{LopezHonorez2018}%
  \BibitemOpen
  \bibfield  {author} {\bibinfo {author} {\bibfnamefont {L.}~\bibnamefont
  {Lopez~Honorez}}, \bibinfo {author} {\bibfnamefont {M.~H.~G.}\ \bibnamefont
  {Tytgat}}, \bibinfo {author} {\bibfnamefont {P.}~\bibnamefont
  {Tziveloglou}},\ and\ \bibinfo {author} {\bibfnamefont {B.}~\bibnamefont
  {Zaldivar}},\ }\bibfield  {title} {\bibinfo {title} {On minimal dark matter
  coupled to the higgs},\ }\href {https://doi.org/10.1007/JHEP04(2018)011}
  {\bibfield  {journal} {\bibinfo  {journal} {JHEP}\ }\textbf {\bibinfo
  {volume} {04}},\ \bibinfo {pages} {011}},\ \Eprint
  {https://arxiv.org/abs/1711.08619} {arXiv:1711.08619 [hep-ph]} \BibitemShut
  {NoStop}%
\bibitem [{\citenamefont {Hisano}\ \emph
  {et~al.}(2011{\natexlab{b}})\citenamefont {Hisano}, \citenamefont {Ishiwata},
  \citenamefont {Nagata},\ and\ \citenamefont {Takesako}}]{Hisano2011}%
  \BibitemOpen
  \bibfield  {author} {\bibinfo {author} {\bibfnamefont {J.}~\bibnamefont
  {Hisano}}, \bibinfo {author} {\bibfnamefont {K.}~\bibnamefont {Ishiwata}},
  \bibinfo {author} {\bibfnamefont {N.}~\bibnamefont {Nagata}},\ and\ \bibinfo
  {author} {\bibfnamefont {T.}~\bibnamefont {Takesako}},\ }\bibfield  {title}
  {\bibinfo {title} {Direct detection of electroweak-interacting dark matter},\
  }\href {https://doi.org/10.1007/JHEP07(2011)005} {\bibfield  {journal}
  {\bibinfo  {journal} {JHEP}\ }\textbf {\bibinfo {volume} {07}}\bibfield
  {number} {\bibinfo  {number} { (07)},\ \bibinfo {pages} {005}},\ }\Eprint
  {https://arxiv.org/abs/1104.0228} {arXiv:1104.0228 [hep-ph]} \BibitemShut
  {NoStop}%
\bibitem [{\citenamefont {Chen}\ and\ \citenamefont
  {Hill}(2020)}]{ChenHill2020}%
  \BibitemOpen
  \bibfield  {author} {\bibinfo {author} {\bibfnamefont {Q.}~\bibnamefont
  {Chen}}\ and\ \bibinfo {author} {\bibfnamefont {R.~J.}\ \bibnamefont
  {Hill}},\ }\bibfield  {title} {\bibinfo {title} {Direct detection rate of
  heavy higgsino-like and wino-like dark matter},\ }\href
  {https://doi.org/10.1016/j.physletb.2020.135364} {\bibfield  {journal}
  {\bibinfo  {journal} {Phys. Lett. B}\ }\textbf {\bibinfo {volume} {804}},\
  \bibinfo {pages} {135364} (\bibinfo {year} {2020})},\ \Eprint
  {https://arxiv.org/abs/1912.07795} {arXiv:1912.07795 [hep-ph]} \BibitemShut
  {NoStop}%
\bibitem [{\citenamefont {Hisano}\ \emph {et~al.}(2015)\citenamefont {Hisano},
  \citenamefont {Ishiwata},\ and\ \citenamefont {Nagata}}]{Hisano2015}%
  \BibitemOpen
  \bibfield  {author} {\bibinfo {author} {\bibfnamefont {J.}~\bibnamefont
  {Hisano}}, \bibinfo {author} {\bibfnamefont {K.}~\bibnamefont {Ishiwata}},\
  and\ \bibinfo {author} {\bibfnamefont {N.}~\bibnamefont {Nagata}},\
  }\bibfield  {title} {\bibinfo {title} {Qcd effects on direct detection of
  wino dark matter},\ }\href {https://doi.org/10.1007/JHEP06(2015)097}
  {\bibfield  {journal} {\bibinfo  {journal} {JHEP}\ }\textbf {\bibinfo
  {volume} {06}},\ \bibinfo {pages} {097}},\ \Eprint
  {https://arxiv.org/abs/1504.00915} {arXiv:1504.00915 [hep-ph]} \BibitemShut
  {NoStop}%
\bibitem [{\citenamefont {Chen}\ \emph {et~al.}(2023)\citenamefont {Chen},
  \citenamefont {Ding},\ and\ \citenamefont {Hill}}]{ChenDingHill2023}%
  \BibitemOpen
  \bibfield  {author} {\bibinfo {author} {\bibfnamefont {Q.}~\bibnamefont
  {Chen}}, \bibinfo {author} {\bibfnamefont {G.-J.}\ \bibnamefont {Ding}},\
  and\ \bibinfo {author} {\bibfnamefont {R.~J.}\ \bibnamefont {Hill}},\
  }\bibfield  {title} {\bibinfo {title} {General heavy wimp nucleon elastic
  scattering},\ }\href {https://doi.org/10.1103/PhysRevD.108.116023} {\bibfield
   {journal} {\bibinfo  {journal} {Phys. Rev. D}\ }\textbf {\bibinfo {volume}
  {108}},\ \bibinfo {pages} {116023} (\bibinfo {year} {2023})},\ \Eprint
  {https://arxiv.org/abs/2309.02715} {arXiv:2309.02715 [hep-ph]} \BibitemShut
  {NoStop}%
\bibitem [{\citenamefont {Bisal}\ \emph {et~al.}(2024)\citenamefont {Bisal},
  \citenamefont {Chatterjee}, \citenamefont {Das},\ and\ \citenamefont
  {Pasha}}]{Bisal2024SD}%
  \BibitemOpen
  \bibfield  {author} {\bibinfo {author} {\bibfnamefont {S.}~\bibnamefont
  {Bisal}}, \bibinfo {author} {\bibfnamefont {A.}~\bibnamefont {Chatterjee}},
  \bibinfo {author} {\bibfnamefont {D.}~\bibnamefont {Das}},\ and\ \bibinfo
  {author} {\bibfnamefont {S.~A.}\ \bibnamefont {Pasha}},\ }\href@noop {}
  {\bibinfo {title} {Radiative corrections to the direct detection of the
  higgsino- (and wino-) like neutralino dark matter: Spin-dependent
  interactions}} (\bibinfo {year} {2024}),\ \Eprint
  {https://arxiv.org/abs/2410.18205} {arXiv:2410.18205 [hep-ph]} \BibitemShut
  {NoStop}%
\bibitem [{\citenamefont {Bahcall}\ \emph {et~al.}(2005)\citenamefont
  {Bahcall}, \citenamefont {Serenelli},\ and\ \citenamefont
  {Basu}}]{Bahcall2005}%
  \BibitemOpen
  \bibfield  {author} {\bibinfo {author} {\bibfnamefont {J.~N.}\ \bibnamefont
  {Bahcall}}, \bibinfo {author} {\bibfnamefont {A.~M.}\ \bibnamefont
  {Serenelli}},\ and\ \bibinfo {author} {\bibfnamefont {S.}~\bibnamefont
  {Basu}},\ }\bibfield  {title} {\bibinfo {title} {New solar opacities,
  abundances, helioseismology, and neutrino fluxes},\ }\href
  {https://doi.org/10.1086/428929} {\bibfield  {journal} {\bibinfo  {journal}
  {Astrophys. J. Lett.}\ }\textbf {\bibinfo {volume} {621}},\ \bibinfo {pages}
  {L85} (\bibinfo {year} {2005})},\ \Eprint
  {https://arxiv.org/abs/astro-ph/0412440} {arXiv:astro-ph/0412440}
  \BibitemShut {NoStop}%
\bibitem [{\citenamefont {Liu}\ \emph {et~al.}(2020)\citenamefont {Liu},
  \citenamefont {Lazar}, \citenamefont {Arguelles},\ and\ \citenamefont
  {Kheirandish}}]{Charon2020}%
  \BibitemOpen
  \bibfield  {author} {\bibinfo {author} {\bibfnamefont {Q.}~\bibnamefont
  {Liu}}, \bibinfo {author} {\bibfnamefont {J.}~\bibnamefont {Lazar}}, \bibinfo
  {author} {\bibfnamefont {C.~A.}\ \bibnamefont {Arguelles}},\ and\ \bibinfo
  {author} {\bibfnamefont {A.}~\bibnamefont {Kheirandish}},\ }\bibfield
  {title} {\bibinfo {title} {{chiaro nu}: A tool for neutrino flux generation
  from wimps},\ }\href {https://doi.org/10.1088/1475-7516/2020/10/043}
  {\bibfield  {journal} {\bibinfo  {journal} {JCAP}\ }\textbf {\bibinfo
  {volume} {10}}\bibfield  {number} {\bibinfo  {number} { (10)},\ \bibinfo
  {pages} {043}},\ }\Eprint {https://arxiv.org/abs/2007.15010}
  {arXiv:2007.15010 [hep-ph]} \BibitemShut {NoStop}%
\bibitem [{\citenamefont {Blennow}\ \emph {et~al.}(2008)\citenamefont
  {Blennow}, \citenamefont {Edsjo},\ and\ \citenamefont
  {Ohlsson}}]{Blennow2008}%
  \BibitemOpen
  \bibfield  {author} {\bibinfo {author} {\bibfnamefont {M.}~\bibnamefont
  {Blennow}}, \bibinfo {author} {\bibfnamefont {J.}~\bibnamefont {Edsjo}},\
  and\ \bibinfo {author} {\bibfnamefont {T.}~\bibnamefont {Ohlsson}},\
  }\bibfield  {title} {\bibinfo {title} {Neutrinos from wimp annihilations
  obtained using a full three-flavor monte carlo approach},\ }\href
  {https://doi.org/10.1088/1475-7516/2008/01/021} {\bibfield  {journal}
  {\bibinfo  {journal} {JCAP}\ }\textbf {\bibinfo {volume} {01}}\bibfield
  {number} {\bibinfo  {number} { (01)},\ \bibinfo {pages} {021}},\ }\Eprint
  {https://arxiv.org/abs/0709.3898} {arXiv:0709.3898 [hep-ph]} \BibitemShut
  {NoStop}%
\bibitem [{\citenamefont {Bauer}\ \emph {et~al.}(2021)\citenamefont {Bauer},
  \citenamefont {Rodd},\ and\ \citenamefont {Webber}}]{Bauer2021}%
  \BibitemOpen
  \bibfield  {author} {\bibinfo {author} {\bibfnamefont {C.~W.}\ \bibnamefont
  {Bauer}}, \bibinfo {author} {\bibfnamefont {N.~L.}\ \bibnamefont {Rodd}},\
  and\ \bibinfo {author} {\bibfnamefont {B.~R.}\ \bibnamefont {Webber}},\
  }\bibfield  {title} {\bibinfo {title} {Dark matter spectra from the
  electroweak to the planck scale},\ }\href
  {https://doi.org/10.1007/JHEP06(2021)121} {\bibfield  {journal} {\bibinfo
  {journal} {JHEP}\ }\textbf {\bibinfo {volume} {06}}\bibfield  {number}
  {\bibinfo  {number} { (06)},\ \bibinfo {pages} {121}},\ }\Eprint
  {https://arxiv.org/abs/2007.15001} {arXiv:2007.15001 [hep-ph]} \BibitemShut
  {NoStop}%
\bibitem [{\citenamefont {Maity}\ \emph {et~al.}(2025)\citenamefont {Maity},
  \citenamefont {Saha}, \citenamefont {Mondal},\ and\ \citenamefont
  {Laha}}]{MaityEtAl2023}%
  \BibitemOpen
  \bibfield  {author} {\bibinfo {author} {\bibfnamefont {T.~N.}\ \bibnamefont
  {Maity}}, \bibinfo {author} {\bibfnamefont {A.~K.}\ \bibnamefont {Saha}},
  \bibinfo {author} {\bibfnamefont {S.}~\bibnamefont {Mondal}},\ and\ \bibinfo
  {author} {\bibfnamefont {R.}~\bibnamefont {Laha}},\ }\bibfield  {title}
  {\bibinfo {title} {Neutrinos from the sun can discover dark matter-electron
  scattering},\ }\href {https://doi.org/10.1103/3f66-nfd5} {\bibfield
  {journal} {\bibinfo  {journal} {Phys. Rev. D}\ }\textbf {\bibinfo {volume}
  {112}},\ \bibinfo {pages} {023025} (\bibinfo {year} {2025})},\ \Eprint
  {https://arxiv.org/abs/2308.12336} {arXiv:2308.12336 [hep-ph]} \BibitemShut
  {NoStop}%
\end{thebibliography}%

\end{document}